\documentclass[reprint, 10pt,english,aps,notitlepage,onecolumn,groupedaddress,shortbibliography,nofootinbib,pra,floatfix]{revtex4-2}
\usepackage{mathtools,amsthm,amsmath,amsfonts,amssymb}
\usepackage{graphicx}
\usepackage{physics}
\usepackage{comment}
\usepackage[colorlinks=true, linkcolor=blue, urlcolor=blue, citecolor=blue, anchorcolor=blue,hypertexnames=false]{hyperref}
\usepackage{pgfplots}
    \usetikzlibrary{positioning}
\usepackage[normalem]{ulem}
\usepackage[noend]{algpseudocode}
\usepackage[dvipsnames]{xcolor}
\usepackage{orcidlink}
\usepackage{subcaption}
\usepackage{tikz}
    \usetikzlibrary{math}
\usepackage{parskip}
\usepackage[newcommands]{ragged2e}

\def\be{\begin{equation}}
\def\ee{\end{equation}}

\def\sJ{\mathcal{J}}

\def\bea{\begin{eqnarray}}
\def\eea{\end{eqnarray}}

\def\qq{\mathfrak{q}}

\def\OO{\mathcal{O}}
\def\tg{\tilde{\gamma}}
\def\ta{\tilde{\alpha}}
\def\td{\tilde{\delta}}
\def\RR{\mathcal{R}}
\def\kf{k_{f}}
\def\kb{k_{b}}
\def\R{\mathbf{R}}
\def\thetami{\theta_{-}}
\def\thetapl{\theta_{+}}
\def\nc{n_{c}}

\DeclareMathOperator{\tF}{\tilde{\mathcal{F}}}
\DeclareMathOperator{\mF}{\mathcal{F}}
\DeclareMathOperator{\lT}{\bf{l}}
\DeclareMathOperator{\ftau}{\tau}
\DeclareMathOperator{\ddif}{\delta}

\pgfplotsset{compat=1.18}

\newcommand{\ictpsaifradd}{ICTP South American Institute for Fundamental Research \\
                           Instituto de F\'{i}sica Te\'{o}rica, UNESP - Univ. Estadual Paulista \\
                           Rua Dr. Bento Teobaldo Ferraz 271, 01140-070, S\~{a}o Paulo, SP, Brazil}
\begin{document}
\title{{Krylov Complexity from  Loschmidt Amplitude}}
\author{Debarghya Chakraborty\,\orcidlink{0009-0000-1078-5348}}
    \email{debarghya.chakraborty@ictp-saifr.org}
    \affiliation{\ictpsaifradd}
\begin{abstract}
Krylov complexity is a powerful diagnostic of quantum dynamics, with clear connections to other measures of quantum chaos and operator growth. One such measure is the Loschmidt amplitude, defined as the overlap of initially identical states evolved under two slightly different Hamiltonians. Its decay in certain systems is controlled by the classical Lyapunov exponent. Using the algebraic properties of the Krylov complexity operator, we express Krylov complexity as the derivative of a Loschmidt amplitude whose perturbation is parameterized by an angular variable $\phi$. This formulation allows us to define a spectral propagator that encodes the entire complexity distribution, which we characterize for specific types of systems. We study the two-dimensional quantum geometry spanned by time and $\phi$ where the original and deformed trajectories reside, demonstrating that Krylov complexity is upper-bounded by its volume. We also express the time derivative of Krylov complexity in terms of a distinct Loschmidt amplitude. Depending on the growth of the Lanczos coefficients, the perturbation term in this amplitude can be truncated. We propose that the strength of this perturbation provides a classification scheme for Krylov complexity dynamics and relate it to the $\phi$-derivative of the spectral propagator. Using this analytical framework, we derive general relations between the time-dependence of the survival amplitude and Krylov space measures.
\end{abstract}
\maketitle

\section{Introduction}
\label{sec:introduction}
There has recently been a lot of interest in understanding quantum dynamics as it unfolds in the orthogonal basis defined by the Lanczos recursion method \cite{Parker_2019} (see \cite{rabinovici2025krylovcomplexity, Nandy_2025_rev} for reviews). The merits of this approach are easy to state. Restricting to the Krylov space means only tracking portions of the Hilbert space available to the evolving state $\ket{\psi(t)}$. Constructing the low-order elements of this special Krylov basis is simple using the Taylor coefficients of an autocorrelation function. In this basis, the Hamiltonian only causes nearest-neighbor hopping and that leads to the two following features. For suitable initial states and Hamiltonians, this basis is ordered by a notion of locality. Additionally, there is a clear connection between the index labeling the orthogonal basis and physical energy/time scales. 

Operationally, this approach is so powerful because the entire dynamics is packaged in a concise but abstract form.  The only input required is an autocorrelation function, or its Fourier transform, which we call the spectral function\footnote{It can be better to talk in terms of the spectral function because the asymptotics of the spectral function translate more directly to the behavior of Lanczos coefficients and Krylov complexity.}. Krylov complexity is defined as the mean position of $\ket{\psi(t)}$ in this orthonormal basis. It provides a clear definition of delocalization of the wavefunction in the Hilbert space. The flip-side is that all the intrinsic information contained in Krylov space measures like Krylov complexity can only reconstruct the system up to the autocorrelation function. This means that two very different physical systems with identical autocorrelation functions will have identical expansions in the Krylov basis. Additional model-specific input is needed to translate features of Krylov basis expansion back to the language of the Hilbert space with a local tensor product structure. Examples of contexts where this translation has been done include \cite{Dymarsky_2021,Jian_2021,Caputa_2021,Kar_2022,Kim_2022, Alishahiha_2023,Bhattacharjee_2022,Caputa_2023,Erdmenger_2023, Aguilar-Gutierrez:2024nau, Ambrosini:2024sre,Bento_2024,Heller_2025,Nandy_2025,Aguilar_Gutierrez_2025,Imani_2025,grabarits2026universalgrowthkrylovcomplexity,graef2026universalityprobecomplexitymathcaln4,baume2026controlledchaos4dscfts}.

\cite{Parker_2019} accomplished this by studying operator dynamics in $k$-local many-body systems. Translated to our language using the precise dictionary set up in Section~\ref{sec:setup_and_general_result}, $\ket{\psi(t)}$ describes a time-evolving operator in the Hilbert space of operators. Its time-evolution generator, known as the Liouvillian, is the adjoint action of the original system's Hamiltonian. Each application of the Liouvillian can grow the operator size by at most $k$, stratifying the Krylov basis by operator size. Using the connections between operator size and the out-of-time-ordered correlator (OTOC), \cite{Parker_2019} proved that at infinite temperature the growth rate of the OTOC is bounded by that of Krylov complexity. \cite{Parker_2019} conjectured a bound---formulated in terms of the autocorrelation function---sharpening the Maldacena-Shenker-Stanford bound at finite temperature \cite{Maldacena_2016}, whose statement was subsequently refined by \cite{Avdoshkin_2020,Avdoshkin_2024}. The exponential growth of the OTOC is known, in certain systems, to be the quantum analog of classical Lyapunov growth governing the divergence of nearby trajectories.   

Meanwhile \cite{Balasubramanian_2022} emphasized the special role of this Krylov basis when it comes to minimizing the spread of the wavefunction. They constructed a wide class of cost functions that measure the spread over an orthonormal basis. The minimization over all possible bases is achieved with the Krylov basis---a result that is completely general. Krylov complexity is the simplest among these ``complexities'', and was dubbed spread complexity in \cite{Balasubramanian_2022} as opposed to the term Krylov complexity, first used for operator dynamics. We stick to calling it Krylov complexity and work with a formalism that applies broadly. We are especially interested in the consequences of defining Krylov complexity using an operator with integer eigenvalues that is designed to annihilate $\ket{\psi(0)}$. 

\begin{figure}[htbp]
\centering
\includegraphics[width=0.89\linewidth]{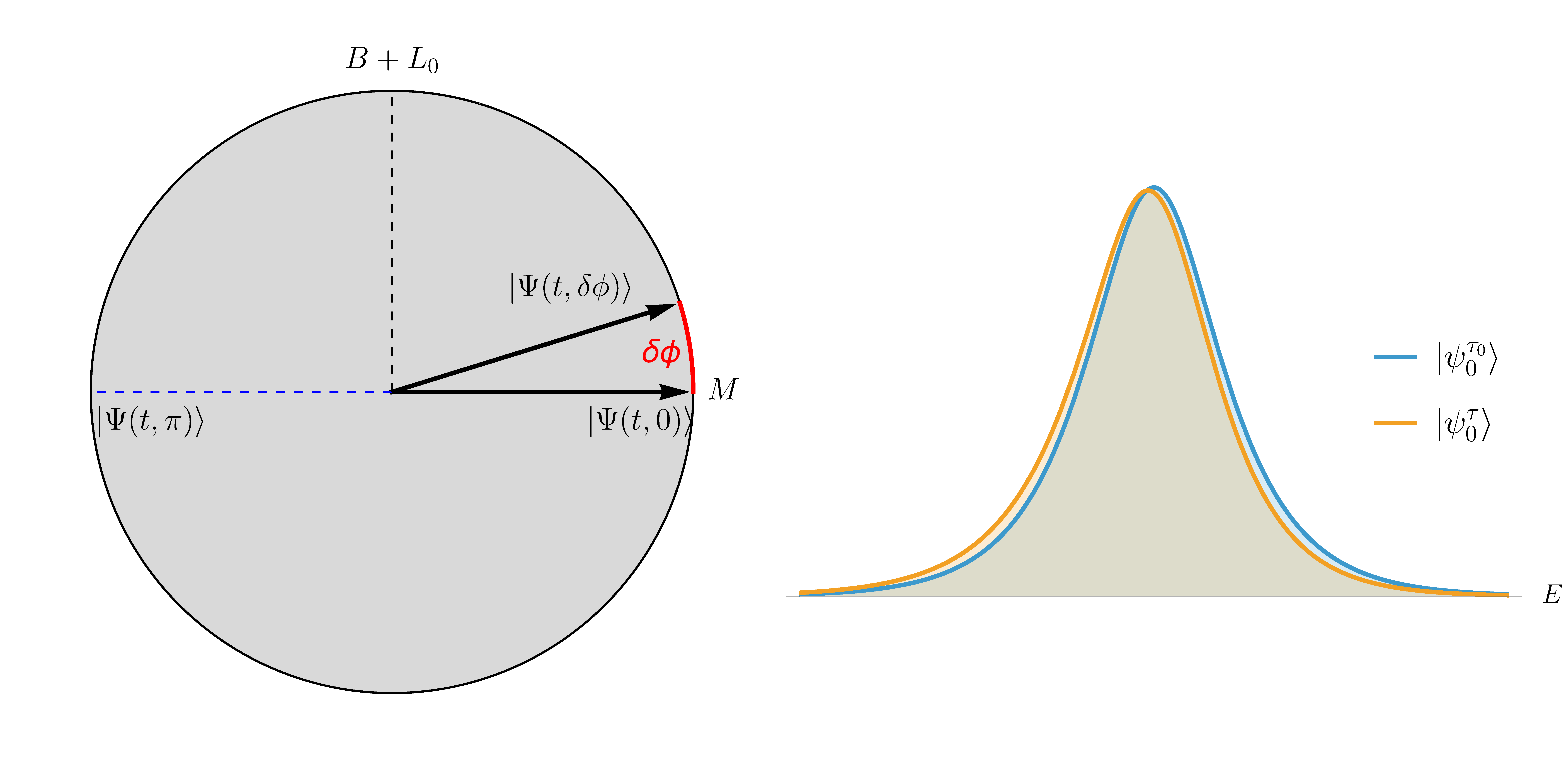}
   \captionsetup{justification=Justified}
\caption{\textbf{Visual guide}: (Left) Krylov complexity is the rate of change in overlap between two trajectories that move apart in time. The horizontal arrow depicts the physical trajectory and the slanted one describes evolution with the $\delta\phi$-deformed Hamiltonian. At $\phi=\pi/2$, the $\phi$-evolved Hamiltonian becomes $B+L_0$. Complexity is bounded by the volume of this disc at time $t$ with a canonical choice of volume element. When the autocorrelation is real, the relation $\ket{\Psi(t, \pi)}=\ket{\Psi(-t, 0)}$ leads to stronger results. (Right) $B$ generates rotations of the Krylov basis when using infinitesimally Euclidean-evolved $\ket{\psi^{\tau}_0} = e^{-(\tau-\tau_0) H} \ket{\psi^{\tau_0}_0}$ as the initial state for the Lanczos algorithm. This allows rewriting the time-derivative of complexity as a Loschmidt amplitude in terms of $\tau$.}
     \label{fig:deformations}
\end{figure}

This special choice of the complexity operator implies the picture in Fig.~\ref{fig:deformations}, which is the subject of Section \ref{sec:Loschmidt}. Krylov complexity at time $t$ is given by the leading-order difference in an amplitude where the initial state is evolved forward with the Hamiltonian of interest and then backwards with a slightly deformed Hamiltonian. Alternatively, Krylov complexity measures the sensitivity of a state's time evolution when the Hamiltonian is kicked in a \textit{specific} direction. Thus Krylov complexity is the rate of change of a Loschmidt amplitude \cite{Peres_1984, Wisniacki_2012, Gorin_2006}. Loschmidt echoes are defined as the modulus squared of Loschmidt amplitudes. Long studied as a measure of stability in quantum dynamics, the decay rate of Loschmidt echoes can be related to the classical Lyapunov exponent via semiclassical techniques \cite{Jalabert_2001, Jacquod_2001,Jacquod_2002}. The particular deformation we study is highly nonlocal, and defined naturally only in the Krylov basis, using an angular variable $\phi$. This angle quantifies how much the spectrum of the deformed Hamiltonian spreads relative to the undeformed one. The rate of this spread is determined by the behavior at $\phi=\frac{\pi}{2}$, the point of maximal delocalization.

There is an operator closely related to the Hamiltonian at $\phi=\frac{\pi}{2}$, which plays a key role in the analysis. It has the following features: (1) It plays the role of a ``momentum'' operator, conjugate to $H$ under the $\phi$-evolution. (2) Its expectation value in $\ket{\psi(t)}$ measures the time derivative of Krylov complexity. (3) It generates the rotation of the Krylov basis constructed using infinitesimally Euclidean-evolved $e^{-\tau H} \ket{\psi(0)}$ as the base for recursion \cite{Dymarsky_Toda}. By combining these ingredients, we rewrite the time-derivative of Krylov complexity in terms of a \textit{distinct} Loschmidt amplitude with a deformation in Euclidean time $\tau$. Though still nonlocal, depending on the growth rate of the Lanczos coefficients, the $\tau$-perturbation could decay, grow or asymptote to a constant. This provides a powerful classification of Krylov complexity according to the strength of the operator sourcing the perturbation at fixed $\tau$. For Lanczos coefficients growing slowly enough---where complexity is expected to grow no faster than quadratically---we show that this perturbation can be well-approximated up to a predictable timescale, by a finite-rank truncation. The classification of the perturbation controlling Krylov complexity resembles the use of adiabatic-gauge potential norm scaling to detect quantum chaos \cite{Pandey_2020}. 

We show a complementary picture of this classification in the energy basis. For the class of perturbations that decay rapidly enough, we are able to impose stringent self-consistency conditions. Using them, we derive a completely general bound for Krylov complexity in terms of the autocorrelation function, with only minimal assumptions about the growth of Lanczos coefficients. We also leverage this analysis to compute complexity. We set up a general formalism in Appendix \ref{sec:Poisson_Kernel} based on a spectral propagator, for computing all moments of the complexity operator in the state $\ket{\psi(t)}$. We write an approximate form of this propagator using the classical results of \cite{szego1959orthogonal}, valid for a generic spectral function strictly vanishing outside an interval. This approximation is expected to be accurate for predicting the late-time behavior of the complexity distribution. In the regime where complexity grows linearly with time, we write a universal formula that yields the leading order growth rate of all the complexity moments directly from the spectral function in Section \ref{sec:universality_complexity_dist}. With this approach, we are able to directly relate the properties of the Krylov distribution with those of the spectral function. 

Section \ref{sec:complexity_geometry} is devoted to understanding the space on which Fig.~\ref{fig:deformations} is set. The physical trajectory is the line at $\phi=0$ and $t$ running radially outward. We organize the nearby states whose overlap gives the Loschmidt amplitude into a geometry equipped with a Riemannian metric to solidify the connection between Krylov complexity and the picture of separating trajectories. We do this using the metric induced from the Fubini-Study metric, and generalize the ``complexity geometries'' of \cite{Caputa:complexity_geometry} to arbitrary systems. The distance between two $\phi$-slices at fixed $t$ is determined by the variance of the Krylov complexity operator. The volume of this two-dimensional geometry up to a fixed $t$-slice places an upper bound on Krylov complexity. This bound is saturated when the algebra between the Krylov complexity operator, the Hamiltonian and the ``momentum'' operator closes. These cases always give rise to spaces of constant curvature; however, tuning parameters at fixed curvature can still yield qualitatively different dynamics. 

Using this geometric picture, we derive a rigorous bound relating the decay of the autocorrelation function and Krylov variance, for autocorrelation functions with vanishing imaginary part. We obtain stronger results under simple, plausible assumptions about correlations between nearby $\phi$-deformed states. This includes a stronger bound relating the Krylov variance to the decay of the real autocorrelation function, and a concrete relation between the inverse-participation ratio and the Krylov variance, that holds more generally. The important conceptual takeaway from Section \ref{sec:complexity_geometry} is that the $\phi$ and $t$ directions of the geometry are highly constrained by one another.    

$\tau$ and $\phi$ described above are auxiliary parameters. The Loschmidt amplitude could simply be viewed as the generating function for complexity, and the latter has already been studied using various techniques, see for example \cite{M_ck_2022,Murugan:2026yyu,Bhattacharjee_2023,Aguilar-Gutierrez:2025pqp}. We point out that \cite{Caputa:complexity_geometry} identified Krylov complexity as a measure of sensitivity for displacements along the same $\phi$-direction, although their discussion relied extensively on the Lie algebraic structure, with $\phi$ appearing as part of the label for a generalized coherent state. 

The goal of the present work is to understand the physical meaning of the kicked trajectory, and the ensemble of adjacent states in general. A key question is how the dynamics generated by the deformed Hamiltonian could be realized in a physical system. One motivation comes from \cite{Chakraborty_2025}, where the Krylov complexity operator was approximately embedded into a simple bilinear operator to construct many-body scars in two copies of the SYK model, building on \cite{Lin_2022,Rabinovici:2023yex,Xu:2024gfm,Aguilar-Gutierrez:2025pqp}. In that case, the $\phi$ direction gets a physical realization as a dimensionless time. 

In other cases where Krylov complexity grows exponentially $\sim e^{\lambda_{K} t}$, its growth rate \textit{could} encode a physical timescale for chaos. The clearest example is operator growth in SYK, where $\lambda_{K}$ and the Lyapunov exponent given by the OTOC coincide in some regimes. Our formulation of Krylov complexity as sensitivity of nearby trajectories makes the Krylov exponent $\lambda_K$ reminiscent of the definition of a classical Lyapunov exponent: 
\be \lambda_{K} =  \lim_{\delta \epsilon \rightarrow 0} \frac{1}{t} \log \abs{ \frac{\delta G(t, \epsilon)}{\delta \epsilon} },  \label{eq:Lambda_k_as_Lyapunov} \ee 
taking $t$ sufficiently large and identifying $\abs{\delta G(t, \epsilon)}$ as the quantum-mechanical analog of the norm of a separation vector. $\epsilon$ could be either $\phi$ or $\tau$ and $G(t, \epsilon)$ is the amplitude whose derivative is taken. If we assume that Krylov variance also grows as $\sim e^{2\lambda_{K} t}$, then we can interpret $\abs{\delta G(t, \epsilon)}$ another way to recover $\lambda_{K}$. We can take $\abs{\delta G(t, \epsilon)}$ to be a true distance in the geometry of Section \ref{sec:complexity_geometry} between two $\phi$-slices at fixed $t$. Given that Krylov complexity always measures spread in Hilbert space, and that it sometimes measures scrambling, we expect that there are some physical contexts where \eqref{eq:Lambda_k_as_Lyapunov} defines a true chaos exponent. Identifying them and establishing the precise connection with chaos requires a careful consideration of the physical content of the autocorrelation function and is beyond the scope of this paper. We refer to conceptually similar discussions emerging from various approaches in \cite{Bueno:2019ajd,Lin_2020,Murugan:2026yyu,balasubramanian2025variationsthemekrylov}.      

In Section~\ref{sec:setup_and_general_result}, we begin by reviewing the Lanczos method and introduce the main technical ingredients. We make extensive use of specific analytical models that are elaborated on in the Appendices. The most useful example appears in the context of the double-scaled SYK model (DSSYK) (see Appendix \ref{sec:dssyk}), where the Lanczos coefficients are a function of $\qq$ between $0$ and $1$. Our primary interest lies in the fact that they specify a family of spectral functions with fixed variance, continuously interpolating between the semicircle distribution at $\qq=0$ and the normal distribution as $\qq \rightarrow 1^{-}$, thereby providing a useful setting for testing many of our general claims.  Our only direct contribution to the physics of the model consists of applying the results of Section \ref{sec:universality_complexity_dist} to DSSYK. In DSSYK, Krylov complexity has a holographic interpretation in terms of a geodesic length in the dual gravitational theory \cite{Lin_2022,Rabinovici:2023yex, Ambrosini:2024sre,Blommaert_2025, Heller_2025}. Our results give a prescription for computing the dominant late-time contributions to the quantum fluctuations of the length.\footnote{A bulk computation for the quantum fluctuations of the length is discussed in~\cite{Fu_2026}.} The implications of these findings for choosing a reference state whose Krylov complexity matches the bulk length will be discussed in upcoming work~\cite{dc_2026}.

Another important class of models is specified by the closure of ``complexity algebra'', covered in Appendix \ref{sec:complexity_algebra}. Despite their special algebraic structure, they appear as limiting cases for various physical systems, notably in the context of operator growth in conformal field theories.

\textbf{Note on v2:} In addition to improvements to the writing, we elaborate on the universal form of the late-time complexity distribution for bounded spectral function in Section \ref{sec:universality_complexity_dist}. All the ingredients for Section \ref{sec:universality_complexity_dist} were present in the previous version, but the answer was only computed for the case of constant Lanczos coefficients corresponding to the semicircle distribution. In the present version, we write the general result for a connected spectral function with support restricted to an interval, and analyze how the higher complexity moments depend on features of this spectral function.
\section{Setup}
\label{sec:setup_and_general_result}
In this work, we will use formalism which treats Krylov complexity for states and operators on an equal footing. This is because the spectral methods used here apply quite generally. Some of our results can be adapted to wider settings, beyond the context of quantum time evolution. In all cases, the fundamental object is an autocorrelation function, or equivalently, survival amplitude that we call $C(t)$ 
\be C(t) := \bra{\psi(0)}\ket{\psi(t)}.  \label{eq:surival_amplitude_def} \ee 
The choice of the initial state in \eqref{eq:surival_amplitude_def} amounts to a coordinate choice in the Krylov space that maps $\ket{\psi(0)}$ to the origin in Fig.~\ref{fig:deformations} and seeds the Lanczos recursion. It also supplies a spectral measure specified by $ \Phi(E)  dE$, where $\Phi(E)$ is the Fourier transform of $C(t)$:
\be  \Phi(E) = \frac{1}{2\pi} \int^{\infty}_{-\infty} e^{i E t} C(t) dt. \label{eq:def_spectral_function} \ee 
The time evolution of the state $\ket{\psi(t)}$ with a time-independent Hamiltonian $H$ is dictated by the Schr\"odinger's equation 
\be i \partial_{t} \ket{\psi(t)} = H \ket{\psi(t)}, \quad \ket{\psi(t)}= e^{-i H t}\ket{\psi(0)}, \label{eq:schrodingers_equation} \ee 
in units where $\hbar=1$. The analogous equation for operators is  
\be \mathcal{L}(H) O(t) := i \partial_{t} O(t)  = -[H, O(t)],~~~O(t) = e^{-i \mathcal{L}(H) t} O = e^{i H t} O e^{-i H t}. \label{eq:heisenberg_equation} \ee
where we defined the Liouville superoperator $\mathcal{L}(H) = -[ H, .]$ and showed how it generates operator time evolution as a consequence of Heisenberg equation of motion. $C(t)$ becomes a two-point function, or a linear combination thereof, when viewing $O(t)$ as a vector in the Hilbert space of operators. This is because for operators, there is a further need of specifying an inner product, unlike for states where the canonical choice is inherited from the standard inner product on $\mathbb C$. This ambiguity is related to the freedom of ordering operators within a two-point function, see Section IV. of \cite{Balasubramanian_2022} for a detailed discussion. The dependence of the inner product on their smooth parameter is governed by integrable dynamics related to the open Toda chain and generalizations \cite{Dymarsky_Toda, Angelinos:2025drf, 2026:K_complexity_under_Hamiltonian_deformations}. 

Having made a choice of operator inner product, it is convenient to map the time evolution of operators to that of states in a doubled-Hilbert space using the Choi-Jamiołkowski isomorphism. It is helpful to introduce the thermofield double state defined as 
\be \ket{\beta}  = \frac{1}{\sqrt{Z(\beta)}} e^{-\frac{\beta}{4}(H_L + H_R) } \ket{0} = \frac{1}{\sqrt{Z(\beta)}} \sum_n e^{-\frac{\beta}{2} E_n} \ket{E_n }\ket{\Theta E_n},~~~ Z(\beta) = \sum_n e^{-\beta E_n}. \label{eq:def_tfd} \ee 
where $H_L$ and $H_R$ are the Hamiltonian acting on the two subsystems $L$ and $R$, and $\ket{0}$ is a maximally entangled state across the two systems. $H_R = \Theta H_L \Theta^{-1}$, which means the left and right Hamiltonians are identical up to a choice of anti-unitary $\Theta$ which makes the isomorphism well-defined. On the doubled system, the Liouville superoperator gets mapped to the ``boost'' operator $\tilde{H} = H_L - H_R$ and different choices of inner product correspond to different $\ket{\psi(0)}$. For example, the Wightman ordered two-point function
\be C(t) = \frac{1}{Z(\beta)} \tr(e^{-\beta H/2} O(t) e^{-\beta H/2}  O ),  \label{eq:wightman_inn_prod} \ee 
is mapped to a survival amplitude of the form \eqref{eq:surival_amplitude_def} for 
\be \ket{\psi(0)} =  e^{-\frac{\beta}{8} (H_L + H_R)} O \ket{ \beta/ 2},~~~ \ee 
and $\ket{\psi(t)}$ satisfies \eqref{eq:schrodingers_equation} with $\tilde{H} = H_L - H_R$, on the doubled Hilbert space. Having set up the map between operators and states, we can talk about the Krylov basis for both in a unified way, and the operator inner-product will simply translate to extra conditions on $H$ and $\ket{\psi(0)}$.

The Lanczos algorithm is a method for recursively performing the Gram-Schmidt orthonormalization of $ \{ H^{k} \ket{\psi(0)} \}$, which yields an orthogonal basis of states. Following conventions, we call it the Krylov basis and denote it by $\{ \ket{\psi_n} \}$ and write its normalized version as $\{ \ket{n} \}$. It is constructed with the recursion    
\be \ket{\psi_{n+1}} = (H - a_n) \ket{\psi_n} - b^2_{n-1} \ket{\psi_{n-1}}, \label{eq:monic_Lanczos_H}  \ee 
or equivalently,
\be
H \ket{n} = b_{n-1} \ket{n-1} + a_{n} \ket{n} + b_{n} \ket{n+1}. \label{eq:orthonormal_Lanczos_H}
\ee 
with boundary conditions $b_{-1}=a_{-1} = 0$, $\ket{\psi_0} := \ket{\psi(0)}$ and $\ket{0}= (\bra{\psi_0}\ket{\psi_0} )^{-1/2} \ket{\psi_0}$. \eqref{eq:orthonormal_Lanczos_H} should be viewed as a recursion relation that allows one to compute $\ket{n+1}$ from knowledge of $\ket{n}$ and $\ket{n-1}$. The Lanczos coefficients $a_n$ and $b_n$ are fixed by orthogonality to be
\be b^2_n = \frac{ \braket{\psi_{n+1}}{\psi_{n+1}} }{\braket{\psi_n}{\psi_n}}, \quad a_n =  \frac{ \bra{\psi_{n}} H\ket{\psi_{n}} }{\braket{\psi_n}{\psi_n}}.  \label{eq:Lanczos_defining_relation} \ee 
The dimension of the Krylov space $N$ is determined by the number of non-degenerate eigenvectors appearing in the  decomposition of $\ket{\psi_0}$. For a finite dimensional $H$, $N$ is always finite.      

A direct, constructive approach to the Krylov basis comes from recognizing that the recurrence \eqref{eq:orthonormal_Lanczos_H} precisely defines a system of orthogonal polynomials of $H$. That means  
\be \ket{n} =  p_n(H) \ket{0},\quad \ket{\psi_n} = P_n(H) \ket{\psi_0}, \label{eq:Orthogonal_Polynomial_Krylov} \ee 
where $p_n(H)$ and $P_n(H)$ are both orthogonal polynomials of $H$ with degree $n$. $P_n(H)$ has leading coefficient $1$ i.e $P_n(H) = H^n + \ldots$, whereas $p_n(H)$ is orthonormal with the measure being induced from the relation $\bra{m} \ket{n} = \delta_{mn}$. The orthogonality measure is specified by the spectral function $\Phi(E)$, which is manifestly nonnegative in the energy eigenbasis:
\be \int \Phi(E) p_n(E) p_m(E) dE = \delta_{mn}, \label{eq:orthogonality_for_polynomials}  \ee 
where the integration range is implicitly given by the support of $\Phi(E)$. Combining \eqref{eq:Orthogonal_Polynomial_Krylov} and \eqref{eq:orthogonality_for_polynomials} we see that the combination $p_n(E) \bra{E}\ket{\psi_0}$ are matrix elements of the unitary implementing the change of basis between the Krylov basis and the energy eigenbasis. For finite $N$, $\Phi(E)$ is to be interpreted as a sum of Dirac delta functions at discrete locations. In the thermodynamic limit, the extent of $\Phi(E)$, especially whether it is supported only on a finite interval of $\mathbb R$, plays an important role and will feature later in our analysis. A useful observation is that $\ket{n}$ and the Lanczos coefficients through $a_n$ and $b_n$ are determined by the moments 
\be \mu_k = \int \Phi(E) E^{k} dE = i^k \frac{d^k}{dt^k} C(t) \Bigg|_{t=0}, \label{eq:moments_def} \ee
for $0 \leq k \leq 2n$. Combining \eqref{eq:heisenberg_equation}, \eqref{eq:moments_def} and \eqref{eq:wightman_inn_prod}, we see additional symmetries of operator time evolution with a symmetrized inner product like \eqref{eq:wightman_inn_prod}. The odd imaginary terms in the expansion of $C(t)$ around $t=0$ vanish, which implies that $\Phi(E)$ is even around $E=0$. It is then straightforward to show that each $\ket{\psi_n}$ will be either symmetric or antisymmetric in energy eigenbasis around $E=0$. From \eqref{eq:Lanczos_defining_relation}, this means $a_n=0$ for all $n$. In general, the even $\Phi(E)$ case is simpler and of particular interest, so we will  analyze cases with $a_n=0$ in greater detail.  

$\ket{\psi(t)}$ is expanded in the Krylov basis as follows:
\be \ket{\psi(t)} = \sum_n \varphi_n(t) \ket{n},  \label{eq:decomposing_psi_in_phi_basis} \ee 
with the coefficients $\varphi_n(t)=\bra{n}\ket{\psi(t)}$ satisfying a discrete Schr\"odinger equation 
\be i \partial_{t} \varphi_n(t) = b_{n-1} \varphi_{n-1}(t)+  a_n \varphi_n(t) +  b_n \varphi_{n+1}(t),  \ee 
with initial condition $\varphi_n(0) = \delta_{n0}$.
Krylov complexity is the mean position on the Krylov chain defined as 
\be   \bra{\psi(t)} \hat{n} \ket{\psi(t)}   = \sum_{n} n | \varphi_n(t)  |^2  \quad \text{where} \quad \hat{n}=\sum_n n \ket{n}\bra{n},\label{eq:K_complexity_definition} \ee 
is the number operator on the Krylov chain. It is a measure of ``complexity'' of dynamics in the sense, that at fixed $t$, a larger value of complexity means that $\ket{\psi(t)}$ has propagated deeper in the Krylov chain, exploring more of the available Hilbert space. By this logic, the choice of the weight proportional to $n$, is only a convention chosen out of convenience. Later we will exploit the special algebraic structure arising from this choice. We will often write complexity as $\expval{\hat{n}(t)}$ and write $\expval{\hat{n}^k(t)}:=\bra{\psi(t)}\hat{n}^k \ket{\psi(t)}$ for higher moments, using Heisenberg picture notation for brevity. For the same reason, we will use $\partial^k_{t} \expval{\hat{n}(t)}$. 
 
First, using \eqref{eq:orthonormal_Lanczos_H} we observe that in the Krylov basis, $H$ is tridiagonal.\footnote{In other words, there is an orthogonal transformation that maps $H$ to $M$. When the Krylov space dimension is smaller than the dimension of $H$, the orthogonal transformation maps $H$ to a block diagonal form, with the block inside Krylov space being $M$. } We call this tridiagonal matrix $M$. We choose to split $M$ 
\be M = L_{+} + L_{-} + L_{0}, \ee 
into a sum of generalized creation $L_{+}$, annihilation $L_{-}$ and a occupation number preserving operator $L_0$:
\be L_{+} \ket{n} = b_n \ket{n+1}, \quad L_{-} \ket{n} = b_{n-1} \ket{n-1}, \quad L_{0} \ket{n} = a_n \ket{n}, \ee 
 $\hat{n}$ commutes with $L_0$, and its adjoint action is simple on $L_{\pm}$
\be  [\hat{n}, L_{\pm}] = \pm L_{\pm}. \label{eq:adjoint_action_of_n_L} \ee 
\eqref{eq:adjoint_action_of_n_L} is identical to the commutation relation between the number operator and ladder operators of a harmonic oscillator. In this analogy, $M$ is like a position operator with $L_0$ acting as a shift in the axes. The ``momentum'' operator
\be B := i [\hat{n}, M] =  i ( L_{+} - L_{-} ), \label{eq:B_definition} \ee 
will also play an important role. Because \eqref{eq:B_definition} is antisymmetric, we get the two boundary conditions 
\be \expval{\hat{n}(0)} = 0,\quad \partial_{t} \expval{\hat{n}(t)} \big|_{t=0} = 0. \label{eq:boundary_conditions_n(t)} \ee
\textit{Note about regimes:}
In a system where $N \gg 1$ but finite, the Krylov basis and Lanczos coefficients show sharply distinct scaling regimes depending on $n$. The focus of this paper is the regime that holds for Krylov index up to $ O(\log(N))$ (i.e, polynomial in system size for local systems), that can be described solely by the coarse-grained spectral function. Here, Krylov complexity can grow rapidly, even exponentially. For both states and operators, this regime gets stretched out in the thermodynamic limit.  In contrast, there is a parametrically longer regime characterized by $O(1)$ values of $n/N$  with gradually decaying Lanczos coefficients that we call the``Lanczos descent'' following~\cite{Kar_2022}, which we refer to for more details. The ``Lanczos descent'' controls the late-time linear growth and its approach to the eventual saturation of Krylov complexity~\cite{Rabinovici_2021,Kar_2022, Balasubramanian:2023_plateau,Chakraborty2024}. While some of our discussion readily applies in that context, due to sharp differences in physical behavior we omit this regime from our analysis. The main technical difference can be encapsulated by stating that the polynomials in \eqref{eq:orthogonality_for_polynomials} must be treated as \textit{discrete} orthogonal polynomials on the lattice of energies. 

In contrast, in Section~\ref{sec:universality_complexity_dist} we directly address the complexity distribution for the Lanczos ``plateau'' regime---ubiquitous in finite systems—which can be viewed as the transition to the Lanczos ``descent''. This regime recovers the extensive energy bounds on $\Phi(E)$ while retaining its description as a coarse-grained continuous function. 
\section{Krylov Complexity and Loschmidt Amplitude}
\label{sec:Loschmidt}
\subsection{The $\phi$ and $\tau$ deformations}
Together~\eqref{eq:B_definition} and~\eqref{eq:adjoint_action_of_n_L} imply a very simple answer for the Heisenberg evolution of $M$ generated by $\hat{n}$
\be M(\phi) := e^{i \hat{n} \phi} M(0) e^{-i \hat{n} \phi} = M \cos(\phi) + B \sin(\phi) + L_0 (1-\cos( \phi) ),  \label{eq:def_Mphi} \ee
or that $M$ and $B$ evolve like position and momentum operators of a harmonic oscillator. The deformed Hamiltonian  $M(\phi)$ precisely defines the generating function for Krylov complexity distribution:
\begin{subequations}
\begin{align}
\expval{\hat{n}^k(t)} &= (-i)^k \frac{\partial^k}{\partial \phi^k} G(t, \phi) \Big|_{\phi=0}, \label{eq:Gt_generator_of_n_dist} \\
G(t, \phi) &:= \bra{0} e^{i M t} e^{- i M(\phi) t} \ket{0} = \sum_n \abs{\varphi_n(t)}^2 e^{i n \phi}.  \label{eq:Gt_def}
\end{align}
\end{subequations}
This brings us to our first conceptual point: Krylov complexity can be thought of as the leading phase response of a Loschmidt amplitude. The \textit{entire} complexity distribution follows from sufficient knowledge of the operator $M(\phi)$. This can be seen concretely through the time-ordered exponential representation:
\be G(t, \phi) = \bra{0} \mathcal{T} \exp \Big(-i \int^{t}_{0} e^{i M(0) s} \big( M(\phi)-M(0) \big) e^{-i M(0) s} ds   \Big) \ket{0}. \label{eq:gt_phi_interaction_picture} \ee
which implies that when $L_0 = 0$, the spectral representation of $B$ is sufficient to determine the entire distribution in \eqref{eq:Gt_generator_of_n_dist}. 
From \eqref{eq:def_Mphi}, we see that $M(\phi)$ is just the Hamiltonian written in a rotated basis. The fact that $M(\phi)$ takes a simple form in the Krylov basis is a special feature of the translation generated by $\hat{n}$. We can also consider the infinitesimal $M(\phi)= M(0)+  B \phi + O(\phi^2)$, sufficient to get 
\be  \expval{\hat{n}(t)} =  -i \partial_{\phi} \bra{0} e^{i M t} e^{- i ( M + B \phi) t} \ket{0} \Big|_{\phi=0}. \label{eq:linearized_Mphi}  \ee
If the amplitude $G(t,\phi)$, or its infinitesimal form were known, the $\phi$-derivative could be readily estimated. However, it is difficult to compute $G(t,\phi)$ directly from only knowing $M(\phi)$ in the Krylov basis, although in Appendix \ref{sec:refined_dyck} we describe the combinatorial method for computing the correlators $\bra{0} M(0)^{k_1} M(\phi)^{k_2}  \ket{0}$ that generalizes the computation of moments as a sum over weighted Dyck paths for $a_n=0$. $G(t,\phi)$ can also be expressed simply as $\bra{0} e^{i \hat{n}(t)} \ket{0}$ in terms of the Heisenberg evolved $\hat{n}(t)$, but this representation is impractical in general, though see, for example the path-integral approaches of~\cite{Murugan:2026yyu, Aguilar_Gutierrez_2025}. An important exception is the case of complexity algebra~\eqref{eq:g_tphi_for_Complexity_Algebra}, for which this representation gives a very convenient way of evaluating $\expval{\hat{n}^k(t)}$ for low orders.

Perhaps the most illuminating expression for $G(t, \phi)$ in general, is the following spectral representation: 
\be  G(t, \phi) = \int \int \sqrt{\Phi(E_1) \Phi(E_2)} e^{i (E_1-E_2) t} K(E_1, E_2, \phi)  dE_1 dE_2,\label{eq:G_from_E1_and_E2}  
\ee
in terms of the bilinear Poisson kernel $K(E_1, E_2, \phi)$ defined precisely in~\eqref{eq:reproducing_Poisson_Kernel}. $K(E_1, E_2, \phi)$ should be interpreted as the propagator for dynamics generated by $\hat{n}$ in the energy eigenbasis. Its Taylor coefficients around $\phi=0$ determine the moments of $\hat{n}$. The evenness of the spectral function around $E=0$ leads to additional identities relating $K(E_1, E_2, \phi)$  and the spectral representation of $B = M(\pi/2)$. In other words, $K(E_1, E_2, \frac{\pi}{2})$ determines the entire distribution $K(E_1, E_2, \phi)$.\footnote{This is also expected because the propagator $e^{-i t(  M \cos(\phi)  + B \sin(\phi) )}$ can be written as a path integral by time-slicing and inserting resolution of the identity in the energy basis.} Below we describe an alternate characterization of $B$, also applicable for $L_0 \neq 0$ which helps us better understand $M(\phi)$ for infinitesimal $\phi$ appearing in~\eqref{eq:linearized_Mphi}.  
To begin, we write the first two time-derivatives of $\expval{\hat{n}(t)}$ as expectation values evaluated in $\ket{\psi(t)}$ as follows:
\be \partial_{t} \expval{\hat{n}(t)} = -\bra{\psi(t)} B \ket{\psi(t)},\quad \partial^2_{t} \expval{\hat{n}(t)} =  \bra{\psi(t)} \mathcal{F} \ket{\psi(t)},\quad\mathcal{F} := i [B, M]   \label{eq:t_derivative_Kt}. \ee 
\eqref{eq:t_derivative_Kt} is useful because the operators appearing on the right hand side are simple in the Krylov basis. Because of the boundary conditions at $t=0$ \eqref{eq:boundary_conditions_n(t)}, either of the first two derivatives specifies $\expval{\hat{n}(t)}$ uniquely. $\mathcal{F}$ is symmetric and tridiagonal:
\be \mF_{n,n+1}= \mF_{n+1,n}= b_n (a_{n+1} - a_n) \quad \text{and} \quad \mF_{nn} = 2 (b^2_{n} - b^2_{n-1} ), \label{eq:b_com_m_Krylov_basis}  \ee 
where we remind the reader that $a_{-1}=b_{-1} =0$. We will shortly use the key observation that the entries of $\mF$ involve differences between consecutive Lanczos coefficients. Treating the Lanczos coefficients as smooth functions of $n$, we can approximate the coarse-grained behavior of the entries of $\mF$ using first derivatives. 

The operators on the right-hand side of~\eqref{eq:t_derivative_Kt} are closely related to the imaginary time-deformations of the seed $\ket{\psi_0}$. We assume a discrete, non-degenerate energy spectrum for clarity. The Krylov space is invariant under the following transformation of the initial state 
\be \ket{\psi_0} \rightarrow \ket{\psi^{\ftau}_0} = \frac{1}{\sqrt{C(-i2\ftau)} } e^{- \ftau H} \ket{\psi_0}. \label{eq:ftau_deformation_def}\ee 
Performing the Lanczos algorithm with the $\ftau$-dependent family of initial states $\ket{\psi^{\ftau}_0}$ gives a smooth family of Krylov bases spanning the same space. The action of $H$ on the rotating basis is specified by the $\ftau$-dependent tridiagonal matrix $M(\ftau)$, whose $\ftau$ dependence is governed by the integrable Toda equations. The Toda equations can be written in matrix form \cite{Dymarsky_Toda}
\be  \frac{d M(\ftau) }{d \ftau} = i [ B(\ftau), M(\ftau) ] \label{eq:dM_dtau}. \ee 
\eqref{eq:dM_dtau} follows from the more fundamental relations
\begin{subequations}
\begin{align}
 B(\ftau) &= i \frac{d \OO(\ftau) }{d \ftau} \OO^{T}(\ftau), \label{eq:B_generator_of_spectral_flow} \\ 
  M(\ftau)&= \OO(\ftau) \Lambda \OO^{T} (\ftau),\label{eq:M_tau}
\end{align}
\end{subequations}
where $\Lambda$ is the diagonal matrix of eigenvalues of $H$\footnote{By construction, only those eigenvalues appear for which $\bra{E} \ket{\psi_0} \neq 0$. Each degenerate eigenvalue appears only once.} and $\OO(\ftau)$ is the orthogonal matrix connecting the Krylov basis to the energy eigenbasis. \eqref{eq:B_generator_of_spectral_flow} can be derived straightforwardly on plugging in the definition of the deformation \eqref{eq:ftau_deformation_def} within \eqref{eq:def_spectral_function}, using this deformed spectral function together with $\tau$-dependent orthogonal polynomials in \eqref{eq:orthogonality_for_polynomials}, and finally, taking $\ftau$ derivatives.  $B(\ftau)$ is tridiagonal and antisymmetric because $M(\ftau)$ traces an orbit which is the intersection of the orbit under the actions of the orthogonal group and the group of upper-triangular matrices with positive diagonals. $B(\ftau)$ is invariant under $\OO(\ftau) \rightarrow \OO(\ftau) P(\ftau_0)$ for constant orthogonal $P(\ftau_0)$\footnote{ \eqref{eq:B_generator_of_spectral_flow} also implies that $M(\ftau) = Q(\tau, \tau_0) M(\tau_0) Q(\tau, \tau_0)^{T}$ where $Q(\tau, \tau_0)= \OO(\tau) \OO^{T}(\tau_0)$. \cite{Dymarsky_Toda} studied $\tau$ deformations $M(\tau)$ of a constant $M(\tau_0)$ defined this way, also obeying \eqref{eq:dM_dtau} }. \eqref{eq:dM_dtau} features the adjoint action of $B(\ftau)$ and is like a Heisenberg equation for the $\tau$ evolution of $M(\ftau)$. To get analogous equations where $B$ generates $\ftau$ evolution by multiplication, we consider an arbitrary $\ket{\psi_{K}}$ defined in the $\tau$-dependent Krylov space, but transformed back into the energy eigenbasis
\begin{subequations}
\begin{align}
    \frac{d}{d\ftau} ( \OO^{T}(\tau) \ket{\psi_K}  ) =  i  \tilde{B}(\ftau) \OO^{T}(\tau) \ket{\psi_K} + \OO^{T}(\tau) \partial_{\ftau} \ket{\psi_K}, & \label{eq:B_tilde_as_connection} \\ 
 \quad \tilde{B}(\ftau)  := \OO^{T}(\ftau) B(\ftau) \OO(\ftau) = i \OO^{T}(\ftau) \partial_{\tau} \OO(\ftau), &  \label{eq:B_tilde_def}
    \end{align}
\end{subequations}
where $\tilde{B}$ is the operator $B$ written in the energy eigenbasis. $\tilde{B}$ can be thought of as a gauge connection defining the covariant derivative $D_{\ftau} :=\big( \frac{d}{d\ftau} - i  \tilde{B}(\ftau) \big)$ through \eqref{eq:B_tilde_as_connection}, such that all the $\tau$-dependent Krylov basis vectors are parallel transported. The connection $\tilde{B}$ is also known as the adiabatic gauge potential \cite{Kolodrubetz_2017,Pandey_2020}. The relation with the $\phi$-deformations discussed earlier can be seen by evaluating the parallel transport equation $D_{\tau} \ket{\psi^{\ftau}_0}=0$ using the definition \eqref{eq:ftau_deformation_def}
\be  \frac{d}{d\ftau}  \ket{\psi^{\ftau}_0} =  i \tilde{B} \ket{\psi^{\ftau}_0} = i \OO^{T} B \ket{0} =  -(\Lambda- \bra{\psi^{\ftau}_0} H \ket{\psi^{\ftau}_0} ) \ket{\psi^{\ftau}_0}. \label{eq:B_tilde_on_psi0}  \ee 
The $\tau$-dependent frames were introduced so that we could better analyze the spectral representation of $G(t,\phi)$:
\be  G(t, \phi) = \bra{\psi_0} e^{i \Lambda t} e^{-i (\Lambda  + \phi \tilde{B} t )}  \ket{\psi_0} = 1 - i \phi \int^{t}_{0}  \bra{\psi_0} e^{i \Lambda s} \tilde{B}(0) e^{- i \Lambda s} \ket{\psi_0} ds  + O(\phi^2). \label{eq:gt_spectral_linear_phi}\ee
\eqref{eq:B_tilde_on_psi0} only gives the action of $\tilde{B}$ on $e^{-i\Lambda s}\ket{\psi_0}$ at $s=0$, and does not simplify \eqref{eq:gt_spectral_linear_phi} for $s>0$. $\tilde{B} e^{-i\Lambda s}\ket{\psi_0}$ is only determined through the spectral representation \eqref{eq:Orthogonal_Polynomial_Krylov} at fixed $\tau$, or if $\OO(\tau)$ was known for infinitesimally small $\tau$. Assuming small  $\phi$ and $\ftau$, we get the relation between flows
\be   \partial_{t} \partial_{\phi} G(t, \phi) =   \bra{0} e^{i M(0) t} \partial_{\ftau} e^{- i M(\ftau) t} \ket{0}, \label{eq:dtau_dphi} \ee
\eqref{eq:dtau_dphi} is to be contrasted with \eqref{eq:gtphi_as_euclidean_ev}. Along with \eqref{eq:def_Mphi}, \eqref{eq:dM_dtau} it reveals fascinating connections between the $\tau$ flow, and the $\phi$-flow defined for fixed $\tau$. This brings us to the key observation that \eqref{eq:dtau_dphi} allows the time derivative of Krylov complexity to be rewritten as a Loschmidt amplitude with a $\tau$-dependent perturbation
\be  \partial_{t} \expval{\hat{n}(t)} = i \partial_{\ftau} \bra{0} e^{i M(0) t} e^{- i M(\ftau) t} \ket{0} \Big|_{\ftau=0} = i\partial_{\ftau} \bra{0} e^{i M(0) t} e^{-i(M(0) + \mF \ftau ) t} \ket{0} \Big|_{\ftau=0}  \label{eq:linearized_Kcomplexity_rate_Loschmidt}, \ee
where we used \eqref{eq:M_tau} to write $M(\ftau) = M(0) + i \mF \ftau + O(\ftau^2)$, reminding the readers that $\mF=i[B, M]$. This means that in principle, Toda solutions $M(\tau)$ can be used to obtain the time derivative of complexity. Beyond knowing the $\ftau$-dependent Lanczos coefficients, one needs a way to compute the associated orthogonal polynomials, or the overlaps $\varphi^{\ftau}_n(t) = \bra{n} e^{i M(\tau) t} \ket{0}$. In case $\varphi^{\ftau}_n$ are known exactly for the entire Toda orbit, this gives an alternate characterization of the rate of Krylov complexity. This amounts to knowing the expansion coefficients \eqref{eq:decomposing_psi_in_phi_basis} with $\ket{\psi(0)} = \ket{\psi^{\tau}_0}$ defined in \eqref{eq:ftau_deformation_def} such that    
\be  \partial_{t} \expval{\hat{n}(t)} \big|_{\tau=\tau_0} =  i \sum_{n} \varphi^{\tau_0}_{n}(-t) \partial_{\ftau} \varphi^{\ftau}_{n}(t) \big|_{\ftau = \ftau_0}. \label{eq:varphi_n_tau_derivative} \ee 
where we now introduced constant $\tau_0$ to allow for the possibility that the physical $\ket{\psi(0)}$ itself is defined as the Euclidean evolution of a reference state $\ket{\psi(0)} \sim e^{-\tau_0 H} \ket{\psi_{\text{ref}}}$. Precisely this happens to be the case for a physically very interesting family of solutions described in detail in Appendix \ref{sec:toda_orbit_example}. For this example, Lanczos coefficients, governed by $SL(2,R)$ algebra, are specified with \eqref{eq:sl2_lanczos} and \eqref{eq:alpha_gamma_sl2_toda}, and the wavefunctions \eqref{eq:sl2_toda_wavefunctions} are known explicitly. In \cite{caputa_proper_momentum}, they find an explicit realization in describing the autocorrelation function for the initial state
\be \ket{\psi(0)}  \sim e^{-\tau H} \mathcal{\hat{O}}(x_0) \ket{\beta}, \label{eq:sl_2_psi0} \ee
where $\ket{\beta}$ is the TFD \eqref{eq:def_tfd} and $H$ is the Hamiltonian of a 2d CFT. $\mathcal{\hat{O}}$ is a primary field of scaling dimension $\Delta=h$ in the CFT inserted at $x_0$. Here $\tau$ acts as a UV regulator giving finite energy $\propto \frac{1}{\tau}$ to the state \eqref{eq:sl_2_psi0}. \cite{caputa_proper_momentum} showed a match between $\partial_{t} \expval{\hat{n}(t)}$ and the proper radial momentum of the massive particle dual to $\mathcal{O}$ in the context of the AdS/CFT duality. The autocorrelation function for physical 2d CFT states dual to distinct AdS$_3$ geometries (or patches thereof) is found by replacing $\ket{\beta}$ in \eqref{eq:sl_2_psi0} by the appropriate generalization. They are all described by the $SL(2,R)$ Lanczos coefficients \eqref{eq:sl2_lanczos} with different parameters. $\tau_0$ corresponds to a radial cutoff (in different coordinate patches) for each of these geometries. It could be promising to study the proper radial momentum through the lens of \eqref{eq:varphi_n_tau_derivative}---as the correlation between Krylov wavefunctions when starting with two different initial states with slightly different radial cutoff. Taking $\tau_0= \frac{\beta}{4}$ recovers the results related to the SYK and 2d CFT primary 2pt function at inverse temperature $\beta$ \cite{Parker_2019, Dymarsky_Toda, Caputa:complexity_geometry}.
\subsection{Classifying Dynamics Through $\tau$ amplitude}
While it is not hard to get $\varphi^{\ftau}_n(t)$ for the first few $n$ using the moments, or $\ftau$-derivatives \cite{Dymarsky_Toda}, since $\varphi^{\ftau}_n(t)$ spread out with $t$, we need a method to reliably truncate in $n$ to truly leverage \eqref{eq:linearized_Kcomplexity_rate_Loschmidt}. For this we return to \eqref{eq:b_com_m_Krylov_basis}, along with the linearized form of $M(\tau)$ in \eqref{eq:linearized_Kcomplexity_rate_Loschmidt}. The perturbation $\mF$ is \textit{qualitatively} different depending on how its entries scale with $n$. They can be estimated by $n$-derivatives of $b^2_n$ and $a_n$. This has two related applications: (1) Classifying Krylov space dynamics in terms of Lanczos coefficients and (2) Characterizing scenarios where the complexity could be well-approximated by finite truncations of $\mF$ in \eqref{eq:linearized_Kcomplexity_rate_Loschmidt}. A finite rank perturbation could, in some instances, turn \eqref{eq:linearized_Kcomplexity_rate_Loschmidt} into a measurable amplitude. 

\begin{figure}[htbp]
     \centering
     \begin{subfigure}[b]{0.45\textwidth}
         \centering
         \includegraphics[width=\textwidth]{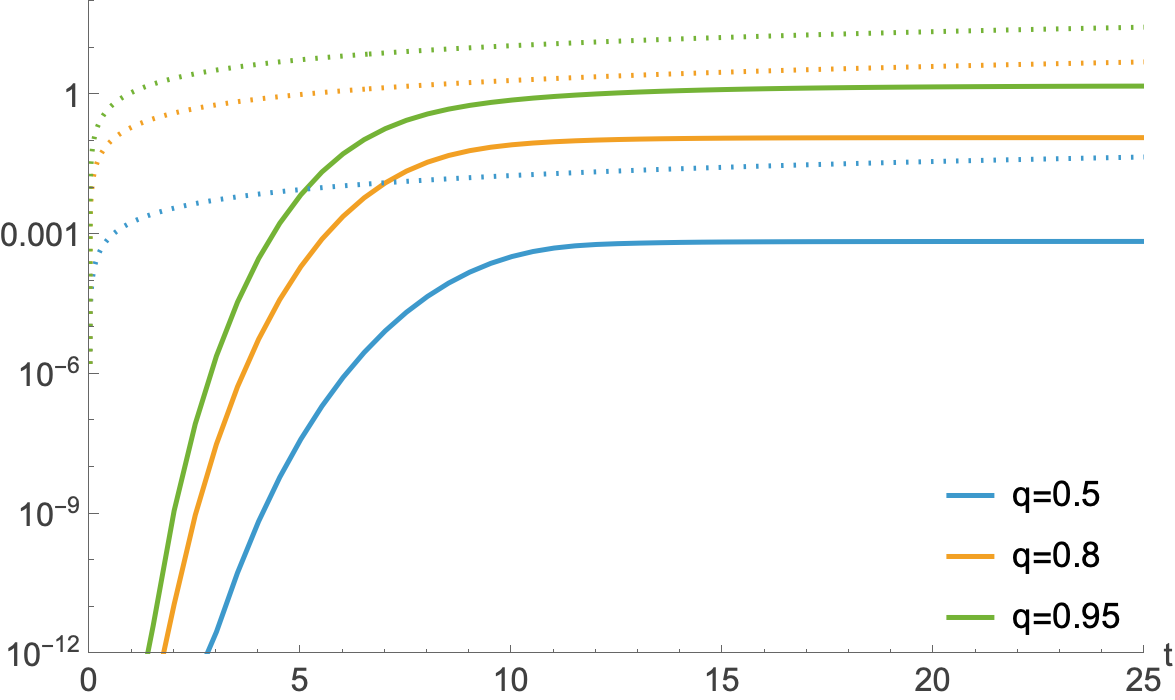}
         \caption{}
         \label{fig:Delta_Kdot}
     \end{subfigure}
     \hfill
     \begin{subfigure}[b]{0.45\textwidth}
         \centering
         \includegraphics[width=\textwidth]{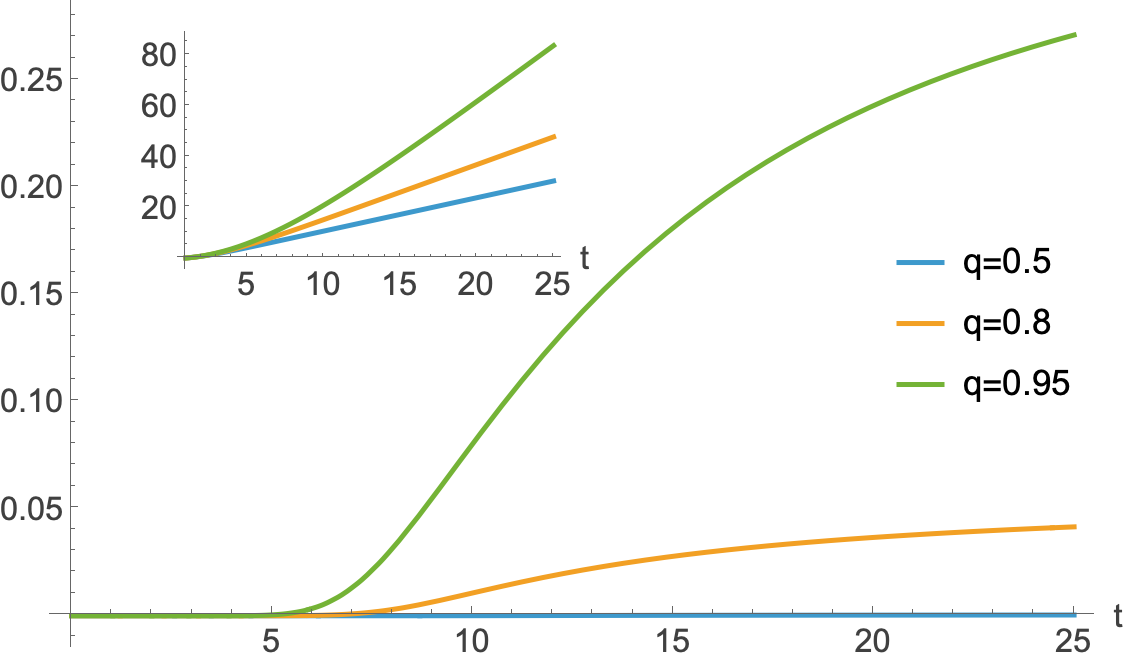}
         \caption{}
         \label{fig:normalized_Delta_K}
     \end{subfigure}
     \centering
      \captionsetup{justification=Justified}
\caption{\textbf{Approximating Krylov complexity from truncation} \eqref{eq:deformed_H_for_physical_amplitude} with $n_{*} =10$. \textbf{(a)} The difference $\ddif(n_{*}, t)$ computed using DSSYK Lanczos coefficients \eqref{eq:dssyk_Lanczos} for varying $\qq$. The dotted lines indicate the worst case error $\RR(n_{*}) t= e^{ n_{*} \log \qq} t$ for the corresponding $\qq$.  \textbf{(b)} The normalized difference $\frac{1}{\expval{\hat{n}(t)}} (\int^{t}_{0}  \ddif(n_{*}, s) ds) $, with the inset showing overall scale of $\expval{\hat{n}(t)}$ for reference. A finite $\tau=10^{-6}$ was used.}
     \label{fig:approximating_kt}
\end{figure}

We begin with when $B$ and $M$ satisfy the Heisenberg algebra of a harmonic oscillator (see Appendix \ref{sec:heisenberg_algebra}) such that $\mF = i[B, M] = 2\alpha^2 \mathbb I$. \eqref{eq:linearized_Kcomplexity_rate_Loschmidt} reduces to $\partial_{t}\expval{\hat{n}(t)} = 2\alpha^2 t$. Next, suppose that $a_n$ and $b_n$ asymptote to constants beyond some $n \sim n_{*}$. Then $\mF$ can be replaced by the truncation $\mF^{(n_{*})}$  with $\mF^{(n_{*})}_{pq} = 0$ for $p,q \geq n_{*}$, in \eqref{eq:linearized_Kcomplexity_rate_Loschmidt} and still exactly recover Krylov complexity. The same strategy works up to some timescale provided the Lanczos coefficients grow slowly enough. Then the time-derivative of complexity can be obtained by computing the amplitude on the rhs using a truncation like: 
\be 
\frac{i}{\tau}( \bra{\psi_0} e^{i H t} e^{-i H(\tau, n_{*})}  \ket{\psi_0} -1), \quad \text{where} \quad H(\tau, n_{*}) = H + 2\tau \sum^{n=n_{*}}_{n=0} (b^2_n - b^2_{n-1}) \ket{p_{n} (H) \psi_0}\bra{p_{n} (H) \psi_0}.  \label{eq:deformed_H_for_physical_amplitude}  \ee 
where we took small $\tau$, $a_n=0$ and translated $\mF^{(n_{*})}$ back into an operator on the Hilbert space. The general case just needs terms like $ b_n (a_{n+1}-a_n)  \ket{n} \bra{n+1}$ up to $n+1=n_{*}$ added to $H(\tau, n_{*})$. This is especially useful whenever  $H^k \ket{\psi_0}$ for $k < n_{*}$ are convenient to access experimentally, or in numerical simulations. The requirement that the entries of $\mF$ in Krylov basis are bounded turns \eqref{eq:deformed_H_for_physical_amplitude} into a controlled approximation. Suppose the $a_n$ vanish, but $b^2_n \sim n^{\kappa}$  for $0<\kappa < 1$, such that $\mF_{nn} \leq \RR(n_{*})$ with constant small $\RR(n_{*}) \sim (n_{*})^{\kappa-1}>0$ when $n > n_{*}$. Then the difference between exact and the truncated propagator amplitude, i.e $i \partial_{\tau} \expval{e^{iMt} ( e^{-i (M + \tau \mF) t} - e^{-i (M + \tau \mF^{(n_{*})}) t})}$ is given by 
\be \delta (n_{*}, t)  = \sum_{n > n_{*}} \mF_{nn} \int^{t}_{0}  \abs{\varphi_{n}(s)}^2 ds \leq \RR(n_{*})  \int^{t}_{0} \sum_{n > n_{*}} \abs{\varphi_{n}(s)}^2 ds    \leq \RR(n_{*}) t, \label{eq:correction_for_kt_der}   \ee 
where we neglect higher order corrections in $\tau$. \eqref{eq:correction_for_kt_der} shows that the local truncation \eqref{eq:deformed_H_for_physical_amplitude} will underestimate $\expval{\hat{n}(t)}$ by at most $\sim\RR(n_{*}) t^2$ where $\RR(n_{*})$ is the small parameter suppressing error. The lhs of \eqref{eq:correction_for_kt_der} shows that this is conservative, especially keeping in mind that for these cases the support of $|\varphi_n(t)|$ in $n$ is expected to grow slowly with $t$. The power-law decay of $\RR(n_{*})$ with $n_{*}$ was just an example. We show the robustness of \eqref{eq:deformed_H_for_physical_amplitude} for the DSSYK Lanczos coefficients \eqref{eq:dssyk_Lanczos} where $\RR(n_{*}) \sim e^{ n_{*} \log \qq}$. In Fig.~\ref{fig:approximating_kt}, $\RR(n_{*})$ is tuned by varying $\qq$ keeping $n_{*}$ fixed. The error in approximation $\delta (n_{*}, t)$ is much better than the linear in $t$ upper bound of \eqref{eq:correction_for_kt_der}.

The arguments can be repeated when the entries of $\mF$ decay modulo a multiple of identity, which can be trivially accounted for if the Lanczos coefficients are known. When $b_n$ grow faster than $\sqrt{n}$, this approach breaks down in general. Thus the growth of Krylov complexity is \textit{intrinsically tied} to how well it can be approximated by the survival amplitude \eqref{eq:deformed_H_for_physical_amplitude} with a finite cutoff $n_{*}$.
A similar analysis can be performed with nonzero $a_n$. The important feature to retain is that $\mF$  is a bounded operator with a finite maximal eigenvalue $\lambda^{\text{max}}_{\mF}$. Then the eigenvalues of $M + \tau \mF$ differ from those of $M$ by at most $\tau |\lambda^{\text{max}}_{\mF}|$.  $M + \tau \mF$ remains close to $M$ as an operator for small $\tau$, such that the Dyson series in $\tau$ converges. This is what underlies the existence of a bound of the form \eqref{eq:correction_for_kt_der}. There is no similar scale when $\mF$ is unbounded---as is the case for operator dynamics in chaotic systems. In fact, there is a double obstruction to \eqref{eq:deformed_H_for_physical_amplitude}: (1) the truncation in $n_{*}$ cannot be bounded as in \eqref{eq:correction_for_kt_der} and (2) there is no scale for $\tau$, below which the finite $\tau$ difference is guaranteed to be close to the $\tau$-derivative. \footnote{It might be worth exploring if there are any connections between the accuracy of truncation for approximating Krylov complexity, and the accuracy of truncation in Krylov space for approximating the adiabatic gauge potential as studied in \cite{Bhattacharjee:2023dik}.}     

Actually, for the sake of computing \eqref{eq:linearized_Kcomplexity_rate_Loschmidt}, the nontrivial part is the component $\mF^{\perp ,M}$ of $\mF= \mF^{\perp ,M}+\mF^{|| ,M} $ which lies outside the kernel of the adjoint action i.e  $[M, \mF^{|| ,M}]=0$. This is assuming that $\mF^{||,M}$ happens to be known simply in terms of $M$. This is seen with the complexity algebra example of Appendix \ref{sec:sl2r_algebra}, where $\mF= \tg M$. In this particular example, $\partial_{t} \expval{\hat{n}(t)} = i \partial_{\tau} C( \tg \tau t )|_{\tau=0}$.
The decay rate of the matrix elements of $\mF$ in the thermodynamic limit provides a more refined classification. $\mF$ is a compact self-adjoint operator as long as its matrix elements eventually decay at large $n$. This implies simplifications at the level of $\varphi^{\tau}_n(t)$ for small $\tau$ and large $n$. To draw useful conclusions, we move to the spectral representation  
\be \tF := i \OO^{T} [B, M] \OO  \label{eq:spectral_repdef_B_and_Bcom},  \ee
and by evaluating the commutators in energy basis we get, 
\be  \tF_{jk} = i (E_k - E_j)  \tilde{B}_{jk} =  -(E_k - E_j)^2  \tilde{n}_{jk},  \label{eq:spectral_relations_n_B_Bcom} \ee 
where $\tilde{n}_{jk} :=(\OO^{T} \hat{n} \OO)_{jk}$, and we phrased everything in the language of discrete energy spectrum for convenient presentation. Analogous statements \eqref{eq:B_from_phi_der}, \eqref{eq:F_from_B_kernel} hold for the continuous kernels related to $\partial_{\phi}K(E_1, E_2, \phi)$ defined in \eqref{eq:def_btilde_kernel}, \eqref{eq:def_ftilde_kernel}. Thus the decay of $\mF$ places strict bounds on the energy dependence of the matrix elements of $\hat{n}$.  

This is best seen first through the subfamily of $\mF$ with bounded Frobenius norm $||\mF||=\sqrt{\tr(\mF^{\dagger} \mF)}$. $||\mF||$ can be well-defined even for growing Lanczos coefficients, such as for $a_n=0, b^2_n \sim n^{\kappa}$ when $\kappa<\frac{1}{2}$. Since the Frobenius norm is unitarily invariant: $||\tF||=||\mF||$, whenever well-defined, it provides a sum-rule for the matrix elements $\tF_{jk}$. The existence of $||\mF||$ constrains the asymptotic growth of complexity through \eqref{eq:spectral_relations_n_B_Bcom}, as shown below in \eqref{eq:upper_bound_frobenius_tighter} and through the analysis of the spectral representation of $\hat{n}$ in Appendix \ref{sec:frobenius_norm_kernel}. 

To derive general relations without any assumptions on $\tilde{n}_{jk}$, we use the sum-rule for finite $|| \mF||$ within the following Cauchy-Schwarz inequality 
\be \abs{\sum_{jk} \tF_{jk} c^{*}_j c_k \lT(E_j, E_k, t) } \leq ||\mF|| \sqrt{ \sum_{j,k} |c_j|^2 |c_k|^2  |\lT(E_j, E_k, t)|^2}, \label{eq:vectorized_cauchy_schwarz} \ee 
where $c_k=\bra{E_k}\ket{\psi_0}$ and $\lT(E_j, E_k, t) $ is the matrix element-independent contribution governing the dynamics. The choice of $\lT$ dictates the order of $t$-derivative of $\expval{\hat{n}(t)}$ we work with. If the lhs of \eqref{eq:vectorized_cauchy_schwarz} stands for $|\partial^2_{t} \expval{\hat{n}(t)}|$, then $\lT(E_j, E_k, t) = e^{i(E_j - E_k) t}$, a pure phase. This gives a flat bound in $t$
\be \partial^2_{t} \expval{\hat{n}(t)} \leq ||\mF|| \implies \expval{\hat{n}(t)} \leq \frac{1}{2} ||\mF|| t^2. \ee 
The inequality for $\expval{\hat{n}(t)}$ can be made much tighter by directly substituting $\lT(E_j, E_k, t)= \int^{t}_{0} ds_1\int^{s_1}_{0} ds_2 e^{i(E_j - E_k) s_2} $ using \eqref{eq:spectral_relations_n_B_Bcom} in \eqref{eq:vectorized_cauchy_schwarz} and performing the sum over energies first:
\be \expval{\hat{n}(t)} \leq ||\mF|| \sqrt{ \int^{t}_{0}\int^{t}_{0} ds_2 ds_1  (t-s_1)  (t-s_2) |C(s_1-s_2)|^2 }. \label{eq:upper_bound_frobenius_tighter} \ee
The equation \eqref{eq:upper_bound_frobenius_tighter} makes no particular assumptions about Lanczos coefficients other than the existence of $||\mF||$, which could be crudely estimated from the moments. This makes \eqref{eq:upper_bound_frobenius_tighter} a widely applicable bound on $\expval{\hat{n}(t)}$ directly from $C(t)$. To estimate the late-time scaling for $C(t)$ that decay to $0$ as $t\rightarrow\infty$, it is sufficient to consider a sharply localized $C(t) \sim \delta(t)$. This gives the asymptotic $\expval{\hat{n}(t)} \lesssim O( ||\mF|| t^{\frac{3}{2}})$. The same upper-bound for time-dependence is also found through independent considerations in Appendix \ref{sec:frobenius_norm_kernel}.

In finite systems with well-defined $||\mF||$, a faster transient $t$-dependence in $\expval{\hat{n}(t)}$ is totally consistent as long as $||\mF||$ is large enough. This happens for local operator complexity in chaotic spin chains, where, for moderate sizes we find the rhs of \eqref{eq:upper_bound_frobenius_tighter} to be vastly greater, despite having a slower $t$ dependence. For short times, tighter bounds can be found by using $|\tF_{jk}| < \lambda^{\text{max}}_{\mF}$, but we did not find a simple way to relate it to decay of $C(t)$.  
\eqref{eq:upper_bound_frobenius_tighter} is useful as an asymptotic (in $t$) statement for sufficiently large $||\mF||$ and predictive for all $t$ provided $||\mF||$ is truly small (in units set by overall scale of Lanczos coefficients). An illustrative analytic example is found in DSSYK. $\expval{\hat{n}(t)}$ transitions from quadratic to asymptotically linear growth--see Section \ref{sec:universality_complexity_dist} for a discussion on time scales. For small $\qq$, \eqref{eq:upper_bound_frobenius_tighter} is found to be tight, in Fig.~\ref{fig:upper_bound} during the quadratic regime. \eqref{eq:upper_bound_frobenius_tighter} predicts $\expval{\hat{n}(t)}$ to be bounded by $\sim \frac{1}{1-\qq^2} t^{\frac{3}{2}}$ after the $O(1)$ time when $C(t)$ has decayed appreciably. Such a bound could be of greater use for cases where $C(t)$ is known with precision, in absence of detailed knowledge of the Lanczos coefficients.

When $N \rightarrow \infty$ and $\tF$ is compact, we can impose stronger consistency conditions on the singularities of $\tilde{n}_{jk}$. These can be translated to statements about the large-$t$ behavior of $\expval{\hat{n}(t)}$, as explored in Appendix \ref{sec:model_for_K}. Using \eqref{eq:spectral_relations_n_B_Bcom}, the matrix elements $\tilde{n}_{jk}$ and $\tilde{B}_{jk}$ can be related. Using \eqref{eq:gt_phi_interaction_picture}, these relations, in principle, can be combined to get the entire distribution $\expval{\hat{n}^k(t)}$ when $\Phi(E)$ is even. In practice, a factorized ansatz for $\tilde{n}$ as in \eqref{eq:leading_singularity_n_frequency_domain} leads to simplifications in the evaluation of the Dyson series to arbitrary order. This observation is especially useful when the low-frequency behavior of $\tF$, or $\tilde{n}$ can be constrained through some means, perhaps using orthogonal polynomial asymptotics. As an example, we work out an approximate late-time form of the complexity distribution in Appendix \ref{sec:model_for_K} for bounded $\Phi(E)$ i.e $\Phi(E)=0$ for $|E|> |E_0|$, using the asymptotics presented in Appendix \ref{sec:SzegoAsymptotics}. This derivation, valid for bounded and \textit{even} $\Phi(E)$, uses only the matrix elements of $\tilde{n}$  in \eqref{eq:Szego_ntilde} to extract $\tilde{B}$. Then, by localizing the nested integrals involving $\tilde{B}$ to their low-frequency contribution, we recover the result presented below in Section \ref{sec:universality_complexity_dist}, which uses more information from the asymptotic distribution \eqref{eq:Szego_Kernel}. This approach could be promising for wider instances of even $\Phi(E)$, if orthogonal polynomial asymptotics are combined with a coarse-grained form of $\mF$, along with the spectral relation \eqref{eq:spectral_relations_n_B_Bcom}.
\subsection{Universality for complexity distribution}
\label{sec:universality_complexity_dist}
Here we consider the case of bounded $\Phi(E)$, with only one connected component. The Lanczos coefficients asymptote to constants $a_n \rightarrow a, b_n \rightarrow b$. These constants are related to the support of $\Phi(E)$, which is fixed to be $[-2b+a, 2b+a]$, with the possibility of additional ``discrete'' states away from this continuum. Here $\mF$ has a well-defined trace in addition to a Frobenius norm, a fact that underpins the large-$n$ universality of the orthogonal polynomials elaborated in Appendix \ref{sec:SzegoAsymptotics}. In what follows, we take $a=0$ and $2b=E_0$, such that the support is $[-E_0, E_0]$.

Such measures are relevant for describing finite systems, say, a collection of qudits.\footnote{We assume $\sim \log N \gg 1$ qudits but restrict to sufficiently small index $n$ such that the Lanczos coefficients are still determined by the coarse-grained $\Phi(E)$. This means that we do not probe the Lanczos descent.} Such a system will have a finite bandwidth in energy, such that $E_0$ is extensive in the system size. From \eqref{eq:spectral_relations_n_B_Bcom}, we see that for such a system $\tilde{n}_{jk}$ will have at most an emergent second-order pole $(E_j-E_k)^{-2}$ from $E_j$ exponentially close to $E_k$. The second-order pole is responsible for the linear growth of $\expval{\hat{n}(t)}$ for $O(N)$ times, which can be matched onto the $N \rightarrow \infty$ asymptotics using \eqref{eq:Szego_ntilde}. The same result was derived in \cite{Alishahiha_2023} using a different method. The contributions in $\tilde{n}(E_1, E_2)$ that lead to parametrically faster complexity growth at shorter times get regularized by finite $N$. 

At sufficiently late times, $\expval{\hat{n}(t)}$ exhibits a linear growth regime that we have argued is universal for bounded $\Phi(E)$. In this regime, we expect
\be \expval{\hat{n}^{k}(t)} = v_{k} t^{k} + O(t^{k-1} ).  \label{eq:ballistic_scaling} \ee
We derive the leading $v_{k} t^{k}$ behavior, including an expression for the constants $v_k$, and then show the $O(t^{k-1})$ scaling of the remainder. We do this by obtaining an approximate form of the kernel $K(E_1,E_2, \phi)$ in \eqref{eq:Szego_Kernel}. Taking a late-time limit in the integration over energies in \eqref{eq:G_from_E1_and_E2} evaluated in Appendix \ref{sec:SzegoAsymptotics}, and selecting the terms where $\phi$ couples linearly to $t$ gives
\be G(t,\phi) \approx \int^{E_{0}}_{-E_0} \Phi(E) e^{i \sqrt{E^2_0-E^2} \phi t} dE,\label{eq:universal_Gt_in_Szego_class} \ee
which is sufficient to extract 
\be 
v_{k} = \int^{E_{0}}_{-E_0} \Phi(E) (E^2_0-E^2)^\frac{k}{2} dE \label{eq:universal_complexity_moments}.
\ee 
The equation \eqref{eq:universal_Gt_in_Szego_class} is truly an asymptotic statement in $t$, whereas the scaling \eqref{eq:ballistic_scaling} can commence sooner; a timescale for self-consistency is analyzed in Appendix \ref{sec:SzegoAsymptotics}. \eqref{eq:universal_Gt_in_Szego_class} should hold for arbitrary $\Phi(E)$ as $t \rightarrow \infty$ provided the assumptions outlined in Appendix \ref{sec:SzegoAsymptotics} are satisfied, at least for the purpose of accuracy of \eqref{eq:universal_complexity_moments} to low orders in $k$. The underlying physical mechanism is that $\hat{n}$ generates sharply-localized motion on the energy spectrum as was identified in \cite{Chakraborty_2025}. \eqref{eq:universal_complexity_moments} gives such a simple expression in terms of energies because the expectation value of $\mF$ vanishes in $\ket{\psi(t)}$ as $t \rightarrow \infty$. Correspondingly, smeared energy eigenstates admit a well-defined momentum $B$ in addition to position $M$. For even measures, this same formula is derived in a different way in Appendix~\ref{sec:model_for_K} by using the representation \eqref{eq:gt_phi_interaction_picture} and applying \eqref{eq:spectral_relations_n_B_Bcom} to the polynomial asymptotics. We show the quality of \eqref{eq:ballistic_scaling} evaluated with \eqref{eq:universal_complexity_moments} for the semicircle distribution $\Phi(E)$ with constant $b_n$ in Fig.~\ref{fig:rmt_ratio}. We use \eqref{eq:Jacobi_measure} with $\alpha=1$, $\beta=\frac{3}{4}$ as a toy model for a $\Phi(E)$ not symmetric about $E=0$, and do a similar numerical check in Fig.~\ref{fig:jacobi_ratio}.

\begin{figure}[htbp]
     \centering
     \begin{subfigure}[b]{0.45\textwidth}
         \centering
         \includegraphics[width=\textwidth]{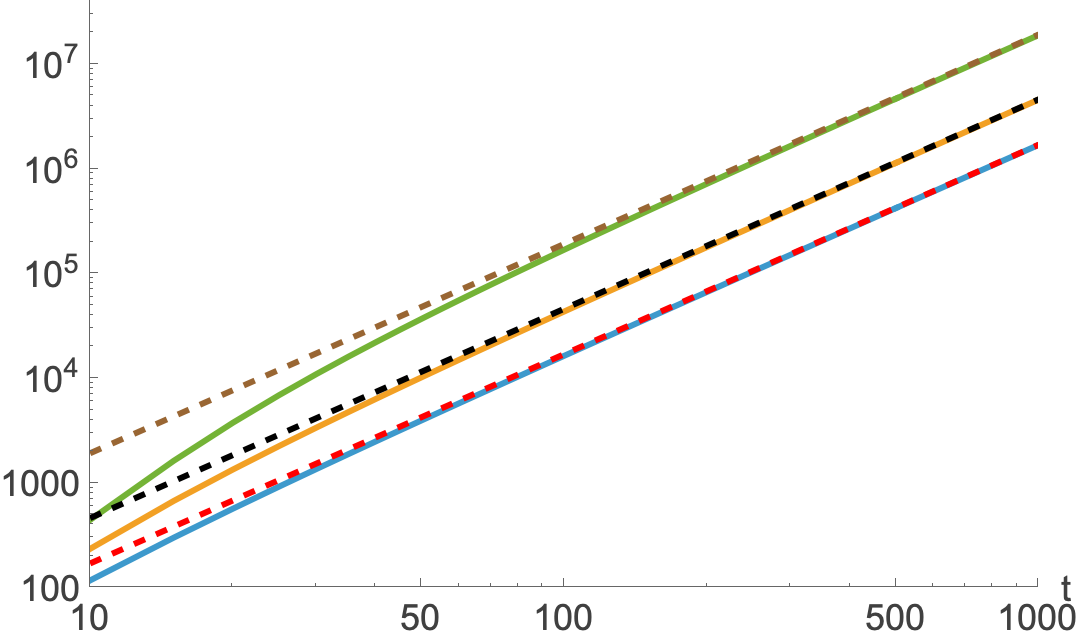}
         \caption{}
         \label{fig:SYK_second_moment_complexity}
     \end{subfigure}
     \hfill
     \begin{subfigure}[b]{0.45\textwidth}
         \centering
         \includegraphics[width=\textwidth]{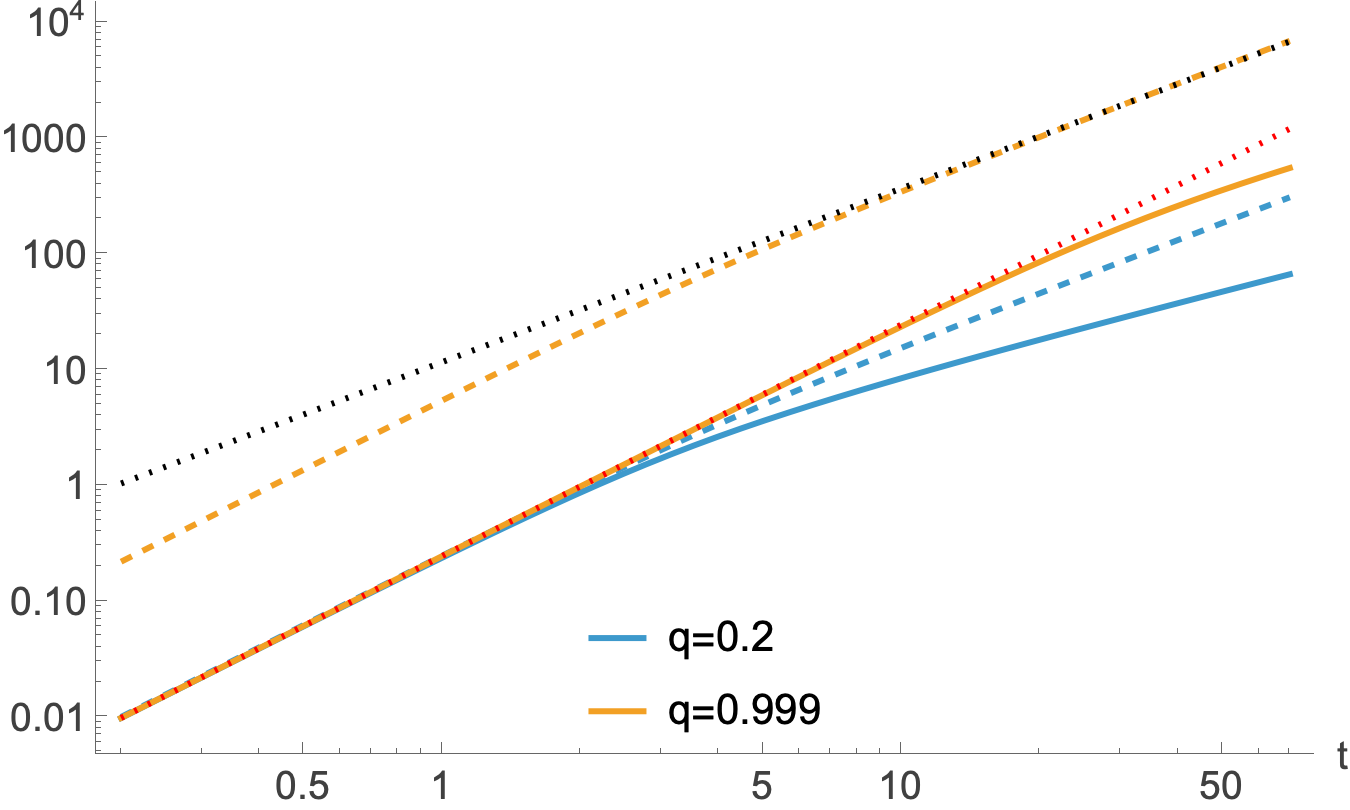}
         \caption{}
         \label{fig:upper_bound}
     \end{subfigure}
     \centering
      \captionsetup{justification=Justified}
\caption{ \textbf{(a) $\expval{\hat{n}^2(t)}$ in DSSYK} for $\qq=0.5$ (blue), $\qq=0.8$ (orange) and $\qq=0.95$ (green). The dashed lines are the prediction $(\frac{1}{1-\qq} -\frac{1}{4}) t^2$ following from \eqref{eq:ballistic_scaling}, \eqref{eq:SYK_v2} and normalization in \eqref{eq:dssyk_Lanczos}. 
\textbf{(b) The bound \eqref{eq:upper_bound_frobenius_tighter} for DSSYK} \eqref{eq:dssyk_Lanczos}. Solid lines show $\expval{\hat{n}(t)}$ for $\qq=0.2$ (blue) and $\qq=0.999$ (orange). Dashed lines are the rhs of \eqref{eq:upper_bound_frobenius_tighter} for the same systems. Dotted lines are power laws $t^2$ (red) and $t^{\frac{3}{2}}$ (black) with coefficients determined from $C(t)$.}
\end{figure}

\begin{figure}[htbp]
    \centering
    \begin{subfigure}[b]{0.48\textwidth}
        \centering
        \includegraphics[width=\textwidth]{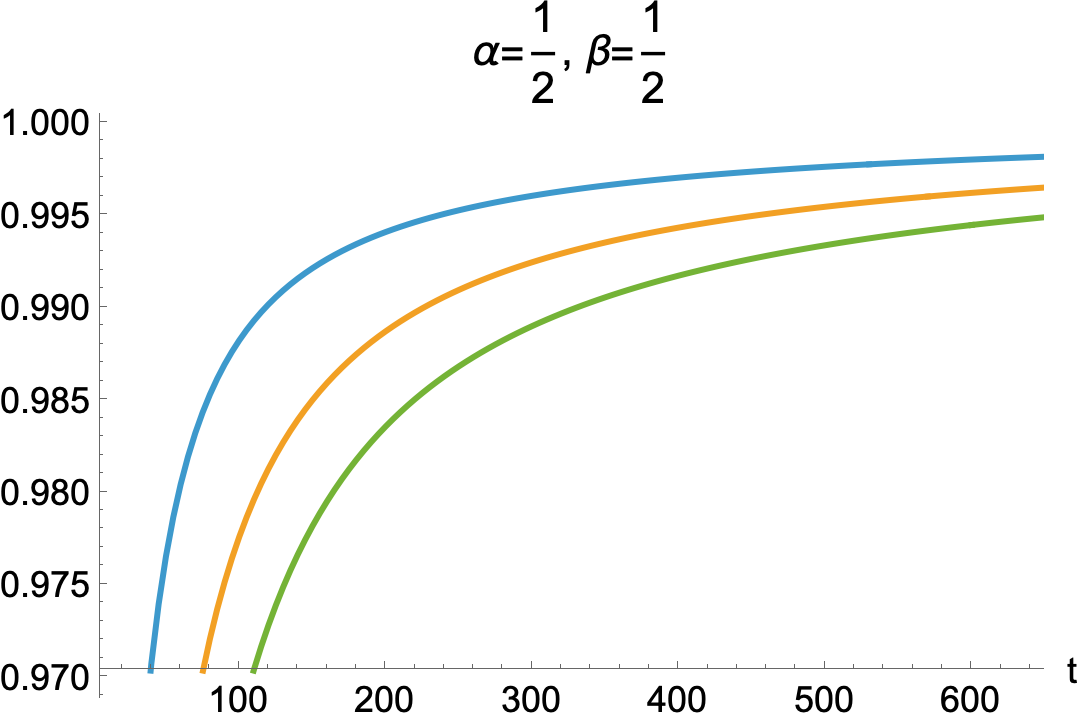}
        \caption{}
        \label{fig:rmt_ratio}
    \end{subfigure}
    \hfill 
    \begin{subfigure}[b]{0.48\textwidth}
        \centering
        \includegraphics[width=\textwidth]{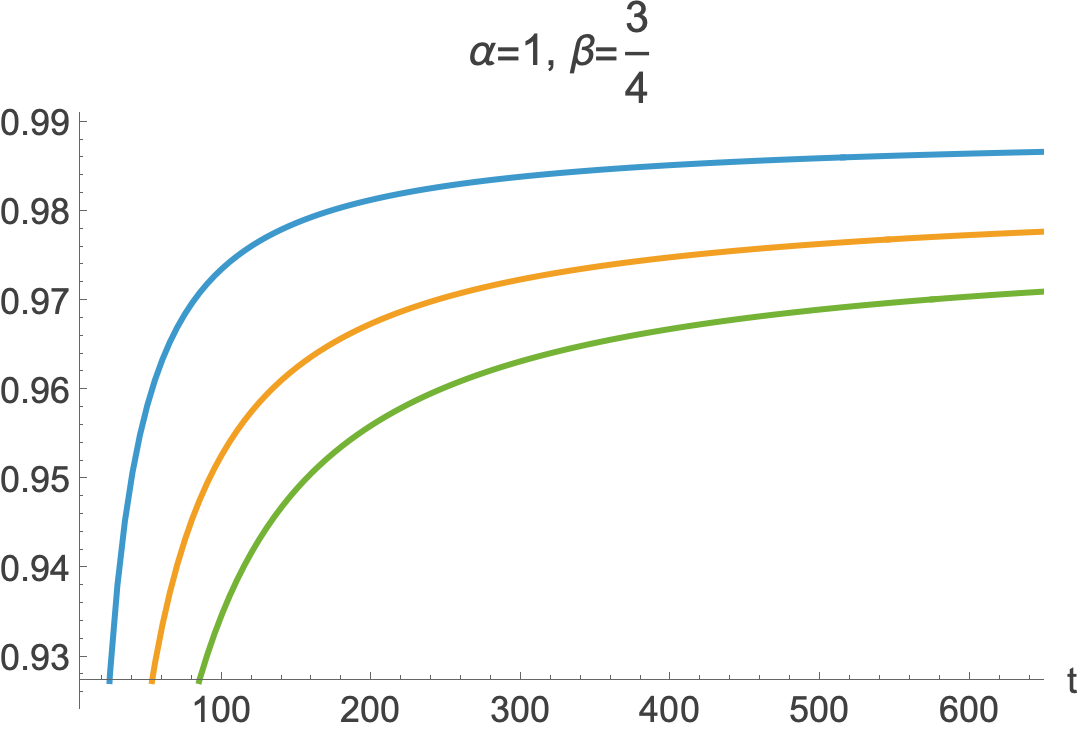} 
        \caption{}
        \label{fig:jacobi_ratio}
    \end{subfigure}
\vspace{0.5cm} 
    \begin{subfigure}[b]{0.48\textwidth}
        \centering
        \includegraphics[width=\textwidth]{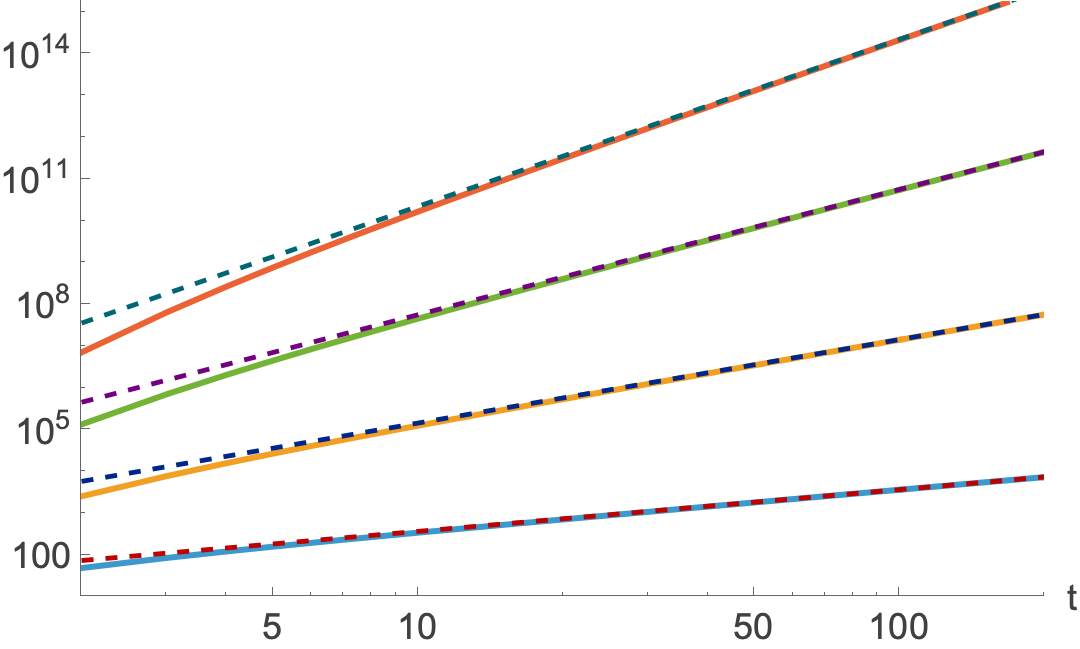}
        \caption{}
        \label{fig:dssyk_with_physical_normalization}
    \end{subfigure}
    \captionsetup{justification=Justified}
    \caption{\textbf{(a) Constant Lanczos coefficients} We plot $\expval{\hat{n}^{k}(t)}/(v_k t^k)$ for $b_n=\frac{1}{2}$ with semicircle $\Phi(E)$ \eqref{eq:rmt_Phi} for $k=1$ (blue), $k=2$ (orange), and $k=3$ (green). The expression for $v_{k}$ is given in \eqref{eq:v_k_for_even_jacobi} setting Jacobi parameters $\alpha=\beta=\frac{1}{2}$. \textbf{(b) Asymmetric $\Phi(E)$} We use the measure \eqref{eq:Jacobi_measure} with values of $\alpha, \beta$ indicated in the figure and numerically compute $\expval{\hat{n}^{k}(t)}/(v_k t^k)$ with the same color scheme as Fig.~\ref{fig:rmt_ratio}. \textbf{(c) Onset of classical behavior} for DSSYK Lanczos coefficients with normalization \eqref{eq:physical_normalization_dssyk} and $\sJ=1, \qq=0.95$. The solid lines are $\expval{\hat{n}^{k}(t)}$ for $k \in \{1,2, 3, 4\}$ (in ascending order vertically) and the dashed lines are corresponding $v_{k} t^{k}$ computed using \eqref{eq:universal_complexity_moments}. Notice how the agreement with dashed lines begins at early times. Compare with Fig.~\ref{fig:SYK_second_moment_complexity} computed with the alternate normalization \eqref{eq:dssyk_Lanczos}. }
\end{figure}

There are several fascinating physical consequences of \eqref{eq:universal_complexity_moments} and \eqref{eq:universal_Gt_in_Szego_class} due to the simple dependence on $\Phi(E)$. Observe that the even coefficients $v_{2m}$ are entirely determined by the \textit{low-order} energy moments $\mu_{2k}$ for $0\leq k\leq m$. This is surprising given that we are discussing a regime where $\ket{\psi(t)}$ has support on sites deep in the Krylov chain with $n \gg 1$, that have constant Lanczos coefficients. We overlay $\expval{\hat{n}^2(t)}$ with $v_2 t^2$ for DSSYK in Fig.~\ref{fig:SYK_second_moment_complexity}, where $v_2$ evaluates to $\frac{1}{1-\qq} - \frac{1}{4}$.

It is straightforward to translate important qualitative features of $\Phi(E)$ to those of the complexity distribution using \eqref{eq:universal_complexity_moments}. The especially remarkable case is when $\Phi(E)$ is a narrow distribution relative to the scale $E_0$. For simplicity, assume that $\Phi(E)$ is sharply peaked around $E=0$. In that case, the onset of \eqref{eq:ballistic_scaling} can, in general, be delayed, see \eqref{eq:t_onset} and surrounding text for a discussion. However, once \eqref{eq:ballistic_scaling} holds, the distribution of $\{ v_k \}$ concentrates around $E^k_0$. This can be seen by expanding 
\be v_{k} = E^{k}_0 \bigg( 1-\frac{k}{2} \frac{\mu_2}{E^2_{0}} + \frac{k(k-2)}{8} \frac{\mu_4}{E^4_{0}} + \ldots  \bigg), \label{eq:v_k_distribution_even_Phi} \ee 
and the expansion continues in $\frac{\mu_{2k}}{E^{2k}_0}$. The statement that $\Phi(E)$ is narrow precisely means $\frac{\mu_{2k}}{E^{2k}_0} \ll 1$. Thus $\expval{\hat{n}^{k}(t)} \approx \expval{\hat{n}(t)}^k$, and the quantum distribution along the Krylov chain becomes effectively classical. When $\Phi(E)$ is centered around the mean $\mu_1$, a similar suppression holds around the value $(E^2_0-\mu_1^2)^{\frac{k}{2}}$ if the width of $\Phi(E)$ is much smaller than $\sqrt{(E^2_0-\mu_1^2)}$, see \eqref{eq:v_k_expansion_central_moments}.   

A concrete physical example of this mechanism can be seen in DSSYK. In our conventions, the energy variance $b^2_{0} = \frac{1}{4}$ and the limiting value $2b = E_0 = \frac{1}{\sqrt{1-\qq}}$. This means that as $\qq$ approaches $1$, $E_0$ diverges relative to its width. The physical system has the overall normalization $\frac{2\sJ}{\sqrt{\lambda_{\qq}}}$ (where $\lambda_{\qq} = -\log \qq$), which we restore to write 
\be b_n = \sJ \sqrt{ \frac{1-\qq^{n+1}}{\lambda_{\qq} (1-\qq)}}. \label{eq:physical_normalization_dssyk}\ee  
The suppression of quantum fluctuations can be read off by examining the first two $v_{2k}$ 
\begin{subequations} \label{eq:dssyk_complexity_slopes}
\begin{align}
 v_{2} &=  \big( \frac{2\sJ}{\sqrt{\lambda_{\qq}}} \big)^2 \bigg( \frac{1}{1-\qq} - \frac{1}{4} \bigg), \label{eq:SYK_v2}    \\
 v_{4} &= \big( \frac{2\sJ}{\sqrt{\lambda_{\qq}}} \big)^4 \bigg( \frac{1}{(1-\qq)^2} - \frac{1}{2} \frac{1}{1-\qq} + \frac{2+\qq}{16} \bigg). 
\end{align}
\end{subequations} 
Equation \eqref{eq:dssyk_complexity_slopes} illustrates how, as $\qq$ approaches $1$, $v_{k} \rightarrow \big( \frac{2\sJ}{\sqrt{\lambda_{\qq} (1-\qq) }} \big)^k$. This makes sense since $\qq \rightarrow 1^{-}$ is a semiclassical limit for the DSSYK model where the quantum fluctuations are expected to vanish \cite{Lin_2023, Blommaert_2025, Aguilar-Gutierrez:2025pqp}. The purpose behind restoring the overall scale was to highlight how the onset \eqref{eq:ballistic_scaling} depends on the nonuniversal part of $\Phi(E)$ towards the center. From the discussion below \eqref{eq:t_onset}, this timescale is given by $t^{(k)}_{\text{on}} \approx \frac{1}{\sJ}$. Through most of our numerics, we chose $\sJ \sim \sqrt{\lambda_\qq}$, which pushes out \eqref{eq:ballistic_scaling} to later times as $\qq$ approaches $1$. In the physical model, $b^2_0 = \mu_2$ is usually taken to be extensive, so $\sJ$ and thus $t^{(k)}_{\text{on}}$ should be $O(1)$. We illustrate this point by showing $\expval{\hat{n}^{k}(t)}$ for $k$ up to $4$ for DSSYK with $\qq=0.95$ in Fig.~\ref{fig:dssyk_with_physical_normalization} by taking $\sJ = 1$.    

Now, consider a generic local system with Gaussian density of states in the bulk of the spectrum and extensive energy variance. Take a state $\ket{\psi(t)}$ such that $\abs{\bra{E}\ket{\psi}}$ is constant or pseudorandom, so that its coarse-grained $\Phi(E)$ matches the density of states. The logic for the Krylov complexity distribution is then identical to that of the $\qq \rightarrow1^-$ limit of DSSYK. The quadratic growth  $\expval{\hat{n}(t)} \sim \mu_2 t^2$ gives way to \eqref{eq:ballistic_scaling} after $O(1)$ times when $t_{\text{on}} \approx \sqrt{E_0/\mu_2}$, where $E_0$ and $\mu_2$ are both extensive. From \eqref{eq:v_k_distribution_even_Phi}, we see that $v_k = E^{k}_0 + O(\frac{1}{\log N})$. Thus only after an $O(1)$ time, $\ket{\psi(t)}$ becomes \textit{classical} in the Krylov basis. 

In many physically interesting contexts, the width of $\ket{\psi}$ does not scale with the system size. In those cases, the fluctuations of $\{|\varphi_n(t)|^2 \}$ are even more suppressed, following \eqref{eq:v_k_distribution_even_Phi}. However, this is only guaranteed after timescales growing with the system size. Consider a local operator in a discrete system satisfying ETH, where operator growth is probed using the Krylov basis constructed from the $\beta=0$ autocorrelation function. Assuming the universal operator growth hypothesis \cite{Parker_2019}, the operator wavefunction in this Krylov basis becomes classical after a timescale logarithmic in the system size.
\section{The Complexity Geometry}
\label{sec:complexity_geometry}
$G(t,\phi)$ in \eqref{eq:Gt_def} is an inner product on states living on a two-dimensional submanifold of the Krylov space generated by moving along the $t$ and $\phi$ coordinates. It is characterized by states $\ket{\Psi(t, \phi)}$, which we require to reduce to the trajectory $\ket{\psi(t)}$ for a specific $\phi$. The inner products $\bra{\Psi(t_1, \phi_1)} \ket{\Psi(t_2, \phi_2)}$ encode the remaining details. More generally, for any set of continuous parameters $\mathbf{x} = (x^{1}, x^2, \ldots x^n)$, the non-negativity of $1-|\bra{\Psi(\mathbf{x})} \ket{\Psi(\mathbf{x} + \mathbf{dx})}|^2$ gives a notion of distance to the submanifold parameterized by $\mathbf{x}$ as follows
\be ds^2 = 1-\abs{\bra{\Psi(\mathbf{x})} \ket{\Psi(\mathbf{x} + \mathbf{dx})}}^2 = \mathcal{Q}_{\mu \nu} dx^{\mu} dx^{\nu}, \ee 
which is induced from the Fubini-Study metric on the ambient Hilbert space. $\mathcal{Q}$ is called the quantum geometric tensor and can be explicitly written as 
\be \mathcal{Q}_{\mu \nu} = \bra{\partial_{\mu}\Psi} \ket{\partial_{\nu} \Psi} - \bra{\partial_{\mu}\Psi} \ket{\Psi} \bra{\Psi} \ket{\partial_{\nu}\Psi}, ~~g_{\mu\nu}  = \Re( \mathcal{Q}_{\mu \nu} ),  ~~ F_{\mu \nu} = 2\Im \mathcal{Q}_{\mu\nu}, \ee
where $\partial_{\mu} := \frac{\partial}{\partial {x^{\mu}}}$. The symmetric part of $\mathcal{Q}$ which contributes to the infinitesimal distance $ds$, is the metric tensor $g$. 

From \eqref{eq:Gt_generator_of_n_dist}, we know that Krylov complexity appears in the amplitude $G(t,\phi)$ purely as a phase. In general, the phase dependence of amplitudes such as $\bra{\Psi(\mathbf{x})} \ket{\Psi(\mathbf{x} + \mathbf{dx})}$ could be beset with ambiguities. A gauge-invariant quantity is the antisymmetric imaginary piece $\frac{1}{2} F_{\mu \nu}$, which measures the phase accumulation along a path \cite{Berry:1984}, and is central to the understanding of the \textit{topological} properties of the parameter space, with applications in quantum Hall effect and topological insulators \cite{Thouless_Hall_Effect, Simon_1984, Qi_2006}. The parameter space we study is a cylinder, so it doesn't quite make sense to integrate $F$ over it, but $F_{\phi t}$ will turn out to be the time-derivative of Krylov complexity. 

Without specifying a family $\ket{\Psi(t, \phi)}$, there is a lot of freedom in choosing a two-dimensional submanifold on which $\ket{\psi(t)}$ lives. We study the simplest nontrivial choice given by $\ket{\Psi(t, \phi)} = e^{i \hat{n} \phi} e^{-i M t} \ket{0}$. The coordinates $t$ and $\phi$ appear in an asymmetric way, with the $\phi$-dependence being particularly simple. The motivation is two-fold. The first is that this choice reproduces the information geometry found in \cite{Caputa:complexity_geometry}. Second, we observe that the Fourier transform of $G(t_1, t_2, \phi_2 + \phi_1) := \bra{ \Psi(t_1, -\phi_1)}\ket{\Psi(t_2, \phi_2)}$ is closely related to the kernel discussed in Appendix \ref{sec:Poisson_Kernel}:
\be \bigg( \frac{1}{2\pi} \bigg)^2 \int^{\infty}_{-\infty} \int^{\infty}_{-\infty} \bra{\Psi(-t_1, -\phi_1)} \ket{\Psi(t_2, \phi_2)} e^{i (E_1 t_1 + E_2 t_2) } dt_1 dt_2  =  \sqrt{\Phi(E_1) \Phi(E_2)} K(E_1, E_2, \phi_2 + \phi_1).  \ee 
The metric components and the Berry curvature are written:
\be 
\begin{split}
g_{tt}  &=\Delta M^2 := b^2_0,   \quad \quad g_{t \phi} = b_0 \text{Re} \bra{\psi(t)} \hat{n} H\ket{\psi(t)},   \\ 
g_{\phi \phi}& = \Delta n(t)^2 :=   \bra{\psi(t)}  \hat{n}^2 \ket{\psi(t)} - \bra{\psi(t)} \hat{n}\ket{\psi(t)}^2, \\ 
F_{t\phi} &= \bra{\psi(t)} B \ket{\psi(t)}.
\end{split}
\label{eq:complexity_geometry}
\ee 
The metric in \eqref{eq:complexity_geometry} reproduces as a special case, the information geometry studied in \cite{Caputa:complexity_geometry} where the algebra between $L_{+}, L_{-}$ and $\hat{n}$ closes.  The agreement between the two approaches can be understood broadly as follows. The metric on the Perelomov coherent states labeled by a continuous complex parameter $z$, is also induced from the Fubini-Study metric. $e^{i \hat{n} \phi}$ translates the phase of this complex $z$. We explicitly show the match with their examples, and the metric given by \eqref{eq:complexity_geometry} in Appendix \ref{sec:complexity_algebra}.

\cite{Caputa:complexity_geometry} obtained the two-dimensional maximally symmetric spaces of constant curvature in the case of operator complexity, where all the $a_n$ vanish. As long as the measure is even, meaning that $a_n$ continue to vanish, $g_{t \phi}=0$ and the metric is diagonal. From \eqref{eq:t_derivative_Kt}, \eqref{eq:B_definition}, combined with the $\phi$ isometry, it simply follows that $F_{\phi t}$ is the time derivative of Krylov complexity.  

Up to the overall units for the Lanczos coefficients inherited from $H$, the volume form associated with $g$ contains intrinsic physical meaning. It gives a measure of the number of sufficiently distinguishable states (in terms of their overlap), per unit volume of the parameter space. The growth of $g_{\phi \phi}$ with time captures the sensitivity of the trajectory to small angular displacements $\delta\phi$ at fixed $t$. We intuitively expect this to be related to Krylov complexity, and in case of the diagonal metric we also expect it to be related to the local volume since the arclength $ \int^{\phi_0+\delta \phi}_{\phi_0} \sqrt{g_{\phi\phi}} d\phi$ contains the nontrivial factor of the volume element i.e. $\frac{1}{b_0} \sqrt{\det g} \delta \phi $. 

The direct connection between the volume and complexity, valid for general $g$ comes from uncertainty bounds. The metric components are (co)variances after all. The Robertson-Schr\"odinger uncertainty relation $\Delta X^2 \Delta Y^2 \geq | \bra{\Psi} \frac{1}{2i} [X, Y] \ket{\Psi}|^2  + | \frac{1}{2} \bra{\Psi} \{ X, Y \} \ket{\Psi} - \bra{\Psi} X \ket{\Psi}  \bra{\Psi} Y\ket{\Psi} |^2$ applied to $X= M$, $Y=\hat{n}$ and $\ket{\Psi}= \ket{\psi(t)}$ is precisely equivalent to 
\be \Delta M^2 (\Delta n(t) )^2  \geq \abs{\expval{\frac{1}{2i} [M, \hat{n}] }}^2 + \Big[ \frac{1}{2} \bra{0} \{ \hat{n}(t),  M \} \ket{0} - \expval{ \hat{n}(t) } a_0 \Big]^2   \implies \det g \geq  \frac{1}{4} (\partial_t \expval{\hat{n}(t)})^2 ,\label{eq:positivity_volume} 
\ee
where, to highlight the nontrivial time-dependence in \eqref{eq:positivity_volume} we switched back to the Heisenberg picture operator $\hat{n}(t)$. We can recognize the RHS of \eqref{eq:positivity_volume} as 
\be  \sqrt{\det g} \geq \frac{1}{2} \abs{ F_{\phi t} } = \Big|\frac{1}{2} \partial_{t} \expval{\hat{n}(t)} \Big|.  \label{eq:volume_form_bounded_byKdot} \ee
The left equation in \eqref{eq:positivity_volume} was derived in \cite{H_rnedal_2022} for $a_n=0$, where it simplifies to  $\Delta n(t) \geq (2b_0)^{-1} \partial_{t} \expval{\hat{n}(t)}$\footnote{See also \cite{Chaudhuri_2026} and \cite{grabarits2026_time_dependent} for structurally similar bounds.}. They did not frame this result in terms of the geometric tensor presented in \eqref{eq:positivity_volume}. 
Since the left-hand side is always $\phi$-independent, we can integrate \eqref{eq:volume_form_bounded_byKdot} and get
\be  \expval{ \hat{n}(t) }  \leq \int^{t}_{0} | \partial_{\tilde{t}} \expval{\hat{n}( \tilde{t} )} | d \tilde{t} \leq 2\int^{t}_{0} \sqrt{\det g}~d \tilde{t} = \frac{2}{2 \pi}  \int^{2 \pi}_{0} d\phi   \int^{t}_{0}  \sqrt{\det g}~d\tilde{t} = \frac{1}{\pi} \text{Vol}_{t}, \label{eq:complexity_volume_bound} 
\ee 
where $\text{Vol}_{t}$ represents the volume enclosed in a shell inside this geometry from time $0$ to $t$. Thus, Krylov complexity is \textit{always} bounded by the volume of this geometry up to a conventional multiplicative factor. 

\eqref{eq:complexity_volume_bound} serves as a consistency relation between expectation values defined in the Krylov basis. Since both sides of the equation are typically difficult to compute, the bound involving the off-diagonal term is equally useful, or similarly limited, in practice. $g_{t\phi}$ is a time-dependent probe of the asymmetry of the spectral function about its mean. It can  be interpreted as a susceptibility to infinitesimal Euclidean evolution as
\be g_{t\phi} = -\frac{1}{2} b_0 \partial_{\ftau}  \bra{\psi(t)} e^{-\tau H} \hat{n} e^{-\tau H} \ket{\psi(t)} \big|_{\tau=0}. \label{eq:gtphi_as_euclidean_ev} \ee 
From the derivation underlying the uncertainty bounds, the second inequality in \eqref{eq:complexity_volume_bound} is only saturated when $(\hat{n}(t) -\expval{\hat{n}(t)} )\ket{0}$ is proportional to $\ket{1}$. \cite{H_rnedal_2022} proved that the closure of the algebra between $L_{\pm}$ and $\hat{n}$--exactly what was studied in \cite{Caputa:complexity_geometry}--is a necessary and sufficient condition for the bound to be saturated when $L_0=0$. In Appendix \ref{sec:complexity_algebra}, we provide an alternate simple derivation extending to $L_0 \neq 0$ case, when the algebra between $L_{\pm}, L_0$ and $\hat{n}$ is closed. We show that this is a sufficient condition for the inequality \eqref{eq:positivity_volume} to be saturated, and assuming the monotonic behavior of $\expval{\hat{n}(t)}$, we find that Krylov complexity matches the rescaled volume of a shell in this quantum geometry. This identification also extends to oscillatory dynamics provided $t$ is suitably restricted with a coordinate redefinition.  

In these examples, although the presence of a nontrivial $L_0$ affects the time-dependence of the complexity, variance and the volume significantly, the curvature is insensitive to it. For $SL(2,R)$ algebra tuning the coefficient of $L_0$ relative to $L_{\pm}$, drives a transition between exponential and oscillatory behavior for $\expval{\hat{n}(t)}$, while the transition point itself exhibits quadratic growth \cite{Balasubramanian_2022}, see \eqref{eq:sl2r_gphiphi}, \eqref{eq:sl2r_off_diag}. This is all while the scalar curvature is rigidly fixed at $-\frac{4}{h}$, where $h$ is the $SL(2,R)$ representation label. This implies that the underlying manifold depends only on the Lie algebra of generators, and the $t$, $\phi$ coordinates obscure the additional symmetries of the constant curvature manifold. It would be interesting to better understand the purely geometric reason behind the saturation of the uncertainty bounds, with an abstract proof along the lines of \cite{Ozawa_2021}. 

Returning to the generic setting, we notice the metric is diagonalized in the new coordinates 
\be \bar{t}= t,~~~ \vartheta= \phi + \int \frac{g_{t \phi}}{g_{\phi \phi}} dt , \implies~~~~g_{\bar{t} \bar{t}} = g_{tt} - \frac{g_{t\phi}^2}{g_{\phi \phi}},~~~ g_{\vartheta \vartheta} = g_{\phi \phi},~~\text{and}~~g_{\bar{t} \vartheta} = 0. \label{eq:metric_new_coordinates} \ee
Lines of constant $\phi$ are no longer geodesics, which means that the wavefunction $\ket{\psi(t)}$ does not move through a geodesic. The geodesics connecting $\ket{\psi(t_1)}$ and $\ket{\psi(t_2)}$ could be interesting to study from the point of view of optimal state preparation protocols and in contexts like \cite{Krylov_Winding}. In two-dimensions, geodesic deviation is solely dictated by the scalar curvature. From the $SL(2,R)$ example, we saw that the latter could obscure the complexity dynamics of the physical trajectory. Nonetheless, there are geodesics which capture details of the physical trajectory. The simplest examples are the geodesics defined by constant $\vartheta$, which we can characterize in terms of $\phi(t)$ through \eqref{eq:metric_new_coordinates}. In the $SL(2,R)$ example the path traced out by $\phi(t)$ winds around when the complexity dynamics is oscillatory. In contrast, $\phi(t)$ asymptotes to a constant at large $t$ when $g_{t\phi}$ grows exponentially, see \eqref{eq:vartheta_sl2} for details.  

It is much simpler to interpret this geometry when $L_0=0$ but for generic $L_{\pm}$. The scalar curvature
\be  R_{\text{curv}} = -\frac{2}{b^2_0 (\Delta n(t)) } \frac{d^2 (\Delta n(t)) }{d t^2}, \label{eq:scalar_curvature_diagonal} \ee 
depends on the convexity of the standard deviation $(\Delta n(t))$, relating intrinsic geometry to delocalization in Krylov basis. \eqref{eq:scalar_curvature_diagonal} can be written in the more suggestive form
\be \frac{d^2 \eta}{ds^2} + \frac{1}{2} R_{\text{curv}} \eta = 0, \quad \eta = \sqrt{g_{\phi \phi}} \delta \phi, \quad s = b_0 t, \ee
which expresses the dynamics of the proper distance between two geodesics angularly displaced by $\delta \phi$ in terms of the arclength $s$, in terms of $R_{\text{curv}}$ and vice-versa. 

\begin{figure}[htbp]
     \centering
     \begin{subfigure}[b]{0.85\textwidth}
         \centering
         \includegraphics[width=\textwidth]{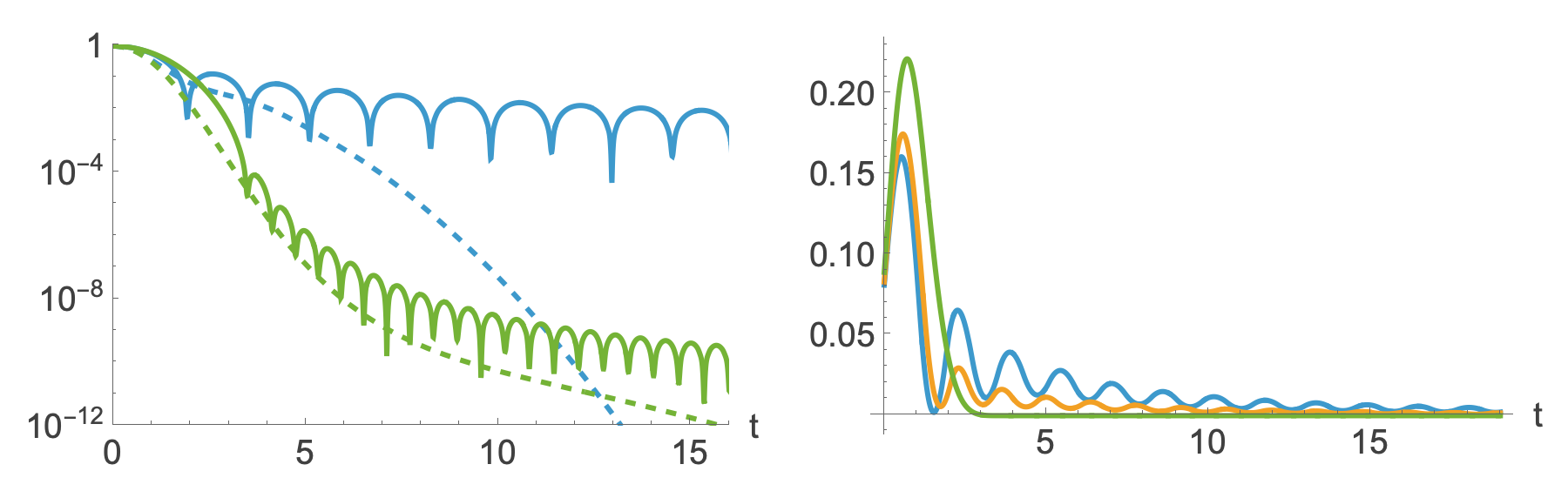}
         \caption{}
         \label{fig:Ct_smearing_SYK}
     \end{subfigure}
     \hfill
     \begin{subfigure}[b]{0.85\textwidth}
         \centering
         \includegraphics[width=\textwidth]{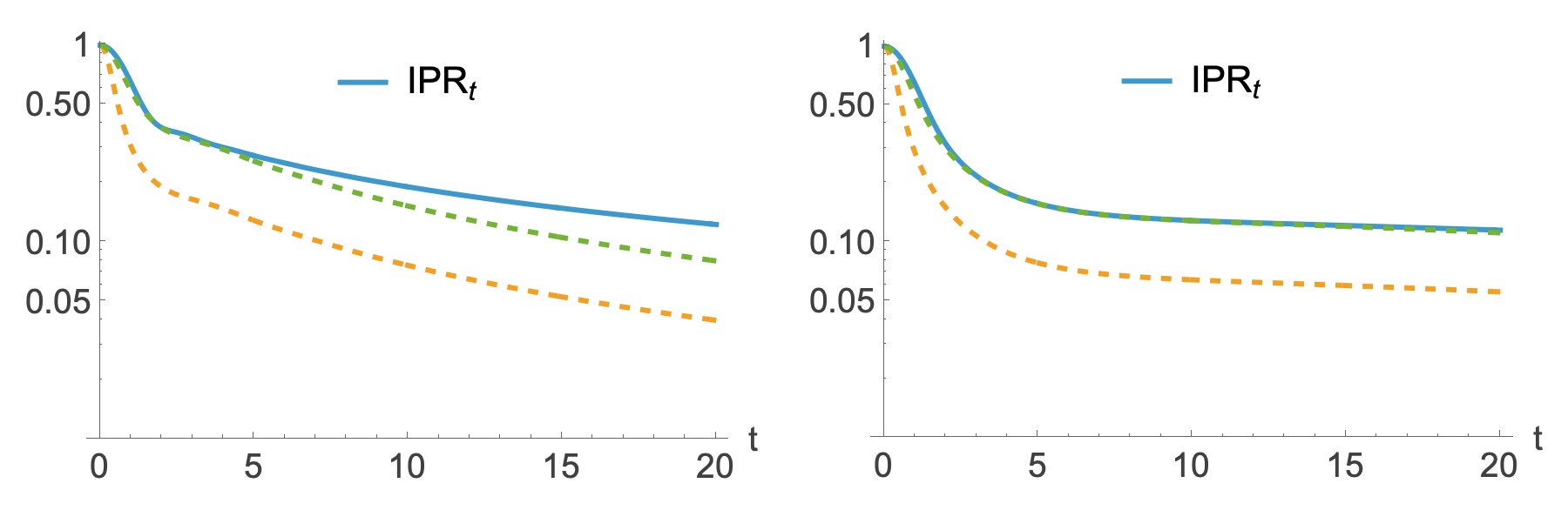}
         \caption{}
         \label{fig:IPR_DSSYK}
     \end{subfigure}
      \captionsetup{justification=Justified}
\caption{ \textbf{(a) Integrated form of bound \eqref{eq:exponentiated_lower_bound_correlator} in DSSYK} The left panel shows $|C(2t)|$ (solid lines) for $\qq=0$ (blue) and $\qq=0.8$ (green). The dashed lines of same colors are the rhs of \eqref{eq:exponentiated_lower_bound_correlator} computed using Lanczos coefficients.  \eqref{eq:exponentiated_lower_bound_correlator} is satisfied away from where $C(2t)$ dips to $0$. The right panel shows the difference smeared over a $\delta t$ window: $\int^{t+\delta t}_{t} \bigg[ |C(2s)| - \exp( -\frac{1}{2}(\Delta n(s))^2 \pi^2 )  \bigg]ds$, with $\delta t= 0.8$ for $\qq=0$ (blue), $\qq=0.2$ (orange) and $\qq=0.8$ (green).  \textbf{(b) $\text{IPR}_{t}$ in DSSYK } for $\qq=0$ (left) and $\qq=0.8$ (right), along with rhs of \eqref{eq:IPR_upperbound_variance} (orange) and \eqref{eq:IPR_stronger_upperbound} (green) computed using $\Delta n(t)$.  }
\end{figure}

The metric $g$ is strictly local data, but there is a powerful global relation for this setup. Because $M(\pi) = -M(0)$, $\ket{\Psi(t, \pi)} =\ket{\Psi(-t, 0)} $ and $G(t, \pi)= C(2t)$. The metric $g$ is induced from the Fubini-Study metric on the ambient Hilbert space, so we can always lower-bound the arclength between any two-points by the Fubini-Study distance. For a path with constant $t$, and $\delta \phi=\pi$, this gives $\arccos |C(2 t)| \leq \Delta n(t) \pi$, which can be inverted in the more useful form
\be \abs{C(2 t)} \geq~~~ \cos( \pi \Delta n(t))\geq ~~~ 1- (\Delta n(t))^2 \frac{\pi^2}{2}. \label{eq:bounding_correlator_rigorous} \ee 
The last inequality in \eqref{eq:bounding_correlator_rigorous} expresses how the infinitesimal equivalence $|G(t, d\phi)| = 1- g_{\phi\phi} d \phi^2$ is promoted to an inequality for finite variation of parameters.  The bound \eqref{eq:bounding_correlator_rigorous} is general but rather conservative, becoming trivial when $\Delta n(t)$ is $O(1)$. Building on the intuition that the wavefunction minimizes spread in the Krylov basis, we hypothesize that physical systems in the thermodynamic limit satisfy an exponentiated version of the bound $\abs{G(t, \phi)} \geq 1 - \frac{1}{2}g_{\phi \phi}  \phi^2 $:
\be \abs{G(t, \phi)} \gtrsim e^{- \frac{1}{2} (\Delta n(t))^2  \phi^2}, \label{eq:exponentiated_bound}  \ee
from which we can deduce
\be \abs{C(2 t)} \gtrsim e^{- \frac{1}{2}(\Delta n(t))^2 \pi^2 }. \label{eq:exponentiated_lower_bound_correlator} \ee
The notation in \eqref{eq:exponentiated_bound}, \eqref{eq:exponentiated_lower_bound_correlator} emphasizes that it is not a true mathematical inequality but a physically motivated one. It is easy to find counter-examples to \eqref{eq:exponentiated_bound}. Due to destructive interference $C(t)$ can have zeros for specific values of $t$, while the right-hand side can only asymptote to $0$ when $\Delta n(t/2)$ is large enough. In practice, the proposed bound \eqref{eq:exponentiated_bound}, holds for a wide variety of examples and is generally rather conservative. In the instances where it fails---such as for the autocorrelation function associated with RMT spectral measure \eqref{eq:rmt_Phi}---it only fails in neighborhoods around the zeros of $C(2t)$. This can be mended by smearing \eqref{eq:exponentiated_bound} in time. It should be sufficient to do this smearing over timescales that are an $O(1)$ fraction of the scale set by $b^{-1}_0$. We demonstrate the validity of this prescription for the DSSYK Lanczos coefficients \eqref{eq:dssyk_Lanczos} in Fig.~\ref{fig:Ct_smearing_SYK}, where we elaborate on the smearing procedure. Here, $\Phi(E)$ is bounded for all $\qq<1$ and goes as $\sqrt{\frac{1}{1-\qq}-E^2}$ near the edges. From large-$t$ asymptotics, we expect $C(t)$ to have zeros around which \eqref{eq:exponentiated_bound} breaks down. \eqref{eq:exponentiated_bound} turns into a useful estimate for $\Delta n(t)$ given knowledge of the autocorrelation function $C(2t)$. 

We can push this kind of approximate reasoning further. Integrating $|G(t, \phi)|^2$ over $\phi$ gives 
\be \text{IPR}_{t} := \sum_n \abs{\varphi_n(t)}^4 =  \frac{1}{2\pi} \int^{2\pi}_{0} |G(t, \phi)|^2 d\phi, \label{eq:IPR_as_mean_value}   \ee 
where the $\text{IPR}_{t}$ stands for the inverse participation ratio of the Krylov wavefunctions $\varphi_n(t)$ at time $t$. Notice that \eqref{eq:IPR_as_mean_value} also holds when the coefficients $a_n$ are nonzero. \textit{Assuming} the inequality \eqref{eq:exponentiated_bound} and integrating $\phi$ from $0$ to $2\pi$, we get
\be  \text{IPR}_{t} \geq \frac{1}{4 \sqrt{\pi} \Delta n(t) } \erf \big(2 \pi \Delta n(t)   \big). \label{eq:IPR_upperbound_variance}  \ee 
The rhs of \eqref{eq:exponentiated_bound} is not periodic in $\phi$ although the lhs is, so the above bound was conservative about the overlap close to $\phi=2 \pi$. We can hope to get a tighter estimate by performing the $\phi$ integral between $-\pi$ to $\pi$ and get 
\be \text{IPR}_{t} \gtrsim \frac{1}{2 \sqrt{\pi} \Delta n(t) } \erf \big( \pi \Delta n(t)   \big). \label{eq:IPR_stronger_upperbound} \ee 
Although both $\Delta n(t)$ and $\text{IPR}_{t}$ measure the spread of $\ket{\psi(t)}$ in the Krylov basis, there is no a priori reason for \eqref{eq:IPR_upperbound_variance} to hold. We have found it to hold quite generally in all considered examples, which helps further constrain the Krylov probability distribution. In contrast, the proposed inequality \eqref{eq:IPR_stronger_upperbound} does not always hold but the rhs provides a tighter estimate for the lhs, see Fig.~\ref{fig:su2_ipr} for an example featuring the $SU(2)$ complexity algebra. We can learn a lot more when both sides of \eqref{eq:IPR_upperbound_variance} (and thus also \eqref{eq:IPR_stronger_upperbound}) have the same time dependence. Surprisingly, this is found to be true in many physically relevant examples, where additionally, \eqref{eq:IPR_stronger_upperbound} is nearly saturated. 

The analytic examples of Appendix \ref{sec:complexity_algebra} are illuminating because $\text{IPR}_{t}$ can be written in closed form.  It is easy to see the saturation of \eqref{eq:IPR_stronger_upperbound} for the Heisenberg-Weyl algebra \eqref{eq:heisenberg_weyl_algebra} using the large $t$ asymptotic of $\text{IPR}_{t}$, given by a Bessel function. In the $SL(2,R)$ case, the representation label $h$ (see Appendix \ref{sec:sl2r_algebra}) plays an important role. For concreteness, we take the family specified by \eqref{eq:sl2_lanczos}, \eqref{eq:alpha_gamma_sl2_toda} with $J>0$ and take $h > \frac{1}{4}$ for which we quote $\text{IPR}_{t}$ in \eqref{eq:sl2r_IPR}. The combination $2 \sqrt{\pi} \Delta n(t) \text{IPR}_{t}$ approaches a $h$-dependent constant larger than $1$ for large $t$, see \eqref{eq:sl2r_IPR_variance_asymptote}. $J$ and $\tau$ control the rate of approach but not the asymptote \eqref{eq:sl2r_IPR_variance_asymptote}. This constant in \eqref{eq:IPR_stronger_upperbound} approaches $1$ as $h$ increases. When this ratio approaches $1$, we can say that $\abs{G(t,\phi)}$ is spread out in the $\phi$ direction like the Gaussian \eqref{eq:exponentiated_bound}.\footnote{A more careful, verifiable statement is that the higher cumulants of the distribution $\{ \abs{\varphi_n(t)}^2 \}$  are suppressed relative to powers of the second cumulant.} The behavior of $\Delta n(t) \text{IPR}_{t}$ is very different for $h<\frac{1}{4}$, where it goes to $0$ asymptotically, still obeying \eqref{eq:IPR_stronger_upperbound}. Similar analyses can be performed for other instances of $SL(2,R)$ and $SU(2)$ complexity algebras. 

The tighter \eqref{eq:IPR_stronger_upperbound} holds for the DSSYK Lanczos coefficients across a range of $\qq$, as shown in Fig.~\ref{fig:IPR_DSSYK}. Here we have a degree of analytic control for late times, using the approximate answer \eqref{eq:gt_spectral_linear_phi}. While it is not enough to analytically pin down $\text{IPR}_{t}$, evaluating \eqref{eq:gt_spectral_linear_phi} by saddle point, suggests that at large $t$ $\text{IPR}_{t}$ goes as $\frac{\log t}{\Delta n(t)}$. This is consistent with the drift away from the bound \eqref{eq:IPR_stronger_upperbound} found for $\qq=0.2$ in Fig.~\ref{fig:IPR_DSSYK}. For $\qq$ closer to $1$, the short-time saturation of \eqref{eq:IPR_stronger_upperbound} can be understood by applying the Heisenberg algebra result as an approximation.

\begin{figure}[htbp]
    \centering
    \begin{subfigure}[b]{0.48\textwidth}
        \centering
        \includegraphics[width=\textwidth]{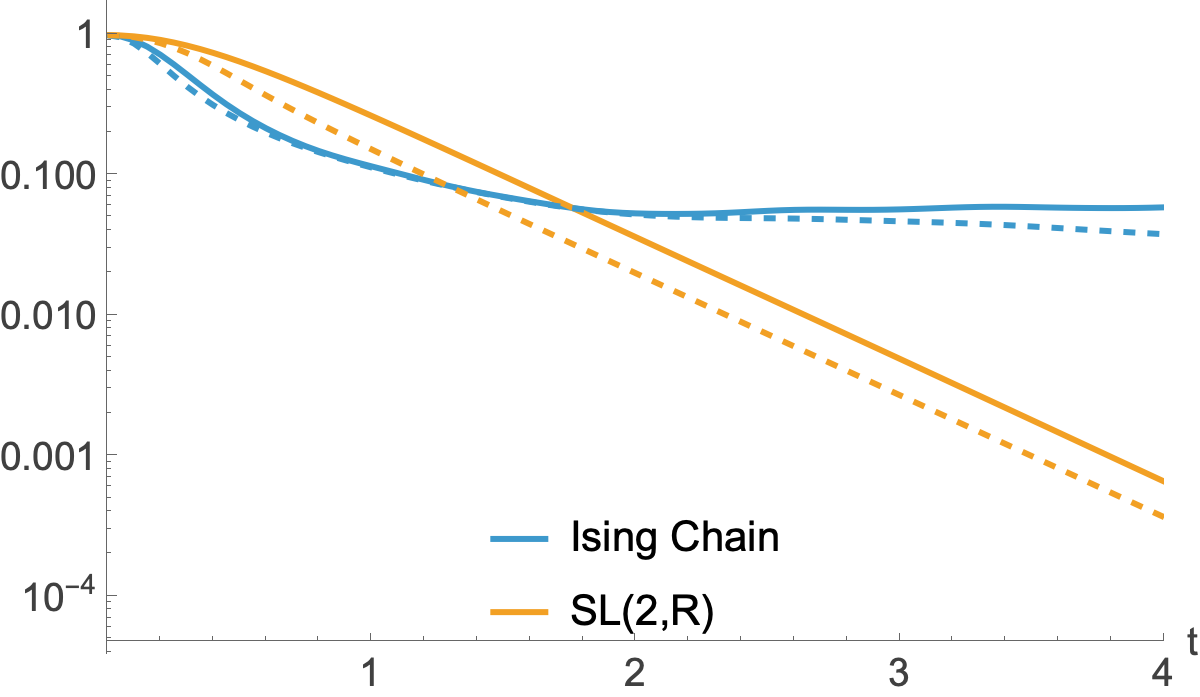} 
        \caption{}
        \label{fig:ipr_ising_sl2}
    \end{subfigure}
    \hfill 
    \begin{subfigure}[b]{0.48\textwidth}
        \centering
        \includegraphics[width=\textwidth]{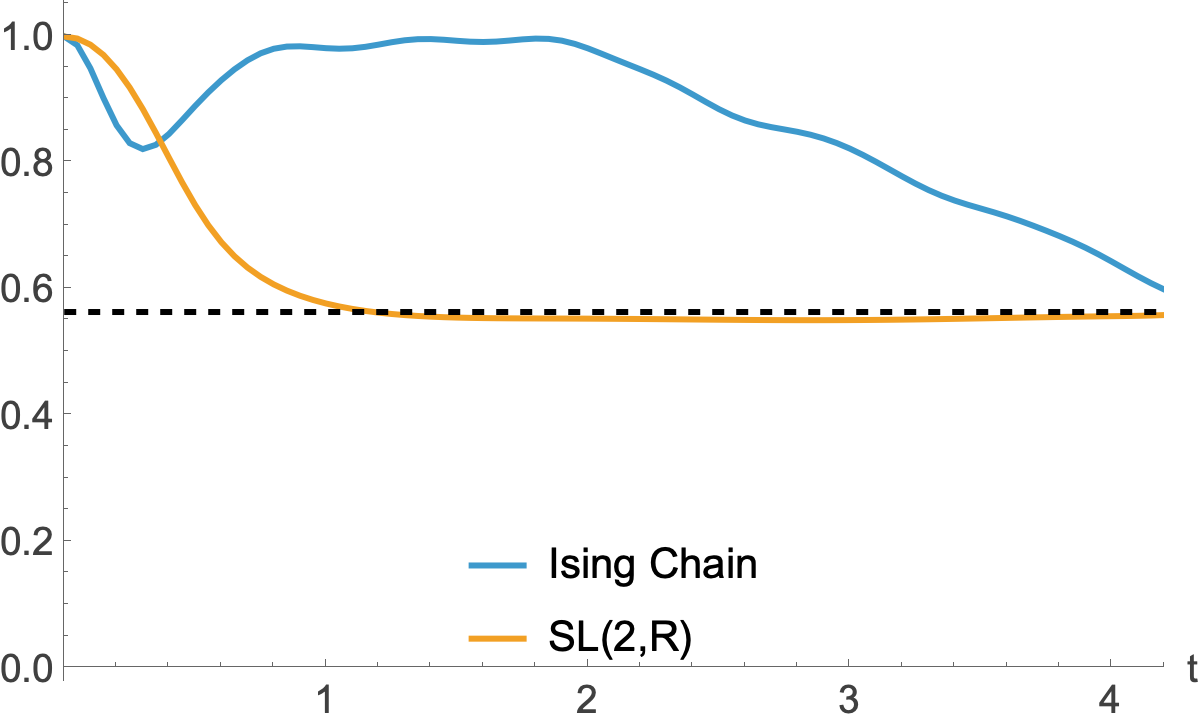} 
        \caption{}
        \label{fig:ipr_ratio}
    \end{subfigure}
\vspace{0.5cm} 
    \begin{subfigure}[b]{0.48\textwidth}
        \centering
        \includegraphics[width=\textwidth]{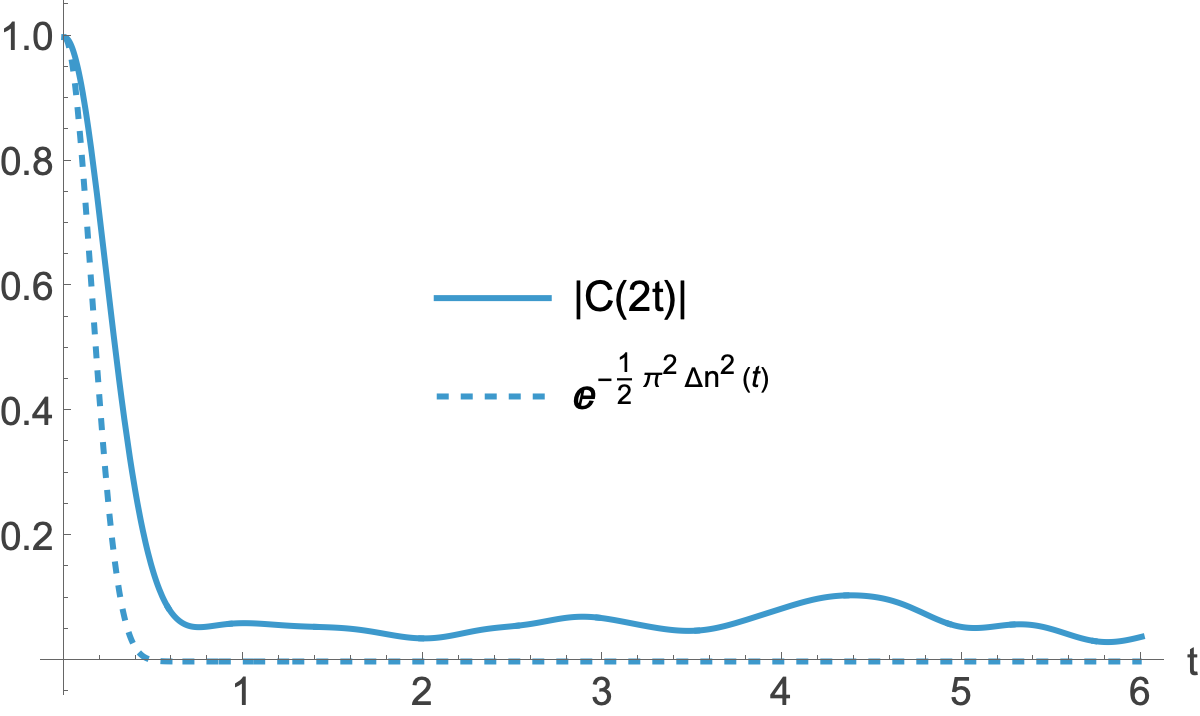}
        \caption{}
        \label{fig:ct_ising}
    \end{subfigure}
    \captionsetup{justification=Justified}
    \caption{\textbf{(a) $\text{IPR}_t$ compared with analytic prediction} the numerically calculated $\text{IPR}_t$ (solid lines) compared against the corresponding \eqref{eq:IPR_stronger_upperbound} in dashed lines of same colors. \textbf{(b) Ratio of the prediction with $\text{IPR}_t$:} The combination $\erf(\pi \Delta n(t) )/(2 \sqrt{\pi}\Delta n(t) \text{IPR}_t )$ for the same systems. The dashed black line is the analytical prediction given by inverse of \eqref{eq:sl2r_IPR_variance_asymptote}. \textbf{Details of the systems} The Ising chain is specified by the Hamiltonian \eqref{eq:finite_ising_spin_chain} and $C(t)$ is the infinite temperature autocorrelation function \eqref{eq:wightman_inn_prod} of the initial operator $\sigma^{z}_{3}$. The $SL(2,R)$ toy model (orange) is specified by the Lanczos coefficients \eqref{eq:sl2_lanczos}, where both the $b_n$ and $a_n$ coefficients have been perturbed by independent positive random variables with mean $0.25$. For the $SL(2,R)$ computation $\alpha=1, \gamma=0.1, h=\frac{1}{2}$. \textbf{(c) Illustration of \eqref{eq:exponentiated_lower_bound_correlator} in spin chain:} Ising chain with same parameters and initial operator as above used. }
    \label{fig:ipr_and_ct_spin_chain_sl2}
\end{figure}

We also find it to generically hold for operator complexity in small-sized chaotic spin chains for time scales that get stretched out in the thermodynamic limit. This property, $\text{IPR}_{t} \sim \frac{1}{2 \sqrt{\pi} \Delta n(t)}$ is thus likely to continue to hold in the thermodynamic limit for a variety of physical systems. At least for some systems, this is because their Lanczos coefficients are ``close'' to the Lanczos coefficients controlled by a complexity algebra. Meaning that the statistical behavior of the probabilities $\{ |\varphi_{n}(t)|^2 \}$ is robust to perturbations that do not alter important details of the system. An example of the latter is the intercept of a linear sequence of $b_n$ which is related to the order of the pole of $C(t)$, when continued to imaginary time. This is precisely what $h$ stands for, when continued to arbitrary positive values, such as for flavor-averaged 2pt function in SYK. We provide evidence for this robustness in Fig.~\ref{fig:ipr_and_ct_spin_chain_sl2} by perturbing the $SL(2,R)$ Lanczos coefficients by independent positive random variables and still finding the agreement to hold. 

Assuming that $\Delta n(t)$ shares the same functional time-dependence as $\expval{\hat{n}(t)}$ yields stronger physical conclusions. This commonly observed behavior can be justified using heuristics and is also found in the analytic examples of Appendix \ref{sec:complexity_algebra}, with the notable exception of Heisenberg algebra. Assuming exponential growth $\Delta n(t) \sim \expval{\hat{n}(t)} \sim e^{\lambda_{K} t}$ gives 
\be  \frac{1}{t} \log \frac{\sqrt{g_{\phi\phi} } \delta\phi}{ \delta\phi} = \frac{1}{t} \log \Delta n(t)= \lambda_{K}, \ee
making $\lambda_{K}$ a measure of exponential separation of trajectories for angular displacements at fixed $t$. Thus this metric concretely relates $\lambda_{K}$ to the ($\log$ of) size of the red circular arc shown in Fig.~\ref{fig:deformations}, relative to the size of the radial $t$ direction. The fixed $t$ path is not a geodesic, but this should not matter when $\delta \phi \rightarrow 0$.

$\Delta n(t) \sim \expval{\hat{n}(t)}$ scaling also relates the classification of Section \ref{sec:Loschmidt} using $\mF$, to properties of the metric. The metric contains information not directly accessible to \eqref{eq:linearized_Kcomplexity_rate_Loschmidt}. Nevertheless, with this assumption, we can argue for a connection for vanishing $a_n$, at large enough $t$. For $b_n$ that asymptote to a constant we know that $\Delta n(t)  \sim t$. Together with the Heisenberg-Weyl algebra, these cases yield locally flat spaces. For generic power-law $b^2_n \sim n^{\kappa}$, $0<\kappa < 2$, the continuum limit estimate \cite{Barb_n_2019,M_ck_2022} $\expval{\hat{n}(t)} \sim t^\frac{2}{2-\kappa}$ gives negatively curved spaces, with curvature decaying as $\sim -\frac{1}{t^{2}}$. Exponentially growing $\expval{\hat{n}(t)}$ generally gives rise to local regions with constant negative curvature. 
\section{Discussion}
In this paper we have studied how Krylov complexity and its time-derivative can be understood as measures of sensitivity to time-evolution with a perturbed Hamiltonian. One formulation used the Hamiltonian translated in $\phi$ by its generator $\hat{n}$, such that all the complexity moments emerge from taking $\phi$-derivatives of a Loschmidt amplitude. This amplitude has a spectral representation in terms of a propagator, which we characterized for specific types of systems where the spectral function $\Phi(E)$ is supported on a finite energy interval. Taking the late-time, small-$\phi$ limit, we reduced the amplitude to a simple one-dimensional integral over the spectral function. This gives a universal formula for the leading late-time growth of $\expval{\hat{n}^k(t)}$, in terms of ordinary energy moments of $\Phi(E)$. In this regime, we show that $\ket{\psi(t)}$ becomes \textit{classical} in the Krylov basis for generic states $\ket{\psi}$ with energy moments small compared to the bandwidth.  

A different parameter $\tau$, related to Euclidean time-evolution, was used to obtain a Loschmidt amplitude giving the time-derivative of Krylov complexity. The scaling of the $\tau$ perturbation offers a classification of Krylov complexity dynamics, with one class amenable to finite-rank approximations. We studied consistency conditions on the spectral representation of this perturbation and related them to derivatives of the $\phi$ propagator. Building upon this analysis, we established a general bound for complexity valid whenever the perturbation has finite Frobenius norm. We derived the precise late-time behavior for all the complexity moments in the important case of constant Lanczos coefficients. 

The joint $\phi$ and $t$-dependence of the states that appeared in the Loschmidt amplitude was studied geometrically using the Fubini-Study metric. The uncertainty bound from covariances translates into an upper bound for complexity in terms of the volume of this geometry. For real $C(t)$, geodesic deviation and curvature relate directly to the Krylov variance, which in turn constrains the decay of $C(t)$. We uncovered a surprising statistical property of the Krylov basis probabilities $\{|\varphi_n(t)|^2\}$, relating their inverse-participation ratio to their standard deviation, that we checked across several physical systems. 

It remains to be seen if the approximate kernel \eqref{eq:Szego_Kernel} accurately reproduces the true complexity distribution at shorter times when using its full structure, including the non-universal phases. An easier task is to obtain the subleading terms in \eqref{eq:ballistic_scaling} by evaluating the approximate kernel to the next order in $1/t$. Extending these results for $\Phi(E)$ with support on the entire real line using the asymptotic results of \cite{deift_1999, deift2001riemann} would be an important next step---see \cite{murugan2026spectraltopologyuniversalkrylov} for recent work in this direction. Such a computation could give $\expval{\hat{n}^k(t)}$ directly from $\Phi(E)$ for some range of $t$, without detailed knowledge of the Lanczos coefficients. That this should be possible is suggested by results of \cite{Balasubramanian:2023_plateau,Chakraborty2024,Ismail:2025pqn} who compute the saturation value of Krylov complexity directly from $\Phi(E)$ in some instances. We sketched a calculation along these lines for specific models of the energy-basis matrix elements of $\hat{n}$.

The finite-rank perturbation written in \eqref{eq:deformed_H_for_physical_amplitude} could be useful for generic systems where $C(t)$ is a complicated function, or only known through numerics. When the Krylov space dimension is smaller than the Hilbert space dimension, an optimal local $H(\tau, n_{*})$ can be designed by judicious choice of projectors orthogonal to the Krylov space of $\ket{\psi_0}$. \cite{Chakraborty_2025} already gave the general recipe to measure the $\phi$-amplitude from many-body scars embedded in a two-copy setup. In that setup, the ability to write down a microscopic interaction term is closely related to how we interpret Krylov complexity physically. This work aims to further stimulate connections between complexity measures and experimentally obtainable amplitudes.

The connections between the $\tau$-deformations and the $\phi$-deformations defined for each $\tau$ are intriguing. We leave a thorough analysis of the $\tau$-dependent kernel $K^{\tau}(E_1, E_2, \phi)$ defined using \eqref{eq:ftau_deformation_def} to work in progress \cite{dc_2026}. The important physical question is whether the state $e^{-i M(\phi) (t-i\tau_2)} \ket{\psi^{\tau_1}_0}$ (in terms of Euclidean time parameters $\tau_1$, $\tau_2$) can be described simply in terms of a complexified time-evolution $\ket{\psi(z)}$ with a certain $z(\phi, t, \tau_1, \tau_2)$. This is exactly true for the Heisenberg-Weyl algebra, and could be a useful approximate starting point more generally \cite{Chakraborty_2025}. This relation is also hinted at by ``Krylov-winding'' introduced in \cite{Krylov_Winding}, which roughly states that complexified time evolution with $z$ factorizes as $e^{i \hat{n} \phi(z)}$ acting on a $z$-dependent reference state with strictly positive-definite $\varphi_n$. By incorporating these additional Euclidean time directions, the complexity geometry of Section \ref{sec:complexity_geometry} could provide a thorough conceptual framework. 

The stretching of the $g_{\phi\phi}$ component of the metric means that $\ket{\psi(t)}$ gets localized in a dual angular basis $\ket{\theta}$. Instances of such dual bases were studied in \cite{Basu:2025mmm,basu2025complexitygrowthkrylovwignerfunction} and \cite{Murugan:2026yyu}, although different mathematical structures were used. The common feature is that $e^{i \hat{n} \phi}$ roughly acts like translations in $\theta$. The geometry of Section \ref{sec:complexity_geometry}, as well as the classification of $\tF$ can be analyzed through the profile of $\ket{\psi(t)}$ in $\ket{\theta}$ basis. This is facilitated by a dictionary between $\ket{\theta}$ and the energy basis. For bounded $\Phi(E)$, the energy basis can be parametrized by an angle which gets translated by the action of $e^{i \hat{n} \phi}$. We exploited this fact to derive the universal late-time complexity distribution. The emergent ``smoothness'' of the Krylov distribution manifests in our work as the general observation $\Delta n(t)\ \text{IPR}_{t} \sim \text{constant}$---a point that deserves further investigation.   

We believe that our emphasis on Krylov complexity as the sensitivity of an amplitude lays the foundation for a deeper understanding of its connection with chaos.    
\section{Acknowledgments}
I thank Diptarka Das, Refat Ismail, Sergio Aguilar-Gutierrez, Onkar Parrikar, Anatoli Polkovnikov, Pratik Rath and Dario Rosa for discussions related to this work. I also thank Nikolaos Angelinos, and especially, Anatoly Dymarsky, for discussions and collaboration during preliminary stages of this work. I am supported by the S\~ao Paulo Research Foundation (FAPESP) through the grant 2024/13100-8.
\bibliography{loschmidt}
\appendix
\section{Poisson Kernel}
\label{sec:Poisson_Kernel}
\subsection{General Formalism}
In this section, having taken the thermodynamic limit first, we consider a continuous energy spectrum which makes the spectral function $\Phi(E)$ continuous and normalized it to be consistent with $\bra{0}\ket{0} = 1$. The Krylov basis is specified by orthogonal polynomials given in  \eqref{eq:Orthogonal_Polynomial_Krylov}. In terms of the orthonormal polynomials $p_n(E)$, we define the Poisson kernel as 
\be K(E_1, E_2, \phi) \ := \sqrt{\Phi(E_1) \Phi(E_2)} \lim_{r \rightarrow 1^{-}}\sum^{n=\infty}_{n=0} r^n e^{ i n \phi }  p_n(E_1) p_n(E_2). 
\label{eq:reproducing_Poisson_Kernel}
\ee 
The factors of $\sqrt{\Phi(E_1) \Phi(E_2)}$ included in \eqref{eq:reproducing_Poisson_Kernel} to ensure the following properties
\be   K(E_1, E_2, 0) = \delta(E_1 - E_2) \label{eq:completeness_in_L2}, \ee 
\be \int K(E_1, E_2, \phi_{1}) K(E_2, E_3, \phi_{2}) dE_2 = K(E_1, E_3, \phi_1 + \phi_2), \label{eq:reproducing_property}  \ee
which follow from \eqref{eq:orthogonality_for_polynomials} assuming \eqref{eq:reproducing_Poisson_Kernel} is well-defined. 
It is useful to think of the formal eigenstates $\ket{E}$, defined by $M \ket{E} = E \ket{E}$. These are not necessarily the physical eigenstates. Rather, they are the objects that help define the unitary map between square summable sequences and $L^2(\mathbb R, \mu)$. That being said, practically, we can treat them as being coarse-grained energy eigenstates. In accordance with usual conventions, we can write the normalization and completeness relations to be 
\be \bra{E_1} \ket{E_2} = \frac{\delta(E_1 - E_2) }{\rho(E_1)}, ~~~\int \rho(E) \ket{E}\bra{E} dE = 1, \ee
where we introduced $\rho(E)$, the actual (undetermined) density of states of the system. Note that the only the combination $\Phi(E) = \rho(E) \abs{\bra{\psi_0}\ket{E}}^2$ can appear for physical quantities we study within this formalism. For example, evaluating the completeness relation inside $\bra{n} \ket{m}$ reduces to the orthonormality condition \eqref{eq:orthogonality_for_polynomials}. 

$K(E_1, E_2, \phi)$ is the propagator for $\hat{n}$ translation written in the $\ket{E}$ basis:
\be K(E_1, E_2, \phi) = \sqrt{\rho(E_1) \rho(E_2)} \bra{E_1} e^{i \hat{n} \phi} \ket{E_2},  \ee 
and then the survival amplitude defined in \eqref{eq:Gt_def} becomes \eqref{eq:G_from_E1_and_E2}.

Using \eqref{eq:G_from_E1_and_E2}, we can express $\expval{\hat{n}(t)}$ in terms of $t$ and $\phi$ derivatives as 
\be \expval{\hat{n}(t)} = \int e^{i (E_1-E_2) t} \sqrt{\Phi(E_1) \Phi(E_2)} \Big( -i  \partial_{\phi} K(E_1, E_2, \phi) \Big|_{\phi=0} \Big)dE_1  dE_2. \label{eq:kphidot_kernel_defining_complexity}  \ee 
We identify $\tilde{n}(E_1, E_2) := -i \partial_{\phi} K(E_1, E_2, \phi) \Big|_{\phi=0}$ as the kernel that stands for the matrix elements $\bra{E_1}\hat{n}\ket{E_2}$ in these conventions. We can define analogous spectral kernels for $B$ and $\mF$:
\be \tilde{B}(E_1, E_2) := \sqrt{\rho(E_1) \rho(E_2) } \bra{E_1} B \ket{E_2}. \label{eq:def_btilde_kernel} \ee 
\be \tF(E_1, E_2) :=\sqrt{\rho(E_1) \rho(E_2) } \bra{E_1} \mF \ket{E_2}. \label{eq:def_ftilde_kernel}\ee 
These satisfy the formal relations analogous to 
\be \tilde{B}(E_1, E_2) =  -i (E_1-E_2) \tilde{n}(E_1, E_2), \label{eq:B_from_phi_der}\ee
\be  \tF(E_1, E_2)  =  -i(E_1 - E_2) \tilde{B}(E_1, E_2) = -(E_1-E_2)^2 \tilde{n}(E_1, E_2),  \label{eq:F_from_B_kernel} \ee 
where \eqref{eq:B_from_phi_der}, \eqref{eq:F_from_B_kernel} need to be interpreted carefully as distributional identities. That means that the we can only expect both sides to agree when integrated over in expressions like \eqref{eq:kphidot_kernel_defining_complexity}. When $L_0=0$ we have another representation for $\tilde{B}$  
\be \tilde{B}(E_1, E_2) = \int E^{\prime} K(E_1, E^{\prime}, \frac{\pi}{2} ) K(E^{\prime}, E_2, -\frac{\pi}{2} ) d E^{\prime} \label{eq:B_from_Kint}. \ee 

In \eqref{eq:G_from_E1_and_E2}, it is often convenient to rotate the variables of integration to  
\be \overline{E} = \frac{1}{2} (E_1 + E_2) ~~ \text{and}~~ \omega = E_1 -E_2, \label{eq:ebar_and_omega_def} \ee 
and write 
\be G(t, \phi) = \int \int \sqrt{\Phi(E + {\omega \over 2}) \Phi( E - {\omega \over 2} )} e^{i \omega t} K(\overline{E}, \omega, \phi ) d \overline{E} d\omega,   \ee  
where to avoid proliferation of new symbols, we will denote the function $f(\overline{E}, \omega)$ with the same letter as $ f(E_1, E_2)$. The domain of integration over $\overline{E}$ and $\omega$ will be coupled whenever the range of energies is finite. For example, if the support of $\Phi(E)$ is $[-E_0, E_0]$, then the integral becomes 
\be G(t,\phi) = \int^{2 E_0}_{-2 E_0} \bigg( \int^{( E_0 -\frac{\abs{\omega}}{2} )}_{-( E_0 -\frac{\abs{\omega}}{2} )} \sqrt{\Phi(E + {\omega \over 2}) \Phi( E - {\omega \over 2} )}  K(\overline{E}, \omega, \phi ) d  \overline{E}  \bigg)  e^{i \omega t}  d\omega. \ee 

\subsection{Modeling the kernels}
\label{sec:model_for_K}
If we can directly estimate  $K(E_1, E_2, \phi)$ through exact results  or asymptotics, then it may be advantageous to compute \eqref{eq:G_from_E1_and_E2} though saddle point or numerical integration and then take derivatives. The derivatives $\partial^k_{\phi} K(E_1, E_2, \phi)$ will typically only be distributional at $\phi=0$. They might only make sense through a regularization for $r$. It may be more convenient to use $\tilde{F}(E_1, E_2)$ (or $ \sqrt{\Phi(E_1)\Phi(E_2)}\tilde{F}(E_1, E_2)$) in some cases, both because it could be simpler to compute and because it maybe better behaved as a function. As a simple example, we use the well-known Mehler kernel for Hermite polynomials orthogonal to $\Phi(E)=\frac{1}{\sqrt{2 \pi \alpha}} e^{-\frac{E^2}{2 \alpha^2}}$:
\be K(E_1, E_2, \phi) = \frac{1}{\sqrt{2\pi} \alpha\sqrt{1-e^{2i\phi}}} \exp\left( \frac{i}{2\alpha^2} \left[ E_1E_2\csc\phi - \frac{E_1^2+E_2^2}{2}\cot\phi \right] \right). \label{eq:mehler_kernel} \ee 
Keeping identical conventions as Appendix \ref{sec:heisenberg_algebra}, it can be checked by direct integration that the kernel reproduces \eqref{eq:gt_for_Heisenberg}, which of course, is much simpler using the Heisenberg algebra. Suppose, one did not know \eqref{eq:mehler_kernel} explicitly but knew its leading order expansion in $\phi$, then using the formal relation \eqref{eq:F_from_B_kernel} could be simpler. This reduces to $2 \alpha^2 \delta(E_1-E_2)$, rather than a more complicated object. A similar fact holds for the Hardy-Hille kernel, where $\tilde{F}(E_1, E_2) \sim \tg E \delta(E_1-E_2)$.

Both these examples display the more general fact that $\tilde{n}(E_1, E_2)$ will be a sharply peaked function around $\omega=0$,  where $\omega= E_1-E_2$ from \eqref{eq:ebar_and_omega_def}. 
When $\tilde{F}$ is square integrable, we expect from \eqref{eq:F_from_B_kernel} that $\tilde{n}(\overline{E}, \omega)$ will have a power-law singularity $\abs{\omega}^{-\eta}$ for $\eta>0$, where we generically expect $\eta \geq 2$. Importantly, we may be able to rewrite \eqref{eq:kphidot_kernel_defining_complexity} as 
\be  \expval{\hat{n}(t)}  \approx \int e^{i \omega t} \big(   \int \Phi(\overline{E} ) \tilde{n}(\overline{E} , \omega) \big) d\omega,  \label{eq:simplification_kt} \ee 
in \eqref{eq:simplification_kt}, we approximated $\Phi(\overline{E} + \omega) \approx \Phi(\overline{E} )$. The point of writing \eqref{eq:simplification_kt} is to show how the $\overline{E}$ dependence of $\tilde{n}(\overline{E}, \omega)$ gets integrated out, such that the time-dependence of $\expval{\hat{n}(t)}$ is largely determined by how singular $\tilde{n}$ is at small $\omega$. From the discussion about bounded Lanczos coefficients \eqref{eq:spectral_relations_n_B_Bcom} and also the asymptotics of Sec. \ref{sec:SzegoAsymptotics}, we can write a schematic model for the low-frequency behavior of $\tilde{n}$ for physical systems where the support of the spectral function scales extensively with system size $\sim \log N$, but is bounded for any finite $N$. We take
\be \tilde{n}(\overline{E}, \omega) \sim \frac{\mathcal{A}_1(\overline{E})}{\omega^2} +  \mathcal{A}^{(N)}_2(\overline{E}, \omega) + \ldots \label{eq:factorization_tilden} \ee 
where $\mathcal{A}^{(N)}_{2}(\overline{E}, \omega)$ is a function which diverges faster at small $\omega$ than $\omega^{-2}$, when $N \rightarrow \infty$. At finite $N$, the $\omega$ dependence of $\mathcal{A}^{(N)}_{2}$ gets cuts off at a scale $\omega^{*}(N)$, which is the inverse of the scale $t^{*}(N)$ controlling the transition of $\expval{\hat{n}(t)}$ from faster asymptotic growth to linear in $t$ growth. For example, if $C(t)$ described the spectral form factor of local spin chains, then $\Phi(E)$ would be a gaussian away from its tails, and $\mathcal{A}^{(N)}_2(\overline{E}, \omega)$ would approach $\frac{1}{\omega^2} \delta(\omega) \sim \frac{d^2}{d\omega^2} \delta(\omega) $ in the infinite $N$ limit. $\mathcal{A}^{(N)}_2(\overline{E}, \omega)$ is responsible for the quadratic growth of complexity. We can pin down the $\omega$-dependence of $\mathcal{A}^{(N)}_2(\overline{E}, \omega)$ more specifically in some cases. Suppose Krylov complexity grows as a power-law $t^{1+\chi}$ for $0<\chi \leq 1$ and $\mF$ is bounded, then we can write
\be \lim_{N \rightarrow \infty} \mathcal{A}^{(N)}_2(\overline{E}, \omega) \sim \frac{\mathcal{A}_2(\overline{E})}{\abs{\omega}^{2+\chi}}.\label{eq:leading_singularity_n_frequency_domain} \ee 
A simple example where this holds is for $a_n =0$ and smooth $b^2_n \sim n^{\kappa}$. Then the continuum limit calculations of \cite{Barb_n_2019, M_ck_2022}  predicts $\chi=\frac{\kappa}{2-\kappa}$. Because $\mF$ is compact we can rule out the asymptotic power-law emerging from more complicated singularity structure involving $\delta$-functions and their derivatives. This is in contrast with the $\chi=1$ example we just presented above for Gaussian $\Phi(E)$.

Restricting to $a_n=0$, using \eqref{eq:B_from_phi_der} and \eqref{eq:gt_phi_interaction_picture}, we could in principle deduce the asymptotic scaling for all the moments $\expval{\hat{n}^k(t)}$. The factorized ansatz \eqref{eq:leading_singularity_n_frequency_domain} gives a particularly simple form of the late-time behavior of complexity distribution. Provided we ignore the $\overline{E}$ dependence from $\mathcal{A}_2(\overline{E})$ in all but the final integral, the time-ordered integral over $\tilde{B}$ in \eqref{eq:gt_phi_interaction_picture} collapses into:  $\text{constant} \times \big( \sin(\phi) \mathcal{A}_2(\overline{E})  \frac{ t^{ (1+\chi)} }{(1+\chi) } \big)^k \frac{1}{k!}$, which after integrating over $\overline{E}$ still gives the asymptotic $\expval{\hat{n}^k(t)} \sim t^{(1+\chi)k}$ at late times. 

In the above argument, we assumed both the factorization \eqref{eq:factorization_tilden} and that we could ignore the mean energy dependence, $\mathcal{A}_2(\overline{E})$ in each of the integrals. The validity of these approximations is not obvious and needs to be evaluated on a case-by basis. There is an important class where this logic can be verified to hold. This is the case where $\mathcal{A}_2(\overline{E})$ coincides with $\mathcal{A}_1(\overline{E})$, when the Lanczos coefficients saturate to a constant. The singular part of $\tilde{n}(E_1,E_2)$ is accurately captured using orthogonal polynomial asymptotics in \eqref{eq:Szego_ntilde}. It certainly does not have the factorized form \eqref{eq:factorization_tilden}. However, for our late-time approximation it is sufficient to extract the coefficient of $\frac{1}{\omega^2}$ in $\tilde{n}$ evaluated at $\omega=0$. This is just $-\tilde{F}(E, \omega)$ at $\omega=0$, which from \eqref{eq:Szego_ntilde} is given by $\mathcal{A}(E)=-\frac{1}{\pi} \sqrt{E^2_0 - E^2}$.

Then on following the steps outlined above, we get the simple integral form of $G(t,\phi)$ 
\be G(t, \phi) \approx \int^{E_0}_{-E_0} \Phi(E) \exp( -i t \Big( (1 - \cos(\phi) )  E - \sqrt{E^2_0-E^2} \sin(\phi) \Big) )  dE. \label{eq:approx_gt_phi_for_RMT} \ee
Expanding $\phi$ to linear order in \eqref{eq:approx_gt_phi_for_RMT} reproduces \eqref{eq:universal_Gt_in_Szego_class}. Doing so can be justified with the expectation that \eqref{eq:ballistic_scaling} holds. It might be worth analyzing if retaining the nonlinear $\phi$-dependence evaluates $\expval{\hat{n}^k(t)}$ to better accuracy. The more important point is whether this method of extracting late-time asymptotics from small $\omega$ behavior of $\tilde{B}$ generalizes to a wider class of systems.  
\subsection{Frobenius Norm Constraint}
\label{sec:frobenius_norm_kernel}
Now we consider a system where $\tilde{F}(E_1, E_2)$ is square-integrable, despite the Lanczos coefficients growing without bound. The relevant integral which gives the square of the Frobenius norm is $ \int \int |\tilde{F}(E_1, E_2)|^2  dE_1 dE_2 $. This can be checked by explicitly plugging in the definition \eqref{eq:def_ftilde_kernel} in terms of matrix elements. Generically, in such cases, we expect $\expval{\hat{n}(t)}$ to grow asymptotically faster than linear in $t$. An example is the $t^{1+\chi}$ behavior that arises from \eqref{eq:leading_singularity_n_frequency_domain}. Assuming the leading singular behavior \eqref{eq:leading_singularity_n_frequency_domain} of $\tilde{n}$ and on using \eqref{eq:F_from_B_kernel}, we get that $\tilde{F}(E, \omega)$ will go as $\frac{1}{\abs{\omega}^{\chi} }$. Demanding that it be integrable in $\omega$ implies that $\int d\omega \frac{1}{\abs{\omega}^{2 \chi}}$ does not diverge due to contributions near $\omega=0$. This gives $\chi < \frac{1}{2}$. The upper bound $\chi =\frac{1}{2}$ gives a late-time asymptotic of $t^\frac{3}{2}$. This matches the asymptotic growth rate predicted by \eqref{eq:upper_bound_frobenius_tighter}, with the additional assumption that $ \lim_{t \rightarrow \infty}C(t) = 0$.   
\section{Asymptotics for spectral function with bounded support}
\label{sec:SzegoAsymptotics}
We now discuss asymptotics for orthogonal polynomials when the spectral function $\Phi(x)$ has support only on a finite subset of $\mathbb R$. Without loss of generality, we take this to be $[-1, 1 ]$ to enhance the readability of the answers. A linear transformation on the spectral variable $x$ recovers the answer for arbitrary intervals. We do not require $\Phi(x)$ to be even, thus the following formula applies also when the $a_n$ coefficients are non-vanishing. We assume the mild regularity condition that $\int^{1}_{-1} (1-x)^{-\frac{1}{2} } \log \Phi(x) dx $ exists, and we parameterize $x = \cos(\theta_x)$. Then, asymptotically \cite{szego1959orthogonal} (see (12.1.8))
 \be \sqrt{\Phi(x) } p_n(x) = \sqrt{\frac{2}{\pi \sin \theta_x}} \cos[ n \theta_x + \gamma_x ] + O( \log n)^{-\lambda} \label{eq:szego_polynomial},  \ee 
where $\lambda >0$ is a constant that depends on $\Phi(x)$. $\gamma(\theta)$ is determined by 
\be  \gamma_x = \frac{1}{4 \pi} \int^{\pi}_{-\pi} \log( \frac{\Phi(\cos \xi  ) \abs{\sin \xi } }{ \Phi (\cos \theta_x ) \abs{\sin \theta_x } } ) \cot \frac{1}{2}( \theta_x-\xi) d\xi. \ee 
This allows us to approximate
\be K(x, y, \phi) \approx \lim_{r \rightarrow 1^{-} } \frac{1}{{2 \pi} \sqrt{\sin \theta_x \sin \theta_y} }   \Big[  \frac{e^{i \gamma_{+} }}{1- r e^{ i (\phi + \theta_x + \theta_y) } }  + \frac{e^{-i \gamma_{+} }}{1- r e^{ i (\phi - \theta_x - \theta_y) } } +   \frac{e^{i \gamma_{-} }}{1- r e^{ i (\phi + \theta_x - \theta_y) } }  + \frac{e^{-i \gamma_{-} }}{1- r e^{ i (\phi  -\theta_x + \theta_y) } }  \Big]. \label{eq:Szego_Kernel} \ee 
where $\gamma_{\pm} = \gamma_{x} \pm \gamma_{y}$. \eqref{eq:Szego_Kernel} shows that $e^{i \hat{n} \phi}$ acts by translating the spectral angle $\theta_x \rightarrow \theta_x \pm \phi$, as manifested in the poles of \eqref{eq:Szego_Kernel}. This structure of $K(x, y, \phi)$ can be understood using the arguments put forward in \cite{Chakraborty_2025} regarding $e^{i\hat{n} \phi}$ generating sharply localized motion on the spectrum. 

We can write the $O(\phi)$ contribution to \eqref{eq:Szego_Kernel} more transparently in terms of the energies $x$ and $y$ if we take the $r \rightarrow 1^{-}$ limit 
\be    
\tilde{n}(x,y) = \frac{1}{\pi (1 - x^2)^{1/4} (1 - y^2)^{1/4}} \frac{  (xy-1) \cos \gamma_x \cos \gamma_y -\sqrt{1-x^2} \sqrt{1-y^2}
   \sin \gamma_x \sin \gamma_y  }{(x-y)^2}, \label{eq:Szego_ntilde}
\ee 
although the right-hand side needs to be properly regulated to obtain $\expval{\hat{n}(t)}$ with correct behavior. It is convenient to switch to the spectral representation of $\tilde{F}$ using \eqref{eq:F_from_B_kernel}, which gives in the time-domain 
\be \partial^2_{t} \expval{\hat{n}(t)} = -\int^{1}_{-1} dx \int^{1}_{-1} dy  \frac{ e^{i (x-y) t} \sqrt{\Phi(x) \Phi(y)} }{\pi (1 - x^2)^{1/4} (1 - y^2)^{1/4} }[  (xy-1) \cos \gamma_x \cos \gamma_y -\sqrt{1-x^2} \sqrt{1-y^2}
   \sin \gamma_x \sin \gamma_y ]. \ee 
In the case of constant $b$ sequence, the kernel \eqref{eq:Szego_Kernel} is exact, with $\gamma_x = \theta_x-\frac{\pi}{2}$. We take the normalization $b_n=\frac{1}{2}$ to be consistent with $\qq=0$ limit of \eqref{eq:dssyk_Lanczos}, such that the density of states is 
\be \Phi(x) = \frac{2}{\pi} \sqrt{1-x^2}, \label{eq:rmt_Phi} \ee 
and the kernel for $\hat{n}$ becomes
\be \tilde{n}(x,y) = - \frac{1}{\pi} \frac{(1-x^2)^{1/4}(1-y^2)^{1/4}}{(x-y)^{2}}. \label{eq:tilden_RMT}~~\ee
It can also be extracted by noticing that only $\mF_{00} = 2b^2_0$ is nonvanishing, such that
\be 
 \partial^2_{t} \expval{\hat{n}(t)} =  \frac{2}{\pi^2} \int^{1}_{-1} dx \int^{1}_{-1}  dy \sqrt{(1-x^2)(1-y^2)} e^{i (x-y) t} = \frac{2}{t^2} J^2_{1}(t),
\ee 
using \eqref{eq:t_derivative_Kt}. This can be used to read off $\tF$ 
\be \tF(x, y) = \frac{1}{2} \sqrt{\Phi(x) \Phi(y)} = \frac{1}{\pi} (1-x^2)^{1/4}(1-y^2)^{1/4}.  \ee 
As a consistency check, using the $\tF$ above in \eqref{eq:F_from_B_kernel}, reproduces \eqref{eq:tilden_RMT}.  

Now we will present a general computation that extracts the dominant term in $\expval{\hat{n}^k(t)}$ for late times, by reducing the two-dimensional integral \eqref{eq:G_from_E1_and_E2}, with the approximate \eqref{eq:Szego_Kernel} to a simple one-dimensional integral. The main observation is that only one of the four terms in \eqref{eq:Szego_Kernel}, the one with the phase $e^{i \gamma_{-}}$, contributes in this regime. 

Let us define $\theta_{\pm} = \theta_x  \pm \theta_y$. The terms with $(1-r e^{ i \phi \pm i\theta_{+}  } )$
 for infinitesimal $\phi$ are localized around $\theta_x, \theta_y$ both close to $0$ or $\pi$. These are the endpoints of the spectrum suppressed by the factor of $\sqrt{\sin(\theta_x)\sin(\theta_y)}$ that appears when changing integration variables from the energy variable $x$ to $\theta$. Now, let us write the entire integral that gives the approximate version of $G(t, \phi)$ as an integral over the variables $\theta_{\pm}$
\be \frac{1}{2 \pi} \int^{\pi}_{-\pi}  \R(t,\theta_{-})  \lim_{r \rightarrow 1^{-}} \Bigg(\frac{e^{i \gamma_{-}}}{1-r e^ { i( \phi + \theta_{-} )}} + \frac{e^{-i \gamma_{-}}}{1-r e^{i( \phi - \theta_{-})} }    \Bigg) d \theta_{-} \label{eq:isolate_theta_minus_integral},  \ee 
in terms of  
\be \R(t,\thetami ) := \frac{1}{2}\int^{2 \pi - \abs{\thetami}}_{\abs{\thetami}} \sqrt{ \Phi(x) \Phi(y) \sin(\frac{\thetapl +\thetami }{2})\sin(\frac{\thetapl - \thetami}{2})} e^{-2 i \sin(\frac{1}{2}\thetapl) \sin(\frac{1}{2}\thetami) t} d\thetapl.   \label{eq:def_rt_minus} \ee 
Since we integrate $\R$ in $\thetami$ restricted to the interval $[-\pi, \pi]$ we can consider its Fourier decomposition in $\thetami$ 
\be \R(t, \theta_{-} )  = \sum_{n \in \mathbb Z} \R_{n}(t) e^{i n \theta_{-}}, \ee 
with $\R_{n}(t)$ defined by 
\be \R_{n}(t) = \frac{1}{2 \pi} \int^{\pi}_{-\pi} \R(t, \theta_{-} ) e^{- i n \theta_{\scriptscriptstyle-}} d\thetami, \label{eq:fourier_coeff_kernel} \ee
\eqref{eq:fourier_coeff_kernel} is a highly oscillatory integral in $\thetami$ for large $t$. The location of the dominant Fourier coefficients can be estimated by finding the stationary phase of the integrand: 
\be t \Big( \sin(\frac{1}{2}\thetapl) \cos(\frac{1}{2}\thetami) + x^{\ast}  \Big) = 0, \label{eq:large_t_saddle_for_Rn} \ee
in terms of $x^{\ast}=n^{\ast}/t$. This scaling allows us to neglect contributions away from the stationary point for $\thetami$ as $t \rightarrow \infty$. Within the bounds of the integral $\cos(\frac{1}{2}\thetami)$ and $\sin(\frac{1}{2}\thetapl)$ are both positive, and thus $x$ is negative. Having established that $\R_{n}(t)$ will be sharply localized around some $x^{\ast}$ defined by \eqref{eq:large_t_saddle_for_Rn}, we can express the integral in $\theta_{-}$ as 
\be \frac{1}{2 \pi} \sum_{m \in \mathbb Z} \R_{m}(t) \int^{\pi}_{-\pi} e^{i m \thetami} \lim_{r \rightarrow 1^{-}} \bigg( e^{i \gamma_{-}} \sum_{n \geq 0} r^n e^{i n (\phi + \thetami)}   + e^{-i \gamma_{-}} \sum_{n \geq 0} r^n e^{i n (\phi - \thetami)}  \bigg)   d\theta_{-}. \label{eq:theta_minus_in_terms_of_Fourier} \ee
Because the Fourier coefficients $\R_{m}(t)$ with $m>0$ do not contribute in this regime, we can neglect the second sum over $n$ in \eqref{eq:theta_minus_in_terms_of_Fourier}. Evaluating the Kronecker deltas $\delta_{m+n } = \frac{1}{2\pi} \int^{\pi}_{-\pi} e^{i (m+n) \thetami }d\thetami$ from the discrete sums gives $e^{i \gamma_{-}} \R(t, -\phi )$. This gives the prescription for turning the kernel in \eqref{eq:Szego_Kernel} into $\delta(\theta_{-}+\phi)$ with the correct normalization. The remaining integral over $\theta_{+}$ can be now performed but it still contains complicated $\phi$ dependence in $\gamma_{\pm}$ and the endpoints. 

To get a simple universal formula, we assume that $\lim_{t\rightarrow \infty} \expval{\hat{n}^k(t)}/t^{k} = v_{k}$ for constant $v_{k}$. With this assumption, if we were to view $G(t,\phi)$ as a function of $\phi \rightarrow 0$, $t \rightarrow \infty$ with $\phi t$ held fixed, we could linearize the exponential time-dependent factor to \eqref{eq:def_rt_minus} to $e^{ 2 i \sin(\thetapl/2) \sin(\phi/2)  t} \rightarrow e^{i \sin(\theta_{+}) \phi t}$ and ignore the $\phi$-dependence from the endpoints, since they do not affect $v_{k}$. Then we use symmetry to write the integral over $\theta$ in the original range: 
\be G(t,\phi) \approx \int^{\pi}_{0}  \Phi(\cos \theta )  e^{i \sin(\theta) \phi t } \sin \theta d\theta =  \int^{1}_{-1} \Phi(E) e^{i \sqrt{1-E^2} \phi t} dE. \label{eq:simplified_ansatz_Gt}  
\ee 
Within the large-$t$ approximation, still using \eqref{eq:Szego_Kernel}, the leading correction is of the form $-i \phi \int^{\pi}_{0}  \Phi(\cos \theta )  e^{i \sin(\theta) \phi t } \partial_{\theta} \gamma(\theta) \sin \theta d\theta$. This gives the scaling of the remainder term in \eqref{eq:ballistic_scaling}. The corrections coming from neglecting the end-points and the $e^{i (\phi + \thetapl)}$ poles come with higher powers of $\phi$ and contribute lower powers of $t$.

When the physical range of energies is $[-E_{0}, E_0]$, the answer becomes  
\begin{subequations}
\begin{align}
    G(t,\phi) &\approx \int^{E_{0}}_{-E_0} \Phi(E) e^{i \sqrt{E^2_0-E^2} \phi t} dE,\\  \label{eq:Gt_ansatz_with_Emax}
    v_{k} &= \int^{E_{0}}_{-E_0} \Phi(E) (E^2_0-E^2)^{\frac{k}{2}} dE. 
    \end{align}
\end{subequations} 
It is hard to rigorously pinpoint the validity of \eqref{eq:simplified_ansatz_Gt} due to the sequence of approximations leading to it. It strongly depends on the approach of the Lanczos coefficients to their asymptotic values. With our present conventions, they are $b_{n} \rightarrow \frac{1}{2}E_0$ and $a_{n} \rightarrow 0$. Let us call this transition scale $\nc$. $\nc$ grows with $ E_0$ and shrinks with the width of $\Phi(E)$. The parametric dependence of $\nc$ on $E_0$ and the width will be non-universal depending on the behavior of $\Phi(E)$ away from the edges. For example, if $\Phi(E)$ is a Gaussian in the bulk with variance $\alpha^2$, $b_n = \alpha \sqrt{n+1}$ for $n \ll \nc$. This gives $\nc \propto E^2_0/\alpha^2$.

The onset of the scaling in \eqref{eq:ballistic_scaling} will also depend on $k$. Let us denote the onset time by $t^{(k)}_{\text{on}}$. It can be estimated using the transient non-universal growth of $\expval{\hat{n}^k(t)}$ by
\be \expval{\hat{n}^{k}(t^{(k)}_{\text{on}} )} := \nc^k. \label{eq:t_onset}\ee
For any fixed $t>0$, $ \expval{\hat{n}^{k}(t)}^{\frac{1}{k}}$ is monotonically increasing in $k$, from Lyapunov's inequality. To prove this one can simply use Jensen's inequality $\expval{ (\hat{n}^{k} )^\frac{k+1}{k}} \geq \expval{\hat{n}^{k}}^\frac{k+1}{k}$, using that $n$ is non-negative and $f(x)=x^\frac{{k+1}}{k}$ is convex. For $\expval{\hat{n}^k(t)}$ growing in $t$, this implies that $t^{(k+1)}_{\text{on}} \geq t^{(k)}_{\text{on}}$. After onset, this behavior persists in physical systems for timescales diverging as $N \rightarrow \infty$, which was assumed throughout this section. For the $b_n = \alpha \sqrt{n+1}$ example above, using \eqref{eq:complexity_for_Heisenberg}, $t^{(1)}_{\text{on}} = \sqrt{\frac{E_0}{\alpha^2}}$, provided there is no additional special structure in the tails of $\Phi(E)$. Using \eqref{eq:gt_for_Heisenberg} and \eqref{eq:t_onset}, one can verify that $t^{(k)}_{\text{on}}$ scales parametrically the same way, regardless of $k$.

Our derivation does not distinguish between cases where $\lim_{t\rightarrow \infty }\frac{1}{v_{k} t^{k}} \expval{\hat{n}^k(t)}$ is exactly $1$ and where it is a constant close to $1$. For the constant $b$-sequence, this ratio approaches exactly $1$ for first few $k$, when checked using Bessel function asymptotics for $\varphi_n(t)$. In all examples we checked numerically, this ratio seems to really asymptote to $1$ for low values of $k$. A separate but related question is the extent to which \eqref{eq:szego_polynomial} remains a good approximation when weighing by factors of $n^k$, as $k$ becomes large. The $k$-dependence of the quality of this approximation is likely related to the rate at which $b_{n}$ approaches its asymptotic value. It is possible that only for specific subclass of $\Phi(E)$, \eqref{eq:Gt_ansatz_with_Emax} truly holds to \textit{all} orders in $k$, while it holds to low-orders as $t\rightarrow\infty$ for all $\Phi(E)$ considered in this section.        

It is useful to isolate the scale-dependence of $v_k$ in $E_0$, by switching to the variable $x=\frac{E}{E_0} \in [-1, 1]$ and defining the rescaled spectral function $\hat{\Phi}(x) = E_0 \Phi(E_0 x)$. Define $\hat{v}_{k}$ as
\be \hat{v}_{k} := \int^{1}_{-1} \hat{\Phi}(x) (1-x^2)^{\frac{k}{2}} dx,\quad v_{k} = E^k_0 \hat{v}_{k}. \label{eq:complexity_moments_asymptotic_slope} \ee
The scaled moments are ordered $\hat{v}_{k+1} \leq \hat{v}_{k} \leq 1$, simply because $(1-x^2)$ always remains smaller than $1$. On the contrary,  ${\hat{v}_{k}}^{\frac{1}{k}}$ is strictly monotonically increasing following the convexity argument below \eqref{eq:t_onset}. When \eqref{eq:complexity_moments_asymptotic_slope} holds to arbitrarily high-orders, $\frac{\expval{\hat{n}^{k}(t)}^{1/k}}{t}$ approaches $E_{0}$ as $k \rightarrow \infty$. These facts can be easily understood together as a simple consequence of how the saddle point in energy moves with $k$ in \eqref{eq:complexity_moments_asymptotic_slope}. Larger $k$ shifts the saddle towards $E=0$, and when $\Phi(E)$ is unimodal centered around $0$, gives the simple estimate 
\be v_{k} \approx E^k_0 \sqrt{\frac{2 \pi}{k}} \Phi(0), \label{eq:c_k_asymptotic_in_k} \ee
The even complexity moments $v_{2m}$ have the explicit binomial expansion 
\be v_{2m} = \sum^{m}_{k=0} (-1)^k \binom{m}{k}  E^{2 (m-k)}_{0} \mu_{2k}, \label{eq:even_complexity_moments_through_even_moments} \ee 
in terms of the even energy moments. These can be expressed in terms of the Lanczos coefficients with indices $n \leq m-1$.  \eqref{eq:even_complexity_moments_through_even_moments} is invertible, giving $\mu_{2k}$ in terms of $v_{2m}$. These relations can be exploited to write the convenient generating function 
\be  \sum^{\infty}_{m=0} v_{2m}\frac{ s^{m}}{m!}  = e^{s E^2_{0}} \int^{E_0}_{-E_0} \rho(E) e^{-s E^2} dE.  \ee 

\eqref{eq:c_k_asymptotic_in_k} already features the interplay between the two dominant scales controlling $c_{k}$, the width of $\Phi(E)$, inverse to $\Phi(0)$ for symmetric $\Phi$, and $E_{0}$. Let us suppose that $\Phi(E)$ is sharply concentrated about its mean $\mu_{1}$. We expand \eqref{eq:Gt_ansatz_with_Emax} for $v_k$ around $E=\mu_{1}$, and get 
\be v_{k} = (E^2_0 - \mu^2_1 )^{\frac{k}{2}} \Bigg( 1 - \frac{k}{2} \frac{ (\mu_2 -\mu^2_1)}{\big( E^2_0 - \mu^2_1 )^2} (E^2_0 - (k-1) \mu^2_1 ) + \ldots \Bigg), \label{eq:v_k_expansion_central_moments}\ee
with the understanding that the series truncates for even $k$. The expansion in \eqref{eq:v_k_expansion_central_moments} is in the central moments $\int \Phi(E) ( E -\mu_1 )^k dE$, weighed by inverse powers of $(E^2_0 - \mu^2_1 )$. Provided $(E^2_0 - \mu^2_1 )$ is large compared to the width of the distribution, $v_{k} \approx (E^2_0 - \mu^2_1 )^{\frac{k}{2}}$ at least to low orders in $k$. The complexity variance is proportional to the energy variance: $\Delta n(t) \approx \frac{\mu_{1}}{\sqrt{E^2_0 - \mu^2_1}} b_0 t$  to leading order in this expansion. For fixed $t$, shrinking the width of $\Phi(E)$ makes the Krylov distribution increasingly classical. The last point is especially clear when $\mu_{1} = 0$ and $\Phi(E)$ is peaked around $E=0$ and the expansion analogous to \eqref{eq:v_k_expansion_central_moments} is presented in \eqref{eq:v_k_distribution_even_Phi}. Here the standard deviation has the expansion $\Delta n(t) = t \big(\frac{\sqrt{\mu_4-\mu^2_2}}{2 E_0} + O(1/E_0 )  \big)$. 

The Jacobi measure for $\Phi(E) \in [-1, 1]$ defined with two parameters $\alpha, \beta >-1$, 
\be 
\Phi(E) = \frac{
   \Gamma (\alpha +\beta +2)}{\Gamma (\alpha +1) \Gamma (\beta
   +1)} \frac{(1-E)^{\alpha } (1+E)^{\beta }}{2^{\alpha + \beta + 1} }, \label{eq:Jacobi_measure}
\ee
provides an example where direct analytical checks can be done. When $\alpha \neq \beta$, $\Phi(E)$ is not even. For simplicity, we only write the Lanczos coefficients when $\beta=\alpha$, 
\be b_n = \sqrt{ \frac{(n+1) (n+2 \alpha +1)}{ (2n + 2 \alpha +1) (2 n + 2\alpha+3)}}. \ee
 For the even Jacobi measures, $v_{k}$ can be evaluated to be
\be v_k = 2^k\frac{ \Gamma (2 \alpha +2)
   \Gamma \left(\frac{k}{2}+\alpha
   +1\right)^2}{\Gamma (\alpha +1)^2
   \Gamma (k+2 \alpha +2)}, \label{eq:v_k_for_even_jacobi} 
\ee 
Setting $\alpha=\frac{1}{2}$ yields the semicircle distribution \eqref{eq:rmt_Phi} with $b_n=\frac{1}{2}$. The late-time complexity distribution for the constant Lanczos coefficients compared with the substitution of \eqref{eq:v_k_for_even_jacobi} in \eqref{eq:ballistic_scaling} is shown in Fig.~\ref{fig:rmt_ratio}.

In general, the width of \eqref{eq:Jacobi_measure} is set by $\alpha$. It can be read off using $\mu_2$
\be \mu_2 = \frac{\sqrt{\pi } \Gamma (\alpha
   +1)}{2 \Gamma \left(\alpha+\frac{5}{2}\right)} = \frac{1}{2} \sqrt{\pi }
   \left(\frac{1}{\alpha
   }\right)^{3/2}+O\left(\frac
   {1}{\alpha^2 }\right)  , \ee
with $\mu_{2k} \propto (\alpha)^{-\frac{2k+1}{2}} $ for large $\alpha$. In the limit of $\alpha \gg 1$, the distribution becomes very narrow around $E=0$. $v_k = 1-\frac{k}{4 \alpha} + O\left(\frac
   {1}{\alpha^2 }\right)$, converges to $1$ for finite $k$ when $\alpha \rightarrow \infty$.

\eqref{eq:simplified_ansatz_Gt} can be evaluated explicitly for the constant $b_n=\frac{1}{2}$-in terms of Struve functions $\pmb{H}$ and Bessel functions $J$,
\be  G(t, \phi) = -2 i \pmb{H}_2(t \phi )+\frac{2 i \pmb{H}_1(t \phi )}{t \phi
   }+\frac{4 i t \phi }{3 \pi }-2 J_2(t \phi )+\frac{2 J_1(t
   \phi )}{t \phi } \label{eq:Gt_for_constant_b}\ee 
Expanding \eqref{eq:Gt_for_constant_b} for large $t$, gives $\abs{G(t, \phi)} \sim \frac{1}{\sqrt{\phi t}}$. This behavior is also directly checked using \eqref{eq:Gt_ansatz_with_Emax} in generality by the saddle point at $E=0$. We could estimate IPR$_{t}$ by integrating this $\abs{G(t, \phi)}^2$. This gives the estimate IPR$_{t}$ $\sim \frac{\log(t)}{\Delta n(t)}$ plus terms which decay more rapidly as $t \rightarrow \infty$.    
\section{Complexity Algebra}
\label{sec:complexity_algebra}
We assume the closure of the algebra between $B$, $M$ and $\hat{n}$:
\be \mF = i[B, M] = \tilde{\alpha} \hat{n} + \tilde{\gamma} M + \tilde{\delta} \mathbb I. \label{eq:closure_ansatz_complexity} \ee 
The expansion for the Heisenberg operator $\hat{n}(t)$,
\be \hat{n}(t) = \sum_k \frac{(it)^k}{k!}  \text{ad}^k_{M} \hat{n} ~~~ \text{where}~~~\text{ad}^k_{M} \hat{n} := \underbrace{[M, [M, \ldots  [M, \hat{n}] }_{k \text{ times}}  \ldots ] \ee 
can now be truncated, using the fact that after hitting $\text{ad}^3_{M} \hat{n} \propto [M, \tilde{\alpha} \hat{n} + \tilde{\gamma} M ]$ and so on. A neat way to solve the system is by noticing that it implies the following differential equation
\be \partial^2_{t} \hat{n}(t) = \tilde{\alpha}  \hat{n}(t) +  \tilde{\gamma} M + \tilde{\delta} \mathbb I \ee 
which can be solved straightforwardly to match the initial conditions $\hat{n}(0) = \hat{n}$ and $\partial_{t}\hat{n}(t) =-B$, assuming $\alpha \neq 0$,
\be \hat{n}(t) = \Big( \hat{n} + \frac{\tilde{\gamma} M + \tilde{\delta} }{\tilde{\alpha}} \Big) \cosh  \sqrt{\tilde{\alpha}} t - \frac{B}{\sqrt{\tilde{\alpha} }} \sinh \sqrt{\tilde{\alpha} } t - \frac{\tilde{\gamma} M + \tilde{\delta}}{\tilde{\alpha}}, \label{eq:Heisenberg_n_for_complexity_algebra}   \ee 
where all the operators on the right hand side are evaluated at $t=0$, and the time-dependence solely lies in the hyperbolic functions. The entire complexity distribution function therefore can be explicitly written in this case to be
\be  G(t, \phi) = \bra{0} \exp{i \phi \Bigg[  \Big( \hat{n} + \frac{\tilde{\gamma} M + \tilde{\delta} }{\tilde{\alpha}} \Big) \cosh  \sqrt{\tilde{\alpha}} t - \frac{B}{\sqrt{\tilde{\alpha} }} \sinh \sqrt{\tilde{\alpha} } t - \frac{\tilde{\gamma} M + \tilde{\delta}}{\tilde{\alpha}}  \Bigg] }  \ket{0}.  \label{eq:g_tphi_for_Complexity_Algebra} \ee
It is very convenient to compute $\expval{\hat{n}^k(t)}$ for small $k$ using \eqref{eq:g_tphi_for_Complexity_Algebra} because the problem reduces to computing the diagonal matrix elements of low-order polynomials of $M$, $B$ and $\hat{n}$ in the state $\ket{0}$. For $k=1, 2$ 
\be \expval{\hat{n}(t)} = \frac{\tg a_0 + \td}{\ta} \Big( \cosh(\sqrt{\ta} t ) -1 \Big), \label{eq:algebra_gen_complexity}\ee 
\be  \expval{\hat{n}(t)^2} = (\cosh \sqrt{\ta}t -1)^2 \frac{(\tg a_0 + \td )^2 + \tg^2 b^2_0}{\ta^2} + \frac{ \sinh^2 \sqrt{\ta}t}{\ta} b_0^2. \ee
The representation \eqref{eq:Heisenberg_n_for_complexity_algebra} is particularly useful, since this tells us that 
\be \hat{n}(t) \ket{0} = \frac{\tg a_0+ \td }{\ta}  \Big( \cosh(\sqrt{\ta} t ) -1 \Big)  \ket{0} +   \frac{b_0 \td}{\ta} \Big( \cosh(\sqrt{\ta} t ) -1 \Big) \ket{1} - \frac{i b_0  \sinh \sqrt{\ta} t}{\sqrt{\ta} }   \ket{1}. \label{eq:n_acting_on0} \ee
Using \eqref{eq:algebra_gen_complexity}, this confirms that in the case of \eqref{eq:n_acting_on0}, $\hat{n}(t) \ket{0} - \expval{\hat{n}(t)} \ket{0}$ is parallel to the vector $\ket{1}$. It also lets us read off
\be g_{t \phi} = \frac{b^2_0 \tg}{\ta} \Big( \cosh(\sqrt{\ta} t ) -1 \Big),~~~\partial_{t} \expval{\hat{n}(t)} = \frac{2 b^2_0}{\sqrt{\ta}} \sinh \sqrt{\ta} t. \ee
\eqref{eq:Heisenberg_n_for_complexity_algebra} is even simpler as $\ta=0$, and its solution simply becomes 
\be \hat{n}(t) = \frac{1}{2} ( \td M + \td) t^2  - B t. \ee 
Taking $\ta \rightarrow 0$ limit in all the equations above, recover the results for the matrix elements.
\subsection{$SU(2)$ algebra}
\begin{figure}
    \centering
    \includegraphics[width=0.8\linewidth]{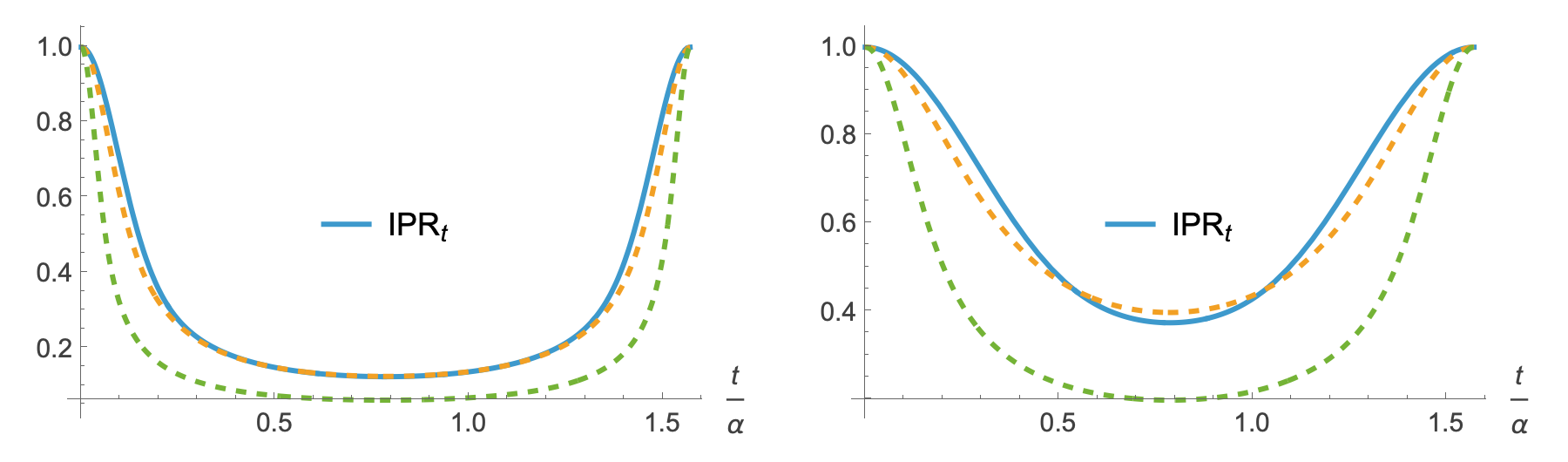}
    \caption{We plot $\text{IPR}_t$ in \eqref{eq:su2_IPR} and compare it against \eqref{eq:IPR_stronger_upperbound} (dashed orange) and \eqref{eq:IPR_upperbound_variance} (dashed green) for spin $j=10$ (left) and $j=1$ (right). On increasing $j$, rhs of \eqref{eq:IPR_stronger_upperbound} merges with $\text{IPR}_t$. }
    \label{fig:su2_ipr}
\end{figure}

For the $SU(2)$ complexity algebra we take $\ket{0}:= \ket{j,-j}$, where the latter denotes the highest weight state in the spin $j$ representation. Following the notation of \cite{Balasubramanian_2022}, we write $M$ as
\be M = \alpha ( J_{+} + J_{-})  +  \gamma J_0 + \delta \mathbb I,~~~ \text{where}~~~[J_0, J_{\pm}] = \pm J_{\pm},~~~[J_{+}, J_{-}] = 2 J_0. \ee 
Comparing with \eqref{eq:closure_ansatz_complexity}, we get  
\be \ta = -( \gamma^2 + 4 \alpha^2),~~~\tg = \gamma,~~~\td=\Big( j ( \gamma^2 + 4 \alpha^2) - \gamma \delta \Big),~~~a_0 = \delta - \gamma j,~~~ b_0 = \alpha \sqrt{2 j},~~~\ee 
and the Lanczos coefficients 
\be a_n = \gamma (n-j) + \delta,~~~ b_n = \alpha \sqrt{(n+1)(2j-n)},  \ee 
which gives the mean and the variance 
\be \expval{\hat{n}(t)} = \frac{8 j \alpha^2}{4\alpha^2+\gamma^2  } \sin^2 \Bigg(\sqrt{ 4\alpha^2+\gamma^2  } \frac{t}{2} \Bigg),~~~ (\Delta n(t) )^2 = \frac{ 8 j \alpha^2 \Bigg(\alpha ^2 \sin
   ^2\left(t \sqrt{4 \alpha
   ^2+\gamma ^2}\right) + \gamma ^2 \sin ^2\left(\frac{1}{2} t
   \sqrt{4 \alpha ^2+\gamma
   ^2}\right) \Bigg)}{ (4 \alpha ^2+\gamma ^2)^2}, \label{eq:su2_mean_variance}\ee 
and the off-diagonal metric component and volume element are given by 
\be g_{t\phi} = \frac{4 j \gamma  \alpha ^2   \sin
   ^2\left(\frac{1}{2} t \sqrt{4
   \alpha ^2+\gamma ^2}\right)}{4
   \alpha ^2+\gamma ^2},~~~ \det g = \frac{1}{4} (\partial_{t} \expval{\hat{n}(t)})^2 = \frac{4 \alpha ^4 j^2 \sin ^2\left(t \sqrt{4 \alpha ^2+\gamma
   ^2}\right)}{4 \alpha ^2+\gamma ^2}, \ee 
from which one finds that $R=\frac{4}{j}$. When $\gamma=0$, the metric is expressed in terms of familiar-looking coordinates on the sphere 
\be ds^2 = 2 \alpha^2 j dt^2 + \frac{1}{2}  j \sin^2 ( 2 \alpha t) d\phi^2. \ee 
We will use below the $\varphi_n(t)$ computed in \cite{Caputa:complexity_geometry}, restricting to $\gamma=0$ for simplicity.
\be \varphi_n(t)  = \tan(\alpha t)^{n} \cos^{2j}(\alpha t) \sqrt{ \binom{2 j }{n} } \label{eq:su2_wave_function}. \ee 
Using $\varphi_0(2t)= \cos(2 \alpha t)^{2j}$ and using \eqref{eq:su2_mean_variance}, shows that \eqref{eq:exponentiated_lower_bound_correlator} holds true on average, except for the zero of $\varphi_0(2t)$ around $t_k=\frac{(2k+1) \pi}{4 \alpha}$. The inverse participation ratio can be written as 
\be \text{IPR}_t = \cos ^{8 j}(\alpha  t) \, _2F_1\left(-2 j,-2 j;1;\tan ^4( \alpha t
   )\right) \geq \frac{\text{erf}\left(\sqrt{2} \pi  \sqrt{j \sin ^2(2 \alpha 
   t)}\right)}{2 \sqrt{2 \pi } \sqrt{j \sin ^2(2 \alpha  t)}}, \label{eq:su2_IPR}\ee 
meaning that it satisfies the inequality \eqref{eq:IPR_upperbound_variance} for all representations $j$. Meanwhile, it does not strictly bound the rhs of \eqref{eq:IPR_stronger_upperbound}, but is almost equal to it, see Fig. \ref{fig:su2_ipr}. 
\subsection{ $SL(2,R)$ algebra}
\label{sec:sl2r_algebra}
We will take $\ket{0} := \ket{h}$, the highest weight state of the representation labeled by $h$. It is an eigenvector of the shifted number operator $L_0 \ket{0} = h \ket{0}$. The raising and lowering operators are constructed using elements of $SL(2,R)$ such that the general $SL(2,R)$ Hamiltonian is 
\be M = \alpha( L_{-1} + L_{1}) + \gamma L_0 + \delta \mathbb I ~~~ \text{where}~~~[L_0, L_{-1}] = \pm L_{\mp 1},~~~[L_{1}, L_{-1}] = 2 L_0.  \label{eq:SL2R_Hamiltonian}  \ee 
and it follows that the Lanczos coefficients are
\be a_n = \gamma (h +n) + \delta,~~~ b_n = \alpha \sqrt{(n+1)(n+2 h)}. \label{eq:sl2_lanczos} \ee 
Comparing with \eqref{eq:closure_ansatz_complexity}, we get the parameters 
\be \ta = (4 \alpha^2- \gamma^2),~~~\tg = \gamma,~~~\td=\Big( h  (4 \alpha^2- \gamma^2) - \gamma \delta \Big),~~~\ee
and the Lanczos coefficients are 
\be \expval{\hat{n}(t)} = \frac{8 h \alpha^2 }{ 4 \alpha^2 - \gamma^2}  \sinh^2 \Bigg(\sqrt{ 4\alpha^2-\gamma^2  } \frac{t}{2} \Bigg),~~~(\Delta n(t) )^2 = \frac{8 \alpha ^2 h \left(\alpha ^2
   \sinh ^2\left(t \sqrt{4 \alpha
   ^2-\gamma ^2}\right)-\gamma ^2
   \sinh ^2\left(\frac{1}{2} t
   \sqrt{4 \alpha ^2-\gamma
   ^2}\right)\right)}{\left(\gamma
   ^2-4 \alpha ^2\right)^2}, \label{eq:sl2r_gphiphi} \ee 
and the off-diagonal metric component and volume element are given by   
\be g_{t\phi} = \frac{4 h \gamma  \alpha ^2  \sinh
   ^2\left(\frac{1}{2} t \sqrt{4
   \alpha ^2 - \gamma ^2}\right)}{4
   \alpha ^2 - \gamma ^2},~~~ \det g = \frac{1}{4} \partial_{t} \expval{\hat{n}(t)}^2 = \frac{4 h^2 \alpha^4 \sinh(t \sqrt{4 \alpha^2 - \gamma^2} )}{4 \alpha ^2 - \gamma ^2}, \label{eq:sl2r_off_diag} \ee 
which on evaluating curvature gives $R =-\frac{4}{h}$. Notice that complexity, and components of the metric are exponentially growing for $4 \alpha^2 > \gamma^2$ and oscillatory for $4 \alpha^2 < \gamma^2$, and power-law at the boundary $\gamma^2 = 4 \alpha^2$. The critical point $\gamma = 2\alpha$ corresponds to the closure of complexity algebra:
\be \mF =  \gamma M - \gamma \delta {\mathbb I}.   \ee 
For $\delta=0$ it is related to the Laguerre Polynomials with the orthogonality measure 
   \be \Phi(E) =  \bigg( \frac{2}{\gamma} \bigg)^h \frac{E^{2h-1} e^{-2 E /\gamma}}{\Gamma(2h)}.  \ee 
We can write the angular coordinate $\vartheta$ that diagonalizes the metric for $a_n \neq 0$, explicitly as 
\be \vartheta = \phi + \arctan \frac{\gamma  \tanh
   \left(\frac{1}{2} t \sqrt{4
   \alpha ^2-\gamma
   ^2}\right)}{\sqrt{4 \alpha
   ^2-\gamma ^2}}, \label{eq:vartheta_sl2} \ee 
which is valid for all values of $\alpha$ and $\gamma$. When $2\alpha>\gamma$, the argument of $\tanh$ in \eqref{eq:vartheta_sl2} is real and $\vartheta(t)$ asymptotes to a constant. In the $\gamma \rightarrow 2\alpha$ limit \eqref{eq:vartheta_sl2} reduces to $\phi + \arctan \alpha t$, and for $2\alpha < \gamma$, the right hand side becomes periodic in $t$.  
\subsubsection{Analysis for Toda orbit}
\label{sec:toda_orbit_example}
We will now analyze in more details a specific family of \eqref{eq:SL2R_Hamiltonian} to illustrate results in the main text. 
\be   \alpha = J \csc{2 J \tau}, \quad \gamma = 2 J \cot{2 J \tau}.  \label{eq:alpha_gamma_sl2_toda} \ee 
The important combination is $\sqrt{4 \alpha^2 -\gamma^2} = 2 J$. For the thermal solution of \eqref{eq:sl_2_psi0}, $J=\frac{\pi}{\beta}$. For reference, we explicitly write $\expval{\hat{n}(t)}$ , $( \Delta n(t) )^2$ and 
\be 
\begin{split}
    \expval{\hat{n}(t)} &= 2 h  \frac{\sinh ^2(J t)}{\sin(2 J \tau)^2}, \quad \partial_{t} \expval{\hat{n}(t)} = 2 h J \frac{\sinh(2 J t )}{\sin(2 J \tau)^2},  \\ 
( \Delta n(t) )^2 &= h  \frac{ \sinh ^2(J t) }{\sin(2 J \tau)^4} (\cosh (2 J t)-\cos (4 J
   \tau)). 
   \end{split}
\ee 
The wavefunctions $\varphi_n(t)$ were explicitly computed in \cite{Dymarsky_Toda, caputa_proper_momentum} and are given by 
\be  \varphi_n(t) = \sqrt{\frac{\Gamma (2 h+n)}{\Gamma (2 h) \Gamma (n+1)}} (-1)^n \Big( \frac{ i\sinh J (t-2 i \tau )}{\sin
   (2 J \tau )} \Big)^{-2 h} 
     \Big(\frac{\sinh J t}{\text{csch}(J (t-2 i \tau ))}
    \Big)^n. \label{eq:sl2_toda_wavefunctions} \ee 
Using this, the generating function can be summed up explicitly to be 
\be G(t, \phi) = \left(\frac{\cos (4 J \tau )-1}{2 e^{i \phi } \sinh ^2(Jt)+\cos (4 J \tau )-\cosh (2 J t)}\right)^{2 h}. \ee
It is convenient to write the inverse participation ratio as a hypergeometric series:
\be \text{IPR}_{t} =   \left(\frac{2\sin ^2(2 J \tau )}{\cosh (2 J t)-\cos (4 J
   \tau )}\right)^{4 h} \, _2F_1\left(2 h,2
   h;1;\frac{4 \sinh ^4(J t)}{(\cos (4 J \tau )-\cosh (2 J
   t))^2}\right). \label{eq:sl2r_IPR} \ee 
To examine the asymptotic behavior of $\text{IPR}_{t}$, it is useful to note that the argument of the hypergeometric series asymptotes to $1$ at large $t$, thus we can approximate the hypergeometric series by the leading singular term
\be \text{IPR}_{t} \overset{\tiny t \rightarrow \infty}{\longrightarrow} \frac{\Gamma(4 h-1 )}{\Gamma(2h)^2} \frac{2^{2-4 h} \sin ^2(2 J \tau )}{\cosh (2 J t)-\cos (4 J \tau
   )}, \ee 
where we assumed $h>\frac{1}{4}$, thus getting 
\be \lim_{t \rightarrow \infty } (2 \sqrt{\pi} \Delta n(t) ) \text{IPR}_{t} = \frac{\sqrt{\pi h} 2^{\frac{5}{2}-4 h}  \Gamma (4
   h-1)}{\Gamma (2 h)^2}, \label{eq:sl2r_IPR_variance_asymptote}  \ee 
and it can be checked using Stirling's approximation that \eqref{eq:sl2r_IPR_variance_asymptote} approaches $1$, on increasing $h$.
Notice that $\gamma = 0$ for $\tau_0=\frac{\pi}{4 J}$ and $\alpha=J$. This solution describe the Lanczos coefficients associated with the leading-order answer for SYK model with $p$-body interactions where $p=\frac{1}{h}$. The two-point function is 
\be C(t) = \frac{1}{\cosh(\alpha t )^{2h}}. \ee 
The mean and the variance of complexity is given by
\be \expval{n(t)} = 2 h \sinh^2(\alpha t) ~~~ \expval{n(t)^2} - \expval{n(t)}^2 = \frac{1}{2} h \sinh(2 \alpha t )^2, \label{eq:sl2_krylov_variance} \ee 
such that the hyperbolic space takes a more familiar form in $t,\phi$ coordinates: 
\be ds^2 = 2 h \alpha^2 dt^2 + \frac{1}{2} h \sinh(2 \alpha t )^2 d\phi^2.  \ee 
This family of solutions is still perfectly well-defined when $J$ takes imaginary values. In  \cite{caputa_proper_momentum},  $J$ was identified as $\frac{\pi}{\beta}$ for the thermal solution. Taking $J$ to be imaginary, physically corresponds to continuing $\beta= i L$, i.e quantizing the CFT on a cylinder of length $L$ such that the holographic dual describes global $AdS_3$. Because $4 \alpha^2 < \gamma^2$, the dynamics is oscillatory. The formulas for complexity moments and wavefunctions \eqref{eq:sl2_toda_wavefunctions} continue to hold with proper analytic continuation but the analysis for the time-dependence of $\text{IPR}_{t}$ changes. 

The $J \rightarrow 0$ limit corresponds to the vacuum, where $\alpha=\frac{1}{2 \tau}$ and $\gamma = \frac{1}{ \tau}$, giving 
\be a_n = \frac{n+h}{\tau}  \quad \text{and} \quad b_n = \frac{1}{2 \tau} \sqrt{(n+1) (n+2h)}.  \ee
\subsection{Heisenberg Algebra}
\label{sec:heisenberg_algebra}
Here we consider the Heisenberg algebra familiar from the harmonic oscillator $[a, a^{\dagger}]=1$. We restrict to the simple case where $M$ and $B$ are exactly like the position and momentum operator. 
\be M = \alpha \Big( a + a^{\dagger} \Big), \quad B = i \alpha  \Big( a^{\dagger} -a  \Big), \quad \mF = 2 \alpha^2 \mathbb I. \label{eq:heisenberg_weyl_algebra}    \ee 
 The time-evolving state $\ket{\psi(t)}$ is a coherent state of the harmonic oscillator. This gives
\be \varphi_n(t) = e^{-\alpha^2 t^2/2} \frac{\alpha^n t^n}{\sqrt{n!}}, \ee
which are just amplitudes for the coherent state in the number basis. Using the algebra \eqref{eq:heisenberg_weyl_algebra} with BCH relations trivially give 
\be G(t, \phi) = e^{-\alpha ^2 t^2 \left(1-e^{i \phi
   }\right)}, \label{eq:gt_for_Heisenberg}\ee
and list the first two moments and the variance
\be \expval{\hat{n}(t)} = \alpha^2 t^2 ,\quad \expval{\hat{n}^2(t)} =  \alpha^4 t^4+\alpha^2 t^2, \quad  (\Delta n(t))^2 =  \alpha^2 t^2. \label{eq:complexity_for_Heisenberg} \ee 
Since both the lhs and rhs of \eqref{eq:exponentiated_lower_bound_correlator} decay as a Gaussian, it is easy to verify \eqref{eq:exponentiated_bound}. The IPR can be written as 
\be \text{IPR}_{t} = e^{-2 \alpha ^2 t^2} I_0\left(2 t^2
   \alpha ^2\right) \geq  \frac{\text{erf}\left( \pi 
  \alpha t\right)}{2
   \sqrt{\pi } \alpha t} , \ee 
showing that the stronger bound \eqref{eq:IPR_stronger_upperbound} holds. Using the asymptotic  $I_0(x) \sim \frac{e^x}{\sqrt{2 \pi x }}$ we  we see that for large $t$, \eqref{eq:IPR_stronger_upperbound} becomes an equality for dynamics governed by Heisenberg algebra. 
Under $\tau$-evolution of the initial state,  $b_n$-remain invariant but $a_n = -2\alpha^2 \tau$. 
\section{Models used in Numerics}
\subsection{Double-Scaled SYK}
\label{sec:dssyk}
The SYK model is specified by the disordered Hamiltonian 
\be  H= i^{\frac{p}{2}} \sum_{1 \leq  j_1 < j_2 <  \ldots < j_p \leq \textbf{N}} J_{j_1 j_2 \ldots j_p}  \psi^{j_1} \psi^{j_2} \ldots \psi^{j_p}, \ee 
where $\psi^{k}$ denotes Majorana fermions at site $k$, and $J_{j_1, j_2 \ldots j_p}$ is a Gaussian random variable specifying the $p-$body interaction with the ensemble averaged first and second moments  
\be \expval{J_{j_1 j_2 \ldots j_p} } = 0, \quad  \expval{J^2_{j_1 j_2 \ldots j_p}} = \frac{\mathcal{J}^2}{ \binom{\mathbf{N}}{p}} \frac{1}{\lambda_{\qq}}, \quad \lambda_{\qq} =\frac{2p^2}{\mathbf{N}}.\ee 
We are interested in the double-scaling limit $\mathbf{N} \rightarrow \infty$, $p \rightarrow \infty$, with the ratio $\lambda_{\qq}$ held fixed and define
\be  \qq := e^{-\lambda_{\qq}}. \ee 
Results for finite $p$-SYK can in some cases be recovered on sending $\qq \rightarrow 1^{-}$. The double-scaled limit is of particular interest to us because of its ensemble-averaged density of states, which we denote by $\Phi^{\qq}(E)$. For $\qq$ strictly smaller than $1$, the range of energies of the ensemble-averaged theory is parametrized by
\be E(\theta) = \frac{2 \sJ \cos(\theta) }{\sqrt{\lambda_{\qq} (1-\qq)}},  \label{eq:dssyk_energies_theta}  \ee 
from which we see that $\Phi^{\qq}$ is compactly supported on the interval $[-E_0, E_0]$ with $E_0=\frac{2 \sJ}{\sqrt{\lambda_{\qq} (1-\qq)}}$. The choice $2\sJ=\sqrt{\lambda_{\qq}}$ is especially convenient for our purpose. With this choice,
$\Phi^{\qq}(E)$ continuously interpolates between the semicircle density of states \eqref{eq:rmt_Phi} at $\qq=0$, and approaches a normal distribution as $\qq  \rightarrow 1^{-}$. Both these limits reflect generic behavior across a variety of systems. With this normalization, the variance of $\Phi^{\qq}(E)$ is rigidly fixed to be $\frac{1}{4}$ as $\qq$ is varied.

We are interested in the Lanczos coefficients associated with the ensemble-averaged spectral form factor amplitude $C(t) = \tr( e^{-i H t} )$. This is equivalent to the survival amplitude on the doubled-space given by $C(t) = \bra{\beta=0}e^{-i (H_L + H_R) t/2}\ket{\beta=0}$, where $\ket{\beta=0}$ is the infinite temperature thermofield double-state. 
The associated Lanczos coefficients are
\be b_n = \frac{1}{2} \sqrt{\frac{1-\qq^{n+1}}{1-\qq}} \quad \text{and} \quad a_n=0 \label{eq:dssyk_Lanczos}. \ee 
$\lim_{n\rightarrow \infty} b_n = \frac{1}{2} \sqrt{\frac{1}{1-\qq}}$. Setting $\qq=0$, gives the constant Lanczos coefficients $b_n = \frac{1}{2}$, for which  
\be \varphi_n(t) = \frac{2(n+1)}{t} J_{n+1}(t), \label{eq:rmt_varphi}\ee
in terms of Bessel functions $J_n(t)$.  While the choice $2\sJ=\sqrt{\lambda_{\qq}}$ is convenient for our purposes, care should be exercised in translating results back to the physical SYK model. As long as $\lambda_{\qq}$ is a finite positive number, this is just setting the unit of time. For discussing the semiclassical $\lambda_{\qq} \rightarrow 0$ limit, we restore $\sJ$ in the normalization of the Lanczos coefficients \eqref{eq:physical_normalization_dssyk}. 

The form of $\Phi^{\qq}$ simplifies when using the $\theta$ parametrization from \eqref{eq:dssyk_energies_theta}, in terms of which 
\be \Phi^{\qq}(\theta) = \frac{(\qq;\qq)_{\infty } \left(e^{-2 i \theta }; \qq\right){}_{\infty
   } \left(e^{2 i \theta };\qq\right){}_{\infty }}{2 \pi }, \ee
such that $\int^{\pi}_{0} \Phi^{\qq}(\theta) d\theta =1.$   
\subsection{Finite Ising spin chain}
For our numerics we used the finite-sized nonintegrable Ising spin chain with the parameters
 \be H=-\sum^{i=6}_{i=1} \sigma^x_{i} \sigma^x_{i+1} + \sum^{7}_{i=1} (0.8 \sigma^{x}_{i} +1.05 \sigma^{z}_{i}), \label{eq:finite_ising_spin_chain} \ee  
in terms of the Pauli matrices $\sigma^{\alpha}_{i}$ at the site $i$. The Krylov space dimension is $N=16257$.
\section{Refined Dyck Paths}
\label{sec:refined_dyck}
Here we will describe a combinatorial method for evaluating $G(t,\phi)$. We take $a_n=0$. The evaluation of the even moments proceeds as 
\be \mu_{2k} =  \sum_{i_1, i_2, \ldots i_{2k}} \bra{0} L^{i_1}_{-} L^{i_2}_{+} \ldots L^{i_{2k}}_{+} \ket{0}, \ee 
where the nontrivial contributions only come from Dyck paths, where there are exactly $k$ factors of $L_{+}$ and $L_{-}$ in addition to the constraint that a string of $L_{-}$ does not annihilate the state on its right. These can be parametrized as 
\be \mu_{2k} = \sum_{h_{1}, h_2 \ldots h_{2k}} \prod^{j={2k-1}}_{j=1} b_{(1+h_i + h_{i+1} )/2},  \label{eq:sum_over_dyck}\ee
where the sum is over all $h_i$ satisfying 
\be h_1 = 0 = h_{2k},~~~h_{i} \geq 0 ~~~ \& ~~~ h_{i + 1} = h_i \pm 1. \label{eq:dyck_parametrization} \ee
The goal here is to show how a similar computation of $G(t,\phi) = \bra{0} e^{iM(0) t} e^{-iM(\phi) t} \ket{0} =\bra{0} e^{iM(-\phi/2) t} e^{-iM(\phi/2) t} \ket{0} $ is accomplished using the correlators
\be \mu_{\kb, \kf} = \bra{0} M(-\phi/2)^{\kb} M(\phi/2)^{\kf} \ket{0}, \label{eq:mu_kb_kf}\ee
in terms of which 
\be G(t, \phi) = \sum_{\kb, \kf}  \mu_{\kb, \kf} (-i)^{\kf - \kb } \frac{t^{\kf + \kb}}{\kf ! \kb !} \label{eq:gt_phi_in_terms_of_mu_forwards_backwards} \ee 
We used $\phi$-invariance of $\ket{0}$ to rewrite $G(t,\phi)$ in terms of $M(-\phi/2)$ and $M(\phi/2)$, to make the counting problem more symmetric across the ``forward'' and ``backward'' paths labeled by $k_f$ and $k_b$. During the forward branch, since $M(\phi/2) = L_{+} e^{i \phi/2} + L_{-} e^{-i\phi/2}$, each application of $L_{+}$ comes with a factor of $e^{i \phi/2}$ and $L_{-}$ comes with  $e^{-i \phi/2}$. The phases are reversed for $M(-\phi/2)$. 

The ladder operators still enforce a sum over Dyck paths of the type \eqref{eq:dyck_parametrization}, but with degeneracy lifted by powers of $e^{i \phi/2 }$. Suppose the number of $L_{+}$ contributing on the forward branch is called $k^u_f$, and the number of $L_{-}$, $k^d_f$. Then we are counting paths weighted with the phase $e^{i \phi (k^u_f -k^d_f)}$, where we used that a similar decomposition for $k_b = k^{u}_{b} + k^{d}_{b}$ has to obey the overall constraint $k^u_f -k^d_f = -(k^u_b -k^d_b)$. 

This is equivalent to cutting open a Dyck path of length $k_{f} + k_b$ and summing over the intermediate height $r=(k^u_f -k^d_f)$ with a weight $e^{i r \phi}$--allowing a representation of $ \mu_{\kb, \kf}$ along the lines of \eqref{eq:sum_over_dyck}. The forward and backward segment each define what is called a ``ballot path'' of height $r$.  With constant weight, there are exactly $\frac{(r+1) }{k_f+1} \binom{k_f+1}{(\kf-r)/2}$ such segments with height $r$ of the same parity as $k_f$, which must also match the parity of $k_b$. Choosing $b=\frac{1}{2}$, we get \be 
\mu_{\kb,\kf}=  \frac{1}{2^{\kf + \kb}}\sum^{\min(\kb, \kf )}_{r \equiv \kf \equiv \kb  \text{mod} 2 } e^{i  \phi r } \frac{(r+1)^2
   }{(\kb+1) (\kf+1) } \binom{\kb+1}{\frac{1}{2}
   (\kb-r)}
   \binom{\kf+1}{\frac{1}{2}
   (\kf-r) }. \label{eq:constant_b_colored_mu} \ee
For this specific case, it is far simpler to write $G(t, \phi) = \sum^{\infty}_{n=0} e^{i n \phi} (2J_{n+1}(t)/t)^2 $ in terms of the Bessel $J_n(t)$ than to use \eqref{eq:gt_phi_in_terms_of_mu_forwards_backwards} in terms of \eqref{eq:constant_b_colored_mu}. The goal would be to develop a path integral technique along the lines of \cite{Avdoshkin_2020} that selects the dominant contribution to $\mu_{\kb, \kf}$ in presence of the weights $e^{\sum_{n  \in \text{Path}} \log b_n}$. The challenging step seems to be handling the delicate phase cancellations while performing the sum over $k_b$ and $k_f$, that arises from working with real time $t$. Such an approach might be better suited for computing complexity distribution for euclidean time-evolution. 
\end{document}